\documentclass[12pt]{iopart}

\usepackage{amssymb,a4}
\usepackage{graphicx}

\usepackage[left=2cm,top=2cm,right=2cm,bottom=2cm]{geometry}
\usepackage{color}

\begin{document}

\newcommand{\red}{\color{red}}
\newcommand{\blue}{\color{blue}}
\definecolor{darkgreen}{rgb}{0,0.8,0.2}
\newcommand{\green}{\color{darkgreen}}
\newcommand{\todo}[1]{\textbf{To do: #1}}
\newcommand{\be}{\begin{equation}}
\newcommand{\ee}{\end{equation}}
\newcommand{\beq}{\begin{equation}}
\newcommand{\eeq}{\end{equation}}
\newcommand{\bea}{\begin{eqnarray}}
\newcommand{\eea}{\end{eqnarray}}
\newcommand{\rar}{\rightarrow}
\newcommand{\lar}{\leftarrow}
\newcommand{\ra}{\right\rangle}
\newcommand{\la}{\left\langle }
\renewcommand{\d}{{\rm d }}
\newcommand{\m}{{\tilde m }}
\newcommand{\p}{\partial}
\newcommand{\nn}{\nonumber }
\newcommand{\ci}{\mathrm{i}}

\newcommand{\fig}[2]{\includegraphics[width=#1]{./figures/#2}}
\newcommand{\Fig}[1]{\includegraphics[width=7cm]{./figures/#1}}
\newlength{\bilderlength}
\newcommand{\bilderscale}{0.35}
\newcommand{\storebilderscale}{\bilderscale}
\newcommand{\bilderskip}{\hspace*{0.8ex}}
\newcommand{\textdiagram}[1]{\renewcommand{\bilderscale}{0.2}\diagram{#1}\renewcommand{\bilderscale}{\storebilderscale}}
\newcommand{\vardiagram}[2]{\newcommand{\bilderscale}{#1}\diagram{#2}\renewcommand{\bilderscale}{\storebilderscale}}
\newcommand{\diagram}[1]{\settowidth{\bilderlength}{\bilderskip\includegraphics[scale=\bilderscale]{./figures/#1}\bilderskip}\parbox{\bilderlength}{\bilderskip\includegraphics[scale=\bilderscale]{./figures/#1}\bilderskip}}
\newcommand{\Diagram}[1]{\settowidth{\bilderlength}{\includegraphics[scale=\bilderscale]{./figures/#1}}\parbox{\bilderlength}{\includegraphics[scale=\bilderscale]{./figures/#1}}}
\bibliographystyle{KAY}

\newcommand{\atanh}
{\operatorname{atanh}}

\newcommand{\ArcTan}
{\operatorname{ArcTan}}

\newcommand{\ArcCoth}
{\operatorname{ArcCoth}}

\newcommand{\Erf}
{\operatorname{Erf}}

\newcommand{\Erfi}
{\operatorname{Erfi}}

\newcommand{\Ei}
{\operatorname{Ei}}

\newcommand{\sgn}{{\mathrm{sgn}}}
\def\e{\epsilon}
\def\l{\lambda}
\def\d{\delta}
\def\o{\omega}
\def\cb{\bar{c}}
\def\Li{{\rm Li}}

\title[Avalanches in the BFM]{Spatial shape of avalanches in the Brownian force model}

\author{Thimoth\'ee Thiery, Pierre Le Doussal, Kay J\"org Wiese}

\address{CNRS-Laboratoire de Physique Th\'eorique de l'Ecole Normale Sup\'erieure\\
24 rue Lhomond, 75231 Paris
Cedex-France}

\date{\today}

\begin{abstract}We study the Brownian force model (BFM), a solvable model of
avalanche statistics for an interface, 
in a general discrete setting. The BFM describes the overdamped
motion of elastically coupled
particles driven by a parabolic well in independent Brownian
force landscapes. 
Avalanches are defined as the collective jump of the particles
in response to an arbitrary
monotonous change in the well position  (i.e.\ in the applied
force). We derive an exact formula 
for the joint probability distribution of these jumps. From it
we obtain the joint density of  local avalanche sizes for
  stationary driving in the quasi-static limit near the depinning
threshold. 
A saddle-point analysis predicts the spatial shape of avalanches
in the limit of large aspect ratios for 
the continuum version of the model. We then study
fluctuations around this saddle point, and obtain the leading corrections to 
the mean shape, the fluctuations around the mean
shape and the shape asymmetry, for finite aspect ratios. Our results are finally  confronted
to numerical simulations.
\end{abstract}

\maketitle

\newpage

\section{Introduction}
A large number of phenomena, as diverse as the motion of domain
walls in soft magnets, fluid contact lines on rough surfaces,
or strike-slip faults in geophysics, have been described by the
model of an elastic interface in a disordered medium \cite{ZapperiCizeauDurinStanley1998,LeDoussalWieseMoulinetRolley2009,DSFisher1998}.
A prominent feature of these systems is that their response to
external driving is not smooth, but proceeds discontinuously
by jumps called ``avalanches". As a consequence of this ubiquitousness,
much effort has been devoted to the study of avalanches, both
from a theoretical and an experimental point of view \cite{BonamySantucciPonson2008,PapanikolaouBohnSommerDurinZapperiSethna2011,DahmenSethna1996,LeDoussalWiese2008c}.
Despite this activity, there are  few exact results for realistic
models of elastic interfaces in random media.

An exactly solvable model for a single degree of freedom, representing
the center of mass of an interface, was
proposed by Alessandro, Beatrice, ~Bertotti  and~Montorsi (ABBM)
\cite{AlessandroBeatriceBertottiMontorsi1990,AlessandroBeatriceBertottiMontorsi1990b}
on a phenomenological basis in the context of magnetic noise
experiments. It describes a particle driven in a Brownian random
force landscape. In \cite{ZapperiCizeauDurinStanley1998,Colaiori2008}
it was shown that for an elastic interface with 
infinite-ranged elastic couplings, the motion of the center of
mass has the same statistics as the ABBM model.

In this article, we study a multidimensional generalization of
the ABBM model, the Brownian force model (BFM). 
This model, introduced in \cite{LeDoussalWiese2011b,LeDoussalWiese2011a,DobrinevskiLeDoussalWiese2011b,LeDoussalWiese2012a},
was shown to provide the correct mean-field theory describing
the full space-time statistics 
of the velocity in a single avalanche for
$d$-dimensional realistic interfaces close to the depinning transition.
Remarkably, restricted to the dynamic of the center of mass, it
reproduces  the ABBM model. This mean-field
description is valid for an interface for $d \geq d_{{\rm uc}}
$ with $d_{{\rm uc}} = 4$ for short ranged elasticity and $d_{{\rm
uc}} = 2$ for long ranged elasticity.

As shown in 
\cite{DobrinevskiLeDoussalWiese2011b,LeDoussalWiese2012a} the
BFM has an exact ``solvability property'' 
in any dimension $d$.
It is thus a particularly interesting model to describe avalanche
statistics, even beyond
its mean-field applicability, i.e.\ for any dimension $d$ and
for arbitrary (monotonous) driving. 
It allows to calculate the statistics
of the spatial structure of avalanches, properties that the oversimplified
ABBM model cannot capture. In Ref. \cite{LeDoussalWiese2012a}
some finite wave-vector
observables were calculated, demonstrating an asymetry in the
 temporal shape.
Very recently the distribution of extension of an avalanche has
also been calculated \cite{DelormeInPrep}.

In this article we study a general discrete version of the BFM
model, i.e.\ $N$ points coupled by an elasticity
matrix in a random medium, as well as its continuum limit. In
the discrete model
each point experiences jumps $S_i$ upon driving.
We derive an exact formula for the joint probability distribution
function (PDF) $P[\{S_i\}]$ of the jumps $S_i$ (the local avalanche
sizes)
for an arbitrary elasticity matrix. In the limit of small driving
this yields a formula for the 
joint density $\rho[\{S_i\}]$ of local sizes for quasi-static
stationary driving near the depinning threshold. This allows
us to discuss the ``infinite divisibility property'' of the BFM
avalanche process.
The obtained results are rather general and
contain  the full statistics of the spatial structure of  avalanches. They are, however, difficult
to analyze in general since they contain many variables, and thus require computing marginals (i.e.\ probabilities where one has integrated over most of the variables)
from a joint distribution. This is accomplished here in
detail for the fully-connected model.  We find that in the limit of large $N$ 
there exist two interesting regimes. The first one corresponds to the 
usual picture from mean-field depinning models \cite{DSFisher1998,DSFisher1985},
whereas the second one is novel and highlights the intermittent nature of the avalanche motion. 

We then  analyze the shape of avalanches, first in a discrete setting by considering  few degrees of freedom. The probability exhibits an interesting saddle-point structure in  phase space. We then study the continuum limit of the model.  We find that the spatial shape of avalanches of fixed total size $S$ and extension $\ell$,
becomes, in the limit of a large aspect ratio $S/\ell^4$,
dominated by a saddle point. As a result, the avalanche shape becomes deterministic, up to small
fluctuations, which vanish in that limit. 
We calculate the optimal shape of these avalanches. We then analyze the fluctuations around the saddle point. This allows us not only to quantify the shape fluctuations seen in numerical experiments, but also to obtain the mean shape for avalanches with smaller aspect ratios. We test our results with large-scale numerical simulations. While our  results are obtained
in the special case of an elastic line with local elasticity ($d=1$)  the method can be extended 
to other dimensions $d$ and more general elasticity. Finally, we  discuss the applicability of our results to  avalanches in realistic, short-ranged correlated disorder. 
The outline of this article is as follows: Section \ref{secBFM} recalls the definition of the BFM model, which is 
first studied in a discrete setting with general, non-stationary driving. The  results of \cite{LeDoussalWiese2011a,DobrinevskiLeDoussalWiese2011b,LeDoussalWiese2012a}
 allow us to obtain the Laplace transform of the PDF of local avalanches sizes. Section \ref{secMain} contains the derivation of the main result: the full probability distribution of the local avalanche sizes. Section \ref{quasi} focuses on the limit of small driving, and how to obtain the avalanche density. Section \ref{secfullyc} contains a detailed analysis of the fully-connected model. Section \ref{discrete} studies  avalanche shapes for interfaces with a few degrees of freedom. Section \ref{secContinuum} contains one important application of our result,
  namely the  deterministic  shape of avalanches with large aspect ratio for an elastic line. Section \ref{s:fluctuations} analyses the fluctuations around this optimal shape.  Section \ref{sec-appli}  discusses the application of our results to short-ranged disorder and quasi-static driving. A series of appendices contains details, numerical verifications and some adjunct results. In particular, in \ref{app-kolmo}, we introduce an alternative method, based on backward Kolmogorov techniques, 
to calculate the joint local avalanche-size  distribution, following a kick in the driving. 

\section{The Brownian force model}\label{secBFM}

\subsection{Model}
We study the over-damped equation of motion in continuous time $t$ of an ``interface", consisting of 
$N$ points with positions $u_{it} \in \mathbb{R}$, $i=1, \dots, N$.
Each point feels a static random force $F_i(u_{it})$ and is elastically coupled to the other points by a time-independent symmetric elasticity matrix $c_{ij}$ with $\sum_{j=1}^Nc_{ij}=0$. Each particle is driven by an elastic spring of curvature $m^2$ centered at the time-dependent position $w_{it}$. The equation of motion reads
\begin{equation}\label{overdampedposition}
\eta \partial_t u_{it} =  \sum_{j=1}^N c_{ij} u_{jt} - m^2( u_{it}-w_{it} ) + F_i(u_{it})
\end{equation}
for $i=1 \dots N$. The $F_i(u)$ are $N$ independent Brownian motions (BM) with correlations
\begin{equation}\label{browniancorrel}
\overline{[F_i(u)-F_i(u')]^2} =  2 \sigma |u-u'| \quad , \quad  \overline{F_i(u)F_j(u')} = 0 \; \mbox{for} \; i\neq j
\end{equation}
and $\overline{F_i(u)}=0$;  the overline denotes the average over the random forces $F_i(u)$. For definiteness we consider \footnote{The model can also be studied in a stationary setting, see e.g. 
\cite{DobrinevskiLeDoussalWiese2011b,LeDoussalWiese2012a}.} a set of one-sided BMs with $u \geq 0$ and $F_i(0)=0$.

We furthermore suppose that (i) the driving is always non-negative: $\forall t,i$, $\dot{w}_{it} \geq 0,$
and (ii) the elastic energy is convex i.e.\ $c_{ij} >0$ for $i \neq j$. 
Under these assumptions, the Middleton theorem \cite{Middleton1992} guarantees that if all velocities are non-negative 
at some initial time: $\exists t_0 \in \mathbb{R} | \forall i$, $\dot u_{it_0} \geq0$, they remain so for all times: 
$\forall i, \forall t \geq t_0$, $\dot{u}_{it}\geq0$. 

\paragraph{Some explicit examples of elasticity matrices:}\label{elastmat}
Throughout the rest of this article, we  sometimes specify the elasticity matrix. The models studied are (where $c$ denotes the elastic coefficient):
\begin{enumerate}
\item The fully connected model: $c_{ij} = c( \frac{1}{N}-  \delta_{ij})  $
\item The elastic line with short-range (SR) elasticity and periodic boundary contitions (PBCs) $c_{ij} =c\left( \delta_{i , j-1} + \delta_{i-1 , j} - 2 \delta_{ij} \right)$ with $i+N \equiv i$
\item The elastic line with SR elasticity and free boundary conditions: 

$c_{ij} =c\left[ \delta_{i , j-1} + \delta_{i-1 , j} - \delta_{ij}(2 - \delta_{i1} -\delta_{iN} )  \right]$
\item The general $d$-dimensional elastic interface with PBCs, where $i \in \mathbb{Z}^d$ and 
$c_{ij}=c ( f(||i-j||) -  \delta_{ij} \sum_{j} f(||i-j||));$ here $||i-j||$ is the Euclidean distance in $\mathbb{Z}^d$
and $f(r)$   the elastic kernel. Long-ranged elasticity (LR) is usually
described by kernels such that $f(r) \sim r^{-(d+\alpha)}$ (i.e.\ $\sim q^\alpha$ in Fourier). 
\end{enumerate}

\subsection{Velocity Theory}
Supposing that we start at rest for $t=0$, $u_{i,t=0}=\dot u_{i,t=0}=0$, then it is more convenient (and equivalent) to study the evolution of the velocity field directly. The equation of motion reads
\begin{equation}\label{overdampedvelocity}
\eta \partial_t \dot{u}_{it} = \sum_{j=1}^N c_{ij} \dot{u}_{jt} - m^2( \dot{u}_{it}-\dot{w}_{it} ) + \sqrt{2 \sigma \dot{u}_{it}}\xi^i_t 
\ ,\end{equation}
where the $\xi^i_t$ are $N$ independent Gaussian white noises, with $\overline{ \xi^i_t \xi^j_{t'}} = \delta(t-t')\delta_{ij}$ 
and $\overline{ \xi^i_t} =0$. Equation (\ref{overdampedvelocity}) is taken in the It\^o sense. 
Note that  we replaced the original quenched noise $\partial_t F_i(u_{it})$ by an annealed one $\sqrt{2 \sigma \dot{u}_{it}}\xi^i_t$, making Eq.~(\ref{overdampedvelocity}) a closed equation for the velocity of the interface. The fact that (\ref{overdampedposition}) and (\ref{overdampedvelocity}) are equivalent (in the sense that disorder averaged observables are the same) is a non-trivial 
exact property of the BFM model. It was first noted for the ABBM model \cite{AlessandroBeatriceBertottiMontorsi1990,AlessandroBeatriceBertottiMontorsi1990b}
and extended to the BFM \cite{DobrinevskiLeDoussalWiese2011b,LeDoussalWiese2012a}. 
It originates from the time-change property of the Brownian motion
$\rmd B (f(t)) \equiv_{\rm in ~law} \sqrt{f'(t)} \rmd\tilde B(t)$ for increasing $f(t)=u_t$, valid  as a consequence of the
Middleton property $\dot u_t \geq 0$. A derivation of this property is recalled in \ref{appMSR}. 

\subsection{Avalanche-size observables}

In this article we focus on the calculation of avalanche-size observables defined in the following way. Starting from rest at $t=0$ as previously described, we apply a driving $w_{i t} \geq 0$ for $t >0$ during
a finite time interval such that $\int_0^{\infty} \rmd t\, \dot{w}_{it} = w_i$ (stopped driving protocol). 
In response to this driving, the points move and we define the local avalanche  size $S_i$ as $S_i = \int_0^{\infty} \rmd t\, \dot{u}_{it}$, that is the total displacement of each point. We adopt the vector notation
\bea
\vec S = (S_1, \dots ,S_N) \quad , \quad \vec w = (w_1, \dots, w_N)\ . 
\eea 
The $S_i$'s are random variables whose statistics is encoded in the Laplace transform, also called generating function $G(\vec\lambda)$, and defined as
\begin{equation}\label{generatingaval2}
G(\vec \lambda) = \overline{ e^{\vec \lambda \cdot \vec S} }
\ .\end{equation}
The BFM possesses a remarkable ``solvability property'' that allows us
to express this functional as \cite{DobrinevskiLeDoussalWiese2011b,LeDoussalWiese2012a}
\bea\label{generatingaval}
G(\vec \lambda) = \overline{ e^{\vec \lambda \cdot \vec S} } = e^{m^2 \sum_{i=1}^N \tilde{u}_i w_i}
\eea
in terms of the solution $\tilde{u}_i$ of the ``instanton" equation. The latter reads
\begin{equation}\label{instantonaval}
\lambda_{i} =- \sigma \tilde{u}_{i}^2 + m^2 \sum_{j=1}^N C_{ij} \tilde{u}_j \ ,
\end{equation}
where we have defined the dimensionless matrix
\begin{equation}\label{dimensionlesselast}
C_{ij}= \delta_{ij} - \frac{1}{m^2} c_{ij}
\ ,\end{equation}
which contains all elastic and massive terms in the instanton equation. 
The solution of Eq.~(\ref{instantonaval}) which enters into Eq.~(\ref{generatingaval}) 
is the unique set of variables $\tilde u_i$ continuous in $\lambda_j$
with the condition that all $\tilde u_i=0$ when all $\lambda_j=0$.
The derivation of this property is recalled in a discrete setting in \ref{appMSR}. 
The instanton equation thus allows us in principle to express the PDF $P(\vec S)$ of the local  avalanche sizes, as 
the inverse Laplace transform of $G(\vec \lambda)$. In the next section we obtain $P(\vec S)$ directly, without solving (\ref{instantonaval}), which admits no obvious closed-form solution. We will note $\langle \dots \rangle$ the average of a quantity with respect to the
probability $P$. Note that the PDF $P(\vec S)$ depends {\it only on the total driving} $w_i = \int_0^{\infty} \rmd t\, \dot{w}_{it}$ and
not on the detailed time-dependence of the $w_{it}$. This is a particularity of the BFM model.

\subsection{The ABBM model} \label{sec-abbm} 
Before going further into the calculation, let us recall the result of Ref. \cite{DobrinevskiLeDoussalWiese2011b,LeDoussalWiese2012a} 
that the statistical properties of the center of mass of the discrete BFM model  is equivalent to that of the ABBM model. To be precise, if we write 
the {\it total displacement} (i.e.\ swept area) ${\sf u}_t = \sum_i u_{it}$ and total drive ${\sf w}_t =\sum_i w_{it}$ then, in law, we have
\begin{equation}\label{ABBM}
\eta \partial_t \dot{{\sf u}}_{t} = - m^2( \dot{{\sf u}}_{t}-\dot{{\sf w}}_{t} ) + \sqrt{2 \sigma \dot{{\sf u}}_{t}}\xi_t\ .
\end{equation}
Here $\xi_t$ is a Gaussian white noise $\overline{ \xi_t \xi_{t'}} = \delta(t-t')$ and $\overline{ \xi_t} =0$. \footnote{Note that
this result uses $\sum_j c_{ij}=0$ and that 
the center of mass obeys the same equation with a noise scaled as $N^{-1/2}$ and driving by $N^{-1}$.}
This equivalence implies that the PDF of the total avalanche size $S = \int_{t=0}^{\infty} \rmd t ~ \dot{{\sf u}}_t = \sum_{i=1}^N S_i$ in the
discrete BFM model,
following an arbitrary stopped driving $\int_{0}^{\infty} \rmd t ~ \dot{{\sf w}}_t = {\sf w}$, is given by the avalanche-size PDF of the ABBM model
\cite{AlessandroBeatriceBertottiMontorsi1990,AlessandroBeatriceBertottiMontorsi1990b,DobrinevskiLeDoussalWiese2011b},
\bea \label{ABBMaval}
\!\! \!\! \!\! \!\! \!\! \!\! \!\! && P_{\rm ABBM}(S)=\frac{ {\sf w} }{2 \sqrt{\pi S_m} S^{\frac{3}{2}}} \exp\left( -\frac{ (S-{\sf w})^2}{4S S_m} \right) 
\quad , \quad ~~~~ S_m = \frac{\sigma}{m^4} \ .
\eea
Here $S_m$ is the large-scale cutoff for avalanche sizes induced by the mass term. This first result on a marginal
of the joint distribution $P(\vec S)$ will provide a useful check of our general formula obtained below for $N>1$.

\section{Derivation of the avalanche-size distribution in the BFM}\label{secMain}
For simplicity we now switch to dimensionless units. We define
\bea \label{dimless} 
v_i = \frac{\sigma}{m^2} \tilde{u}_i \quad , \quad \tilde{w_i} = \frac{w_i}{S_m} \quad , \quad \tilde{\lambda}_i = S_m \lambda_i
\quad , \quad \tilde{S}_i= \frac{S_i}{S_m}\ ,
\eea
where $S_m = \frac{\sigma}{m^4}$. The instanton equation (\ref{instantonaval}) now reads
\begin{equation}\label{instantonaval2}
\tilde{\lambda}_{i} =-v_{i}^2 + \sum_{j=1}^N C_{ij} v_j\ .
\end{equation}
The generating functional is given by 
\bea \label{laplaceG} 
G(\vec \lambda) =\tilde G(\vec{\tilde{\lambda}}) = \overline{e^{\sum_{i=1}^N \tilde{\lambda}_i \tilde{S}_i }} = e^{\sum_{i=1}^N v_i \tilde w_i}\ .
\eea 
In the following we drop the tildes on dimensionless quantities to lighten notations, and explicitly indicate when we restore units. For the ABBM model, it was possible to explicitly solve the instanton equation for the generating function $G(\lambda)$. The inverse Laplace transform was then computed, leading to (\ref{ABBMaval}). Here this route is hopeless because Eq.~(\ref{instantonaval2}) admits no simple closed-form solution. 
We instead compute directly the probability distribution $P(\vec S)$ using a change of variables in the inverse Laplace transform (ILT):
\begin{eqnarray} \label{ILT} 
\fl P( \vec S)  && = \left(\frac{1}{2\ci \pi}\right)^N \int_{\cal C} \rmd^N \vec \lambda \exp\left( - \vec \lambda \cdot \vec S \right)  G(\vec \lambda) \\
\fl && = \left(\frac{1}{2 \ci \pi}\right)^N \int_{-\ci\infty}^{\ci\infty} \rmd v_1 \cdots \int_{-\ci\infty}^{\ci\infty} \rmd v_N \det \left( \frac{\partial  \lambda_i}{\partial v_j} \right) \exp\left( - \sum_{i=1}^N ( -v_{i}^2 + \sum_{j=1}^N C_{ij} v_j)  S_i +\sum_{i=1}^N  v_i  w_i \right), \nonumber 
\end{eqnarray}
where ``$\ci$" denotes the imaginary unit number to avoid confusion with  indexes. The first formula is   
the ILT where 
we left unspecified the multi-dimensional contour of integration $\cal C$. In the
second line we used the expression of $\lambda_i$ in terms of $v_j$ from 
(\ref{instantonaval2}), as well as the dimensionless version of  (\ref{generatingaval}). 
Changing   variables from $\lambda_i$ to $v_j$, the contours of integration are chosen to obtain a convergent integral, see second
line of Eq.~(\ref{ILT}). This makes this derivation an educated guess, which however is verified in \ref{appcheckmainformula}. We also give another derivation for a special case in \ref{app-kolmo}. To pursue the derivation, the Jacobian is written using Grassmann variables as
\begin{equation}
\det \left( \frac{\partial  \lambda_i}{\partial v_j} \right) = \int \prod_{i=1}^N \rmd \psi_i \rmd \bar \psi_i \exp\left(  \sum_{i,j=1}^N \bar \psi_i (-2 v_i \delta_{ij} +  C_{ij} ) \psi_j \right)
\ .\end{equation}
Reorganizing the order of integrations and changing $v_i \to  \ci v_i$, we write
\begin{eqnarray}
\!\!\!\!\!\!\!\!\!\!\!\!\!\!\!\!  P(\vec S) =  \left(\frac{1}{2\pi}\right)^N \prod_{i=1}^N \int \rmd \psi_i \rmd \bar \psi_i \prod_{i=1}^N && \int_{\mathbb{R}} \rmd v_i   \exp\left(  - \sum_{i=1}^N ( v_{i}^2 + \sum_{j=1}^N \ci C_{ij}  v_j)  S_i \right .\nonumber \\
&& \left.  +\sum_{i=1}^N \ci v_i  w_i  + \sum_{i,j=1}^N \bar \psi_i (-2 \ci v_i \delta_{ij} +  C_{ij} ) \psi_j   \right) .
\end{eqnarray}
Integrating on $v_i$ leads to
\begin{eqnarray}
\!\!\!\!\!\!\!\!\!\!\!\!\!\!\!\!\!\!\!\!\!\!\!\!\!\!\!\!\!\!\!\!  \left(\frac{1}{2\pi}\right)^{\!N} \prod_{i=1}^N \int \rmd \psi_i \rmd \bar \psi_i (\pi)^{(N/2)}\left(\prod_{i=1}^N S_i \right) ^{\!\!-\frac{1}{2}} \exp && \left( -\frac{1}{4} \sum_{i=1}^N \frac{( w_i-2\bar{\psi_i}\psi_i-\sum_{j=1}^N C_{ij}  S_j)^2}{S_i} \right. \nonumber \\
 && \left.  +\sum_{i,j=1}^N \bar{\psi_i} C_{ij} \psi_j \right).
\end{eqnarray}
Finally, using $\psi_i^2 = \bar \psi_j^2 =0$, the integration over the Grassmann variables can be expressed as a determinant, leading to our main result 
\bea \label{mainform}
&& \! \! \! \! P( \vec S ) 
= \left(\frac{1}{2\sqrt{\pi}}\right)^{\!\!N} \left(\prod_{i=1}^N  S_i \right) ^{\!\!-\frac{1}{2}} \exp { \left(-\frac{1}{4} \sum_{i=1}^N \frac{( w_i-\sum_{j=1}^N  C_{ij}  S_j )^2}{ S_i} \right)} \det\left(  M_{ij}  \right)_{N \times N}  \\
&&  \! \! \! \! M_{ij} = C_{ij} + \delta_{ij} \frac{ w_i- \sum_{k=1}^N  C_{ik}S_k}{ S_i} \quad , \quad C_{ij} = \delta_{ij} -\frac{1}{m^2} c_{ij} \nonumber \ .
\eea
Here $c_{ij}$ is the elasticity matrix. This is the joint distribution expressed in dimensionless
units (\ref{dimless}). The expression in the original units is recovered by
substituting $S_i \to S_i/S_m$, $w_i \to w_i/S_m$ and $P \to S_m^N P$ in  (\ref{mainform})
while keeping $C_{ij}$ fixed \footnote{Note that this formula can be generalized to the case of site-dependent masses and disorder strengths, $m_i, \sigma_i$: the expression in the original units is 
obtained by the substitution $S_i \to S_i/S_m^i$, $w_i \to w_i/S_m^i$ and $P \to \prod_i S_m^i P$ in (\ref{mainform})
with $S_m^i = \frac{\sigma_i}{m_i^4}$ and $C_{ij} = \delta_{ij} - \frac{1}{m_i^2} \frac{ \sigma_i m_j^2}{\sigma_j m_i^2} c_{ij}$.}.

Note that for zero coupling, $c_{ij}=0$, Eq.~(\ref{mainform}) becomes $ P( \vec S ) = \prod_{i=1}^N  P_{\rm ABBM}( S_i)$: the different points are decoupled and one retrieves $N$ independent ABBM models. Non-trivial tests of the formula are performed in \ref{appcheckmainformula}.
One general property is that the average local size is $\langle S_i \rangle = \sum_{j=1}^N C^{-1}_{ij} w_j$. 
This average gives the shape of the interface in the large-driving limit. 
When $w_i \gg 1$ uniformly in $i$, it is easy to see by expansion of the above formula 
that $S_i = \langle S_i \rangle + O(\sqrt{w_i}) \eta_i$
where $\eta_i$ are (correlated) Gaussian random variables. 

We show in \ref{app-kolmo}, using different methods, that when the driving is in the form of kicks, $\dot{w}_{it} = w_i \delta(t)$
\footnote{This is sufficient, since we noted above that the result does not depend on the detailed time-dependence of the
driving.} 
$P( \vec S )$ satisfies the exact equation
\bea\label{kolmo}
\sum_{\alpha =1}^N  \left( - \frac{ \partial P}{\partial w_{\alpha} } \sum_{j=1}^N C_{\alpha j} w_j + \frac{ \partial^2 P}{\partial w_{\alpha}^2 } w_{\alpha} - w_{\alpha} \frac{ \partial P}{\partial S_{\alpha} } \right) =0\ .
\eea
We also show that (\ref{mainform}) solves this equation. This alternative derivation support our result (\ref{mainform}) ans shed some light on its structure.
\bigskip

{\it Interpretation:} Some features of our main result can be understood as follows. Consider the equation
of motion (\ref{overdampedvelocity}). Upon integration from $t=0$ to $t=\infty$ we obtain

\begin{equation}
0 = \sum_{j=1}^N c_{ij} S_{j} - m^2(S_{i}-w_{i} ) + \int_0^{\infty} \rmd t \sqrt{2 \sigma \dot{u}_{it}}\xi^i_t \ .
\end{equation}
If we could replace the sum of white noises by a gaussian random variable
\bea \label{rep}
\int_0^{\infty} \rmd t \sqrt{2 \sigma \dot{u}_{it}}\xi^i_t  \to \sqrt{ 2 \sigma \int_0^{\infty} \rmd t \dot{u}_{it} } ~ \Xi_i 
= \sqrt{ 2 \sigma S_i} ~ \Xi_i  \ ,
\eea 
then we would obtain (\ref{mainform}), but with a slightly different determinant 
given by the replacement $\delta_{ij} \to \frac{1}{2} \delta_{ij}$ in $M_{ij}$ in (\ref{mainform}). 
However, the replacement (\ref{rep}) is not legitimate because the variables $\dot u_{it}$ are correlated in time. The determinant in (\ref{mainform}) takes care of that correlation.

\paragraph{Probability distribution of the shape} Even if it is far from being obvious on Eq.~(\ref{mainform}), we know from 
Section \ref{sec-abbm} that the probability distribution of $S = \sum_{i=1}^N S_i$ is given by (\ref{ABBMaval}) with ${\sf w} = \sum_{i=1}^N w_i$. This allows us to define the probability distribution of the shape of an avalanche, given its total size $S$: Consider $s_1, \ldots , s_{N}  \in [0,1]$ with $s_{N} = 1 - \sum_{i=1}^{N-1} s_i$, such that $S_i = S s_i$. The probability distribution of the $s_i$ variables, given that the avalanche has a total size $S=\sum_{i=1}^N S_i$ is
\begin{equation}\label{probashape}
P(\vec s | S) = 2 \sqrt{\pi} \frac{S^{N+\frac{1}{2} } }{ \sf{w} }   \exp\left( \frac{ (S-\sf{w})^2}{4S} \right) P( S \vec s)  \quad , \quad \sum_{i=1}^N s_i = 1\ .
\end{equation}

\section{Avalanche densities and quasi-static limit} \label{quasi}

The goal of this section is to define and calculate avalanche densities. These
allow us to describe the intermittent motion of the interface in the 
regime of small driving, $w_i$ small. The dependence of the PDF, $P_{\vec w}(\vec S)$, on the driving is denoted by a subscript $\vec w$. We first study the jumps of the center of mass 
described by the ABBM model. 

\subsection{Center of mass: ABBM}
For the ABBM model (and for the total size $S=\sum_{i=1}^N S_i$ in the BFM model) the avalanche-size PDF is given by
\begin{equation}
P_{\sf w}(S)=\frac{ \sf w }{2 \sqrt{\pi} S^{\frac{3}{2}}} \exp\left( -\frac{ (S-\sf w)^2}{4S} \right)
\ ,\end{equation}
where ${\sf w}=\sum_{i=1}^N  w_i$ is the total driving. The limit of small driving $\sf w$ is very non-uniform. In the sense of distributions, its limit is a delta distribution at $S=0$,
\bea
P_{\sf w}(S) \to_{\sf w \to 0} \delta(S) \ .
\eea
However, this hides a richer picture and a separation of scales between typical small avalanches $S \sim \sf w^2$ and
rare large ones $S \sim 1$. If one defines $S={\sf w^2} s$, the PDF of $s$ has a well-defined $\sf w \to 0$ limit given by
\begin{equation}\label{smallscalecutoff}
p_{0}(s)=\frac{1}{2 \sqrt{\pi} s^{\frac{3}{2}}} \exp\left( -\frac{ 1}{4 s} \right) \ ,
\end{equation}
which is indeed normalized to unity $\int \rmd s\, p_0 (s) = 1$. Hence avalanches
of sizes $S \sim \sf w^2$ are typical ones. However, all positive integer moments of $p_0$ are infinite. This indicates that these small avalanches, though typical, do not contribute to the moments of $P_{{\sf w}}$, which are finite and controlled by rare but much larger
avalanches which we now analyze. In the limit of small $\sf w$, there remains a probability of order $\sf w$ to observe an avalanche of order $1$. For fixed $S=O(1) \gg\sf w^2$ one has
\begin{equation} \label{defrho}
P_{\sf w}(S) = {\sf w} \rho(S) + O({\sf w}^2)  \quad , \quad \rho(S) 
= \frac{1}{2 \sqrt{\pi} S^{\frac{3}{2}}} \exp\left( -\frac{ S}{4} \right) \ .
\end{equation}
This defines the density (per unit $\sf w$) of avalanches. These are the ``main" avalanches with $S \gg \sf w^2$, which are also called ``quasi-static" avalanches (see below and Section \ref{sec-appli}). 
The density is not normalizable because of the divergence at small $S$, but all its integer moments are finite and contain all the weight in 
that limit, i.e.\ $\langle S^n \rangle = {\sf w} \int \rmd S \rho(S) S^n + O({\sf w}^2)$. In particular, $\langle S \rangle ={\sf w}$ implies
$\int \rmd S \rho(S) S=1$.

We now show that the avalanche density contains more information and controls the moments even for finite ${\sf w}$, a property that follows as a consequence of $P_w(S)$ being the PDF of an infinitely divisible process. This is best seen on its Laplace transform
\bea \label{lapabbm}
G_{\sf w}(\lambda) = \int \rmd S e^{\lambda S} P_{\sf w}(S)  = e^{\sf w Z(\lambda)} \quad , \quad Z(\lambda) = \frac{1}{2} (1 - \sqrt{1-4 \lambda}) \ .
\eea
The ``infinite-divisibility property'' indeed follows: $ \forall m$ and $\forall \sf w = \sf w_1 + \cdots + \sf w_m$ such that $\sf w_i >0$
\bea\label{infinitedivisibilityABBM}
G_{\sf w}(\lambda) = \prod_{i=1}^m G_{\sf w_i}(\lambda)  \quad , \quad  P_{\sf w}(S) = \left(P_{\sf w_1} \ast \cdots \ast P_{\sf w_m} \right)(S) \ ,
\eea
where $\ast$ denotes the convolution operation. Hence $S$ is a sum of $m$ independent random variables for all $m$. The ABBM avalanche process can thus be interpreted as a Poisson-type jump 
process (a Levy process) with jump density $\rho(S)$ \cite{duchon}. In general the density can  be defined as
$\rho(S) = \frac{d P_{\sf w}(S)}{d \sf w} |_{\sf w=0}$ for fixed $S>0$ (i.e.\ it does not hold in the
sense of distributions), and the relation between $Z(\lambda) :=\frac{d G_{\sf w}(\lambda)}{d \sf w} |_{\sf w=0} $ and $\rho$ is
\bea \label{Zrho} 
Z(\lambda) = \int \rmd S (e^{\lambda S} -1) \rho(S)\ .
\eea
The $-1$ takes care of the divergence at small $S$. This allows us to write the relation between $P_{\sf w}$ and $\rho$, expanding (\ref{lapabbm}) in powers of $\sf w$, as
\bea \label{decomp1}
 \int \rmd S e^{\lambda S} P_{\sf w}(S) = \sum_{n =0} ^{\infty} \frac{{\sf w}^n}{n !} \int \rmd s_1 \cdots \rmd s_n ( e^{\lambda s_1} -1) \cdots ( e^{\lambda s_n} -1)\rho(s_1) \cdots \rho(s_n)\ . 
\eea
Taking derivatives w.r.t.\ $\lambda$, this decomposition shows that the (positive integer) moments of $P_{\sf w}$ are 
entirely controlled by $\rho$, for arbitrary fixed ${\sf w}$ (beyond the small-${\sf w}$ limit). 
In this sum the term of order ${\sf w}^n$ can be interpreted as the contribution to the total displacement $S$ of the
interface (after a total driving ${\sf w}$) of a $n$-avalanche (quasi-static avalanche) event (of order $O(1)$).  
The convolution structure in (\ref{decomp1}) shows that these events are statistically independent 
in the ABBM model. In this model however, this interpretation only holds at the level of  moments. 
The accumulation of infinitesimal jumps, manifest in the non-normalizable divergence of
$\rho$ at small $S$ prevents us to extend this interpretation 
to the probability itself, see \ref{app-levy} for a discussion. 

\subsection{BFM}
In the BFM, ``the infinite-divisibility property'' of the avalanche process is even richer, since avalanches occur at different positions
along the interface. Let us define the $j$-th ``elementary" driving which applies only to site $j$, 
i.e.\ $w_i = w_j \delta_{ij}$, and denote the corresponding size-PDF as $P_{w_j}(\vec S)$. 
Consider now the PDF for the general driving, $P_{\vec w }(\vec S)$. From the 
structure of its LT, see (\ref{laplaceG}), as
a product of exponential factors linear in the $w_i$, this PDF can
be written as a convolution for $\vec w=(w_1,...,w_N)$,
\bea
\!\!\!\!\!\!\!\!\!\!\!\!\!\!\!\!\! P_{\vec w }(\vec S) = P_{w_1}(\vec S) \ast \cdots \ast  P_{w_N}(\vec S)\ . 
\eea
An avalanche in the BFM can thus be understood as a superposition of $N$ avalanches {\it independently generated} by each local driving $w_j$. 

\medskip

As for the ABBM model (center of mass), the structure of the LT of the PDF $ P_{w_j}(\vec S)$ shows that each of these elementary jump processes is infinitely divisible. We define the avalanche density generated by the driving on the $j$-th point as
\bea
\rho_j( \vec S ) := \frac{d P_{\vec w}(\vec S)}{d w_j} |_{\vec w =0} = \frac{d P_{w_j}(\vec S)}{d w_j} |_{w_j =0\ ,}
\eea
where as in the previous case, this equality is to be understood point-wise in the $\vec S$ variables.
Consider the functions $v_j$ of $\vec \lambda$ 
which appear in Eq.~(\ref{laplaceG}) and satisfy  Eq.~(\ref{instantonaval2}). It is the analogue of $Z(\lambda)$ appearing in (\ref{lapabbm}) for the ABBM model and we thus conjecture the generalization of (\ref{Zrho}),
\bea \label{vjconj} 
v_j  = \int \rmd ^N \vec S \left( e^{ \vec \lambda \cdot \vec S} -1\right) \rho_j(\vec S ) \ .
\eea
This allows us to write an equation relating $P_{w_j}( \vec S )$ to $\rho_j( \vec S )$ similar to (\ref{decomp1}) (see \ref{app-levy}). The subtleties linked with the accumulation of small avalanches and the non-normalizability of $\rho_j( \vec S )$, are the same as in the previous case, which is also reminiscent of the fact that the limit of small driving of $P_{ \vec w}( \vec S)$ is very non-uniform, as we now detail. Consider $w_i = w f_i$ with $w \to 0$ and $f_i$ fixed: the limit of $P_{ \vec w}( \vec S)$ is again given (in the sense of distributions) by $\prod_{i=1}^N \delta(S_i)$. More precisely, in this small-$w$ regime, almost all avalanches are $O(w^2)$: $S_i = w_i^2 s_i$ with the $s_i$ distributed according to
\begin{equation}
p_0( \vec s) = \prod_{i=1}^N p_0(s_i) \ ,
\end{equation}
as can be seen from an examination of (\ref{mainform}) in that regime. The PDF $p_0$ was defined
in (\ref{smallscalecutoff}). One sees that the regime $S_i \sim w^2$ contains all the 
probability, and that for these very small avalanches the local sizes are statistically independent.

The remaining $O(w)$ probability to observe large avalanches $S_i = O(1)$ 
is encoded in the densities $\rho_j( \vec S)$,
\begin{equation}
P_{ \vec w}( \vec S) = \sum_{j=1}^N w_j \rho_j( \vec S) + O(w^2) \ .
\end{equation}
As before, the positive integer moments are entirely controlled by $\rho_j$. A more
general expression, which illustrates that these large avalanches
occur according to a Poisson process, is given in \ref{app-levy}. 

We now give exact expressions for these densities. 
For a general elasticity matrix, the expression of $\rho_j$ is obtained from Eq.~(\ref{mainform}), and contains a determinant. Remarkably, one can compute this determinant in various cases, leading to the following result
\begin{equation}
\!\!\!\!\!\!\!\!\!\!\!\!\!\!\!\!\!\!\! \rho_j(\vec S) = \left(\frac{1}{2\sqrt{\pi}}\right)^{\!\!N} \frac{  S_j } {(\prod_{i=1}^N  S_i ) ^{\frac{1}{2}}} K(\vec S) \exp { \left(-\frac{1}{4} \sum_{i=1}^N \frac{(\sum_{j=1}^N  C_{ij}  S_j )^2}{ S_i} \right)}\ ,
\end{equation}
where $K(\vec S)$ depends on the chosen elasticity matrix:
\begin{itemize}
\item{Fully connected model: $K(\vec S) = (\frac{c}{N m^2})^{N-1}  \frac{(\sum_{i=1}^N S_i)^{N-2}}{ \prod_{i=1}^N S_i}$}
\item{Linear chain with periodic boundary conditions: $ K(\vec S) = (\frac{c}{m^2})^{N-1}  \sum_{i=1}^N  \frac{1}{S_i S_{i+1} }$}
\item{Linear chain with free boundary conditions: $ K(\vec S) = (\frac{c}{m^2})^{N-1}  \frac{1}{S_1 S_N}$}
\end{itemize}

\paragraph{PDF of the shape in the small-driving limit} As we just detailed, the small-driving limit of $P_{\vec w}(\vec S)$ exhibits a complicated structure due to the accumulation of small avalanches. The situation is very different for the PDF of the shape of the interface 
conditioned to a given total size $S =O(1)$ (\ref{probashape}). This conditioning naturally introduces a small-scale cutoff 
that simplifies the small driving limit $w_i = w f_i$ with $w \to 0$ which reads
\begin{equation}\label{probashapequasistat}
\!\!\!\!\!\!\! \rho(\vec s | S ) = \lim_{w \to 0} P(\vec s | S ) = 2 \sqrt{\pi} \frac{S^{N+\frac{1}{2} } }{ \sum_i f_i}   \exp\left( \frac{ S}{4} \right) \sum_{j=1}^N f_j \rho_{j}(S \vec s) \ .
\end{equation}
This limit holds in the sense of distributions, and $\rho(\vec s | S )$ defines a normalized probability distribution. 
This indicates that the only small-scale divergence present in $\rho_j$ originates from 
the direction $S_j \sim S \to 0$ uniformly in $j$, in agreement with the conjecture (\ref{vjconj}). 

\section{Fully-connected model}\label{secfullyc}
In this section we use our result (\ref{mainform}) and analyze it for the fully-connected model with uniform driving. Most calculations are reported in \ref{app-fully}, where we also consider driving on a single site, $w_i = w_1 \delta_{i1}$.
\paragraph{Structure of the PDF and marginals}
In the fully-connected model with homogeneous driving $w_i=w$, it is shown in \ref{app-fully} that our main result (\ref{mainform}) has the simple structure
\bea \label{pfully}
P(\vec S) &&  =  \frac{w}{ w + c S/N} \prod_{i=1}^N p_{w , S/N} (S_i) \ .
\eea
We defined
\bea
p_{w , z }(S_i) = \frac{w + c z}{2 \sqrt{\pi} S_i^{3/2}} \exp\left(- \frac{ \left(w + c z- (1+c) S_i  \right)^2}{4 S_i} \right) \ .
\eea
For each $w, z>0$, it is a probability distribution, that corresponds to the (dimensionless, with $m^2=1$) PDF of the avalanches of one particle in a Brownian force landscape (ABBM model), interacting with {\em one parabolic well} through the force $m^2(w - u_i)$ {\it and with another parabolic well} through the force $c(z - u_i)$. Formula (\ref{pfully}) is thus reminiscent of the fact that the various sites interact with one another only through the center of mass of the interface. This simple structure permits a direct evaluation of various marginals of (\ref{pfully}) of the type $P( \{S_1 , \dots , S_p \} , S )$ (local sizes on $p<N$ sites and total size). This is done in \ref{app-fully}. Here we focus on the joint PDF of the total size $S$, and the single-site local avalanche size $S_1<S$. Its explicit form
is
\bea \label{fullyjoint1}
\! \! \! \! \! \!  && \! \! \! \! \! \!  \! \! \! \! \! \!  P(S_1 , S) = \frac{w}{2 \sqrt{\pi}S_1^\frac{3}{2}}(N-1) \frac{w + c S/N}{2 \sqrt{\pi} (S -S_1)^{3/2}} \exp\left(-  \frac{ \left(w + c S/N - (1+c) S_1  \right)^2}{4 S_1} \right)   \\ 
&& ~~~~~~~~ \times \exp\left(- \frac{ \left((N-1)\left(w + c S/N \right) - (1+c) (S -S_1)  \right)^2}{4 (S-S_1)} \right) \ .   \nonumber
\eea
Of interest is the participation ratio $s_1=S_1/S$ of a given site to the total motion. 
Its average is $\overline{s_1}=1/N$. Its second moment, conditioned to the
total size $S$, is easily extracted from (\ref{fullyjoint1}),
\bea \label{pr} 
\mathbb{E}(s_1^2|S) =  \frac{1}{N}-\frac{\sqrt{\pi } (N-1) e^{\frac{(c S+N w)^2}{4 S}} (c S+N w) 
{\rm erfc}\left(\frac{c S+N w}{2 \sqrt{S}}\right)}{2 N^2 \sqrt{S}} \ .
\eea 
We now study the limit of a large number of sites $N$ in Eq.\ (\ref{fullyjoint1}). There are (at least) two relevant regimes depending on how the driving $w$ scales with $N$. 

\paragraph{First regime: $w = O(1)$ (``many avalanches"):}
Consider the case $N \to \infty$ with $w$ fixed.
In this case, typical values of $S = \sum_{i=1}^N S_i$ are of order $O(N)$. Consider $\bar{S} = \frac{\sum_{i=1}^N S_i}{N}$ (empirical mean avalanche-size $S_i$), which is distributed according to
\bea \label{probaempiricmean}
P(\bar S)=\frac{\sqrt{N} w }{2 \sqrt{\pi} \bar S^{\frac{3}{2}}} \exp\left( -\frac{ N (\bar S-w)^2}{4\bar S } \right) \rightarrow_{N \to \infty} \delta( \bar S - w)\ .
\eea
The joint probability $P(S_1 ,\bar S)$, is given by Eq.\ (\ref{fullyjoint1}) (with the change of variable $S \to N \bar S$), and admits the  large-$N$ limit
\bea \label{fullycmeanfield}
&& P(S_1 ,\bar S) \simeq_{N \to \infty}  \frac{w + c \bar S}{2 \sqrt{\pi}S_1^\frac{3}{2}} \exp\left(- \frac{1}{4} \frac{ \left(w + c \bar S - (1+c) S_1  \right)^2}{4 S_1} \right) P ( \bar S) \nonumber \\
&&~~~~~~~~~~~~  \simeq \frac{w(1+c)}{2 \sqrt{\pi}S_1^\frac{3}{2}} \exp\left(- \frac{1}{4} \frac{(1+c)^2  \left(w  - S_1  \right)^2}{4 S_1} \right)  \delta( \bar S - w)\ .  
\eea
Hence the jump of the center of mass becomes peaked at $\bar S=w$, while the individual sites  keep a broader jump distribution.
The local avalanche statistics is the same as the one for  a particle submitted to the parabolic driving force $m^2(w - u_i),$ and 
to the elastic force from the center of mass of the interface, $c( \bar S - u_i)$. This observation extends to any number of particles $n_{{\rm part}}=O(1)$ with respect to $N$: in the large-$N$ limit, the particles become independently distributed according to the law (\ref{fullycmeanfield}). 
This picture is the ``mean-field" regime usually studied in fully-connected models \cite{DSFisher1998,DSFisher1985}, and here derived in a rigorous way. Note that in this case, due to a cancellation in (\ref{pr}), the participation ratio scales as $\mathbb{E}(s_1^2|S)  = O (1/N^2)$ which shows that $s_1$ is typically of order $1/N$. 

\paragraph{Second regime: small driving $w=O(1/N)$ (``single avalanche")} 
We now focus on the regime $w = \hat w /N$ with $\hat w$ fixed. 
In this case $S = \sum_{i=1}^N S_i$ is typically of order 1 and is distributed according to
\bea
P(S)=\frac{ \hat w }{2 \sqrt{\pi} S^{\frac{3}{2}}} \exp\left( -\frac{ (S-  \hat w)^2}{4 S } \right) \ .
\eea
We now compute, using (\ref{fullyjoint1}), the joint PDF of $S$ and $S_1$ in the scaling regime $S_1=O(1)$ fixed,
\bea
\fl && P(S_1 , S) \simeq_{N \to \infty}  \label{fullycotheregime} \\
\fl &&  ~~~~~~~~~~~~  \frac{ \hat w /N }{2 \sqrt{\pi}S_1^\frac{3}{2}} \exp \left( \frac{-(1+c)^2 S_1}{4} \right) \frac{ \hat w + c S}{2 \sqrt{\pi}(S-S_1)^\frac{3}{2}} \exp\left(- \frac{1}{4} \frac{ \left( \hat w + c S - (1+c)(S- S_1)  \right)^2}{4(S- S_1)} \right) \ . \nonumber
\eea
The first factor is reminiscent of the density of avalanches 
and contains a non-normalizable divergence
$\sim S_1^{-3/2}$ . However (\ref{fullyjoint1})
implies a cutoff on small $S_1$ of order $\frac{1}{N^2}$. The scaling $w = \hat w /N$ allows to isolate single (quasi-static) avalanches (in the interpretation of the BFM avalanche process as a Levy process discussed above) and the factor of $1/N$ is the probability that the site $i=1$ is part of the avalanche. In this regime, the fluctuations are large and the participation ratio scales as $\mathbb{E}(s_1^2|S)  = O (1/N)$.

\section{Spatial shape in small systems $N=2,3$.}\label{discrete}

In this section we analyze the PDF of the spatial avalanche shape  in the small-driving limit, $w_i = w \to 0$, mostly for $N=2,3$. It already exhibits a saddle-point  which allows us to discuss the general-$N$ case below. The analysis can be repeated for finite $w_i$. Similarities and differences give insight into the link between the quasi-static distribution and finite driving. This is done in \ref{app-smallN}.

\begin{figure}[h]
\centerline{\fig{8.5cm}{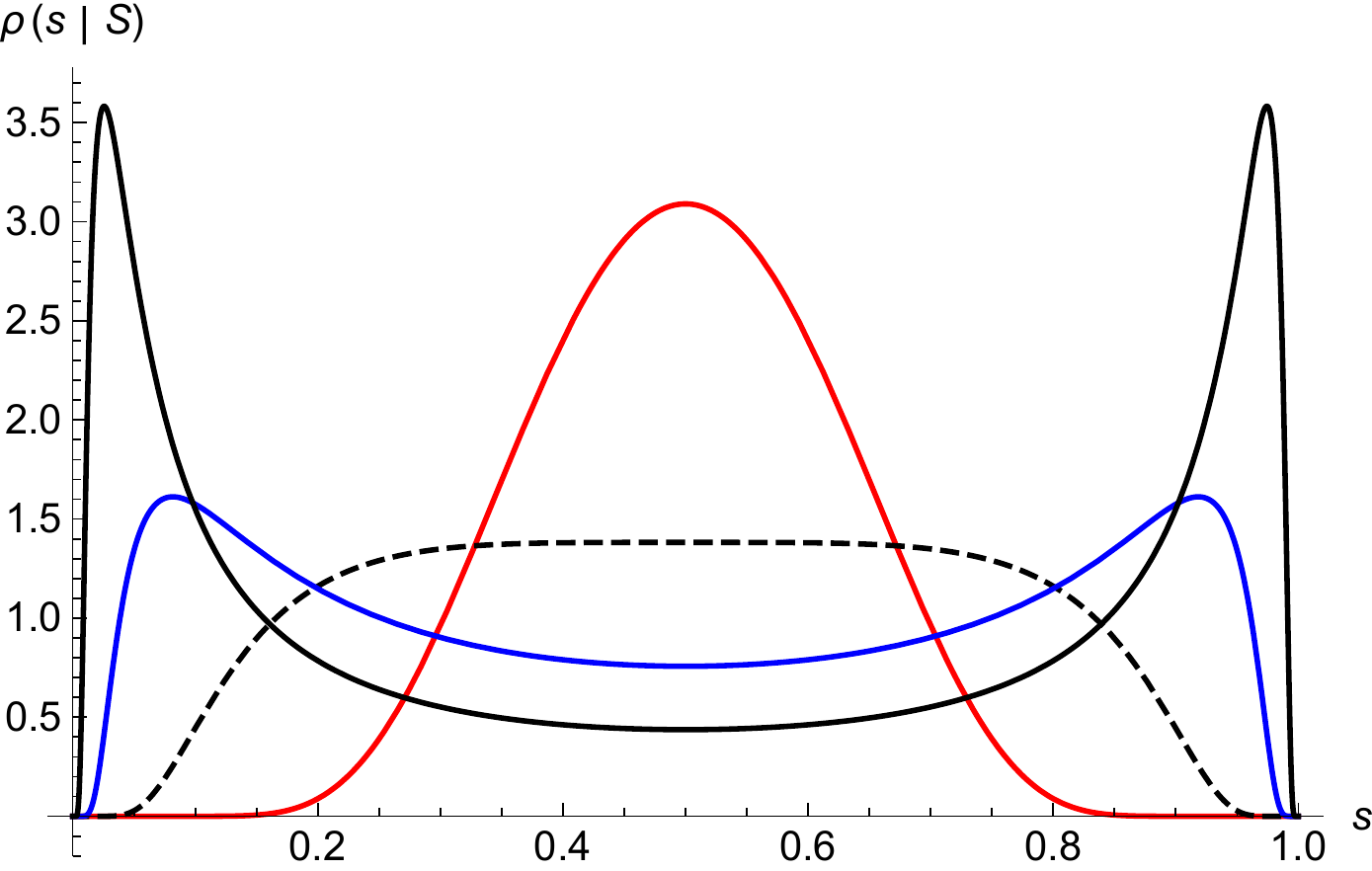}} 
\caption{ Shape transition of the quasi-static PDF (\ref{exact2}) for $N=2$ and $c=1$ in the linear chain with PBCs. For $S = 0.1 S_c $ (black, solid curve)  and $S = 0.3 S_c $ (blue), the distribution has two symmetric maxima. For $S=5 S_c$, the distribution is peaked around $s= \frac{1}{2}$ (red, upper curve). The transition occurs at $S=S_c=3/8$ (black, dashed curve).}
\label{bfmN2}
\end{figure}

\paragraph{N=2,3} We start with $N=2$, for which the different models we considered are all equivalent. To fix notations, we study the linear chain with PBCs (see Section \ref{elastmat}) and $m=1$. The quasi-static PDF of the shape (\ref{probashapequasistat}), conditioned on the total size $S,$ reads
\begin{equation} \label{exact2} 
\rho(s|S) =  \frac{2 c}{4 \sqrt{\pi} (s(1-s))^{\frac{3}{2}}} e^{ - c^2 S\frac{(1-2s)^2}{ s(1-s)}} \ .
\end{equation}
We noted $s=s_1=S_1/S$, the shape variable of the first site. The behavior of this PDF is summarized on Figure \ref{bfmN2}. For small $S$, typical avalanches are mainly distributed on one site. As $S$ increases, the most probable avalanches become more homogeneously distributed over the two sites, and for $S$ larger than $S_c=\frac{3}{8 c^2}$, the probability distribution is peaked around $s =\frac{1}{2}$ and the avalanche is extended over the whole system. We call this phenomenon the {\em\ shape transition}: For small total size, the most probable avalanches have ${\rm max}(s_i) \simeq 1$, whereas for large avalanches $\max(s_i) \simeq 1/N = 1/2$.

\begin{figure}
\centerline{
  \fig{.35\textwidth}{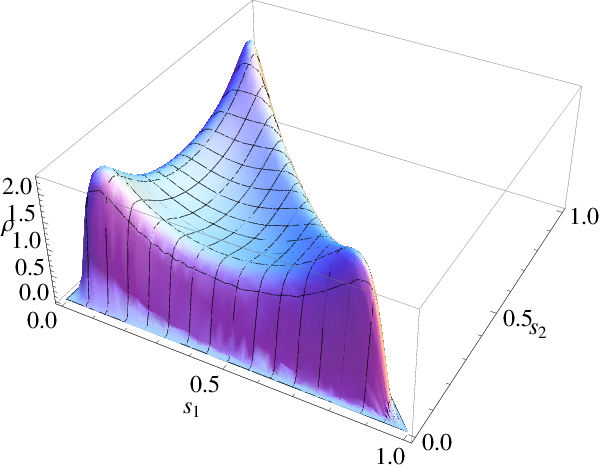}
   \fig{.35\textwidth}{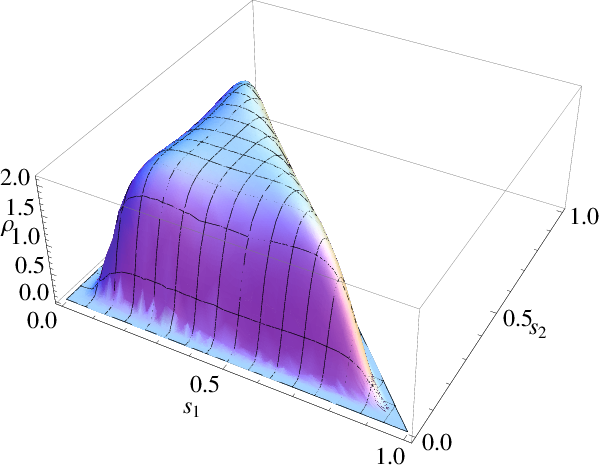}
    \fig{.35\textwidth}{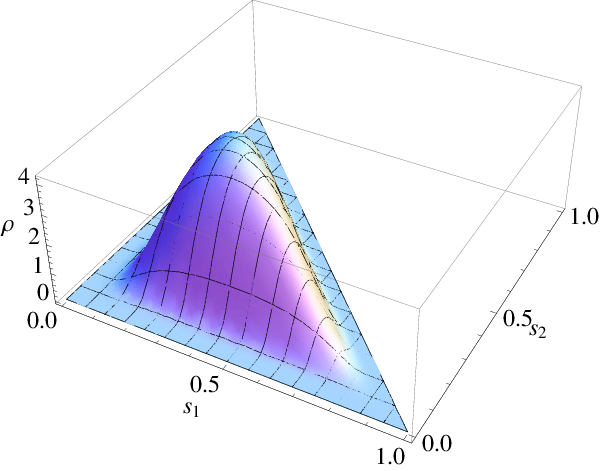}}
 \caption{Shape transition of the quasi-static shape distribution for $N=3$ and $c=1$. From left to right: $S=0.5 ; 1 ; 2$.}
\label{bfmN3}
\end{figure}

The case $N=3$ for a linear chain with PBC  is similar. For $S < \frac{1}{c^2}$, the quasi-static density distribution of the shape $\rho(s_1,s_2,s_3=1-s_1-s_2 | S)$ has three symmetric maxima corresponding to avalanches  mainly centered on a given site, whereas for $S > \frac{1}{c^2}$ there is only one maximum at $s_i = \frac{1}{3}$. This can be seen on Figure \ref{bfmN3}.

\smallskip
\paragraph{General $N$} This study already gives some insight into the structure for generic $N$: the quasi-static distribution of the shape $\rho( \vec s | S)$ exhibits different saddle-points, whose positions and stabilities depend on the value of $S$. For small $S$, avalanches are preferentially located on a single site $j$ and ${\rm max}(s_i)\simeq 1$. As one increases $S$, the most probable avalanches are more and more extended. The analytical calculation of the properties of these saddle points is difficult. However, we can  generalize the shape transition observed for $N=2,3$: The symmetric configuration defined by $\forall i$, $s_i = \frac{1}{N}$ (a situation corresponding to infinitely extended and uniformly distributed avalanches) is always a saddle-point of translationally invariant models. This saddle-point is only stable for $S>S_c(N)$, which is computed in \ref{appuniform} for the fully connected model, and for the linear chain with PBC. The result is
\bea
& S_c^{{\rm fc}} (N ) = \frac{3 N}{c^2} , \\
& S_c^{{\rm PBC }}(N)  \sim_{N \to \infty} \frac{1}{16 c^2 \pi^4} (N^5 + 12 N^4  + O(N^3) )  . \nonumber
\eea
This critical value gives the scaling of the total size above which most probable avalanches are uniformly distributed on all the interface. Below this scaling they adopt a more complex structure (e.g. they are localized on several sites, possess maxima, etc.). Let us already mention that other saddle-points of the shape PDF are numerically studied in \ref{appshapedisc}, where the results are compared to the one obtained in the next section for the most probable avalanche shape in a continuum model.

\section{Continuum limit: avalanches of an elastic line and typical shape of  avalanches with large aspect ratio}\label{secContinuum}
\subsection{Avalanche size PDF and density in the continuum limit}
We now study the generalization of the previous result to the continuum Brownian-force model with short-ranged elasticity for a line of length $L$ 
\begin{equation}\label{bfmeqn}
\eta \partial_t \dot{u}_{xt} = \nabla^2 u_{xt} - m^2( \dot{u}_{xt}-\dot{w}_{xt} ) + \sqrt{2 \sigma \dot{u}_{xt}} \xi_{xt} \ .
\end{equation}
Here $\xi_{xt}$ is a gaussian white noise with $\overline{\xi_{xt} \xi_{x't'} } = \delta(x-x') \delta(t-t')$ and the boundary conditions are either free or periodic. Starting from rest at $t=0$ and imposing a driving $\dot{w}_{xt} \geq 0$ for $t\geq 0$ such that $\int_t \dot{w}_{xt} = w_x$, we note the total displacement of the interface $S_x = \int_{t \geq 0} \dot{u}_{xt}$. The method used in the discrete case can be extended to derive the PDF of avalanches in the continuum. Another route is to consider the continuum model as the appropriate $N \to \infty$ limit of the discrete model, as is detailed in \ref{appcontinuum1}. Both procedures give the same result, which, for the dimensionless PDF of continuum avalanches, includes a functional determinant
\bea\label{continuumproba}
P[S_x] \sim \left(\frac{1}{\prod_x S_x}\right)^{\frac{1}{2}} \det\left(  M \right) \exp{ \left( - \int_{0}^L \rmd x \frac{(w_x-S_x + \frac{1}{m^2} \nabla^2 S_x )^2}{4 S_x} \right)}, \\
M(x,y) = -\frac{1}{m^2} (\nabla^2)_{xy} +   \delta(x-y)\left( 1 + \frac{w_x-S_x + \frac{1}{m^2}\nabla^2 S_x}{S_x} \right). \nonumber
\eea
Here $\nabla^2$ is the usual Laplacian, $(\nabla^2)_{xy} = \delta''(x-y)$. 
Dimensions can be reintroduced as in the discrete case using $S_m = \frac{\sigma_c}{m^4}$.
$S_m$ is the avalanche-size scale of the continuum theory. The first factor $(\frac{1}{\prod_x S_x})^{\frac{1}{2}}$ also comes from a determinant and could be included in the definition of the operator $M$.

As in the discrete case, 
the mean displacement $\langle S_x \rangle$ satisfies $- \nabla^2 \langle S_x \rangle + \langle S_x \rangle = w_x$.
For instance, if the driving is only at one point, $w_x=w \delta(x)$, one has $\langle S_x \rangle = \frac{w}{2} e^{-|x|}$.
The case of a general $w_x$ is obtained by superposition. This is consistent with the discussion in Section \ref{quasi}.
As in the discrete case, the mean displacement gives the avalanche shape in the limit of
large driving (plus an $O(\sqrt{w})$ Gaussian noise). 

One can also study the homogeneous quasi-static limit: $w(x) = w \to 0$ 
and $S(x) = O(1)$ uniformly in $x$. Then $P[S] \simeq w \rho[S]$ with $\rho[S]$ the quasi-static density 
of sizes of continuous avalanches, also obtained as the limit of the discrete ones,
\begin{equation}\label{continuousdensity}
\rho[S_x] \sim  \frac{ (\int_0^L   \rmd x S_x) BC[S_x] } {(\prod_x S_x) ^{\frac{1}{2}}} \exp{ \left(-  \int_{0}^L \rmd x \frac{(S_x - \nabla^2 S_x )^2}{4 S_x} \right)} .
\end{equation}
From now on we set $m=1$ (by a rescaling of $x$). The term $BC[S_x]$ depends on the chosen boundary conditions with $BC[S_x] = \int_{0}^L \frac{\rmd x }{S_x^2} $ (resp. $BC[S] = \frac{1}{S_0 S_L} $) for the periodic case (resp. free case).

\paragraph{Other continuum models} Our discrete setting allows us to obtain the avalanche-size PDF 
of various continuous models, Eq.~(\ref{continuumproba}) being  generalizable to an interface 
of internal dimension $d$. One may also consider an arbitrary elasticity matrix $c_{xy}$ 
by changing $\nabla^2 u_x \to \int \rmd y c_{xy} u_y$. The continuum limit of the formula
for the PDF of the shape conditioned to the total size, either at finite $w$, see Eq.~(\ref{probashape}), 
or for $w \to 0$ (quasi-static limit), see Eq.~(\ref{probashapequasistat}), are also easily derived.

\subsection{Rewriting the probability measure on avalanche sizes} \label{sec-shape}

We now wish to determine the most probable shape of quasi-static avalanches, in the limit $L \to \infty$ \footnote{ In general the shape of avalanches depends on the driving. 
However, an avalanche following an arbitrary driving (in particular in a quasi-static setting more usual for experiments, see Sec.\ \ref{sec-appli}) in the BFM is a sum of quasi-static avalanches (Sec. \ref{quasi}), whose spatial structure is, by definition, independent of the driving.}. 
To render the problem  well defined, one needs to specify two scales. 
A natural choice is the total size $S=\int_x \rmd x\, S_x$ 
and the spatial avalanche extension (or length) $\ell$, i.e.\ the size of the support of $S_x$. 
While the  avalanche-size PDF $P(S)$ is given by the ABBM result (\ref{ABBMaval}),  the existence of a finite extension $\ell$ (i.e. local avalanche sizes being strictly zero outside a finite interval) 
is non-trivial\footnote{In a mathematical sense it may be a peculiarity of the BFM in $d=1$ with short range elasticity. 
Of course rapid decay in space is expected more generally beyond some support region of
extension $\ell$, and often obtained in numerical 
simulations.}.
Here it naturally arises in the search for saddle-points of the shape PDF: 
we only found solutions which vanish outside of an interval. This property
was also shown recently in \cite{DelormeInPrep} where the PDF of the extension $P(l)$ is computed.

\medskip

In the following we study the shape distribution at fixed $S$ and  $\ell$. We do not take into account the term implementing boundary conditions in (\ref{continuousdensity}) since it should not play a role in the bulk (this hypothesis is explicitly checked on the discrete model in \ref{appshapedisc}). So we write the density of continuum avalanches $S_x$ as
\bea \label{defH}
\rho[S] \prod_x {\rmd S_x} \sim \prod_x \frac{\rmd S_x}{\sqrt{S_x}} \left(\sum_x S_x\right) \rme^{-{\cal H}[S]} \\
{\cal H}[S] = \int_x \frac{[S_x -\nabla^2 S_x]^2}{4 S_x} =  \int_x  \frac{S_x}{4}  + \frac{[\nabla^2 S_x]^2}{4 S_x}\ .
\eea
To eliminate  the factor of $(\prod _x S_x)^{-1/2}$ in the  measure, we set  
\be
S_x = \Phi^2(x)\ .
\ee 
The integration $\int_0^\infty \frac{\rmd S_x}{\sqrt S_x} =\int_{-\infty}^\infty \rmd \Phi(x) $, thus the integral over  $\Phi(x)$ runs from $-\infty$ to $\infty$.
To further simplify the calculations, we note that the problem is invariant by translation. We thus impose  the center of the support to be at $x=0$. This leads to the definition of the reduced shape $s(x)=\phi^2(x)$
\bea \label{scaleds} 
\!\!\!\!\!\!\!\!\!\!\!\!\!\!\!\!\!\!\!\!\!\! S_x = \frac{S}{\ell}s(x/\ell)= \Phi^2(x) =\frac{S}{\ell} \phi^2(x/\ell) \quad , \quad \int_{- \frac{1}{2}}^{\frac{1}{2}} \rmd x\,   \phi^2(x) = 1  \quad , \quad |x| \geq \frac{1}{2} \Rightarrow  \phi(x)=0 \ .
\eea 
Note that to study fluctuations around the saddle point it is more convenient to use $\phi(x)$,
but the saddle point itself can be obtained equivalently using $s(x)$ or $\phi(x)$. 
Below we use $\phi(x)$, but also indicate the corresponding 
formulas for $s(x)$ when these are simpler.

We  search for the most probable shape in the limit of small
driving, at fixed size $S$ and extension $\ell$. The path integral takes the form
\bea \label{optimizationreduced}
&& \prod_x {\rmd \phi(x)}  \exp\left(-\frac{S }{4} - \frac{S}{\ell^4} {\cal H}_{{\rm el}}[\phi] \right)  
\nn\\
&& {\cal H}_{{\rm el}}[\phi] = \int_{-\frac 12}^{\frac12} \phi ''(x)^2+\frac{\phi '(x)^4}{\phi
   (x)^2}+\frac{2 \phi '(x)^2 \phi ''(x)}{\phi (x)} \,  \rmd x\ . 
\eea
The boundary conditions are $\phi( \frac{1}{2}) =\phi( -\frac{1}{2}) =\phi'( \frac{1}{2}) =\phi'( -\frac{1}{2}) = 0$ and 
\be\label{constraint}
\int_{-\frac{1}{2}}^{ \frac{1}{2}} \rmd x \,\phi^2(x) = 1\ .
\ee Note the appearance of the factor of $\frac{S}{\ell^4}$ in front of the ``elastic'' energy.

\subsection{The saddle point for large aspect ratio $S/\ell^4$}

The path integral (\ref{optimizationreduced}) is for large $S/\ell^4$ dominated by a saddle-point.  To enforce the constraint (\ref{constraint}), we minimize ${\cal H}_{{\rm el}}[\phi] - {\cal A} \int_{-1/2}^{1/2}\rmd x\, \phi^2(x)$, with Lagrange multiplier $\cal A$, leading to the saddle-point equations
\footnote{{The saddle point equation has a simpler form
in terms of $s(x)$. It reads: $ \frac{1}{2} [s''(x)/s(x)]'' - \frac{1}{4} [s''(x)/s(x)]^2 = {\cal A}$.
Hence $s''(x)/s(x)$ is a Weirstrass function which diverges
as $\sim (x \pm x_c)^{-2}$ at the boundaries.}}.
\be \label{SP-eq}
{\cal A} \, \phi(x)  = \frac12 \frac{\delta {\cal
H}_{\rm el}[\phi]}{\delta \phi (x)} =     \phi ^{(4)}(x)+\frac{5 \phi
   '(x)^4}{\phi (x)^3}-\frac{10 \phi '(x)^2 \phi ''(x)}{\phi (x)^2}  .
\ee
In order to find the solution $({\cal A}_0, \phi_0(x))$ of (\ref{SP-eq}) satisfying the properties written in (\ref{scaleds}), we first obtain numerically, using a shooting method, another solution $({\cal A}_1, \phi_1(x))$ of (\ref{SP-eq}). We impose ${\cal A}_1 =  2.5\times 10^{5}$, $\phi_1(0)=1$, $\phi_1' (0)=\phi_1 ''' (0)=0$, and look for the correct shooting parameter $\phi_1'' (0)$ such that the numerical solution has a support of finite size $[-x_c,x_c]$ with the desired behavior at the boundary, i.e. $ \phi_1'(-x_c)=\phi_1'(x_c)=0$. The obtained (unique) solution has the following properties: $\phi_1'' (0)=-276.797090676018$, $x_c = 0.162713 $, $\sqrt {\phi_1 (x)} \simeq 7.85883 (x_{c}-x)$  for $x\to x_{c}$ and $S_1 := \int_{-x_c}^{x_c}\phi_1^{2}(x) \rmd x  = 0.106289$.
We now take advantage of rescaling, setting 
\begin{equation}\label{k6}
\phi_0 ( x) := \sqrt{\frac{2 x_c}{S_1}}\,\phi_1  (2 x_c x)\ , \quad {\mbox{and } s_0(x) = \phi_0^2( x) \ .}
\end{equation} 
This function is  automatically  a solution of (\ref{SP-eq}) with a different Lagrange multiplier ${\cal A}_0 = (2 x_c)^4 {\cal A}_1$, and the desired properties (\ref{scaleds}). By multiplying (\ref{SP-eq}) by $\phi_0(x)$ and integrating for $x \in [-\frac{1}{2}, \frac{1}{2}]$ (using $\phi_0'(\pm \frac{1}{2})=0$), we obtain the relation ${\cal H}_{{\rm el}}[\phi_0] = {\cal A}_0$. Numerically we find
\begin{equation} \label{elastic-energy-col}
{\cal E}_0:= {\cal H}_{{\rm el}}[\phi_0] = {\cal A}_0 = (2 x_c)^4 {\cal A}_1 = 2803.8 \pm 0.2 \ .
\end{equation}
An estimate of the numerical accuracy is given. The error is mostly due to the imprecision in determining $x_c$.

\begin{figure}
\fig{9cm}{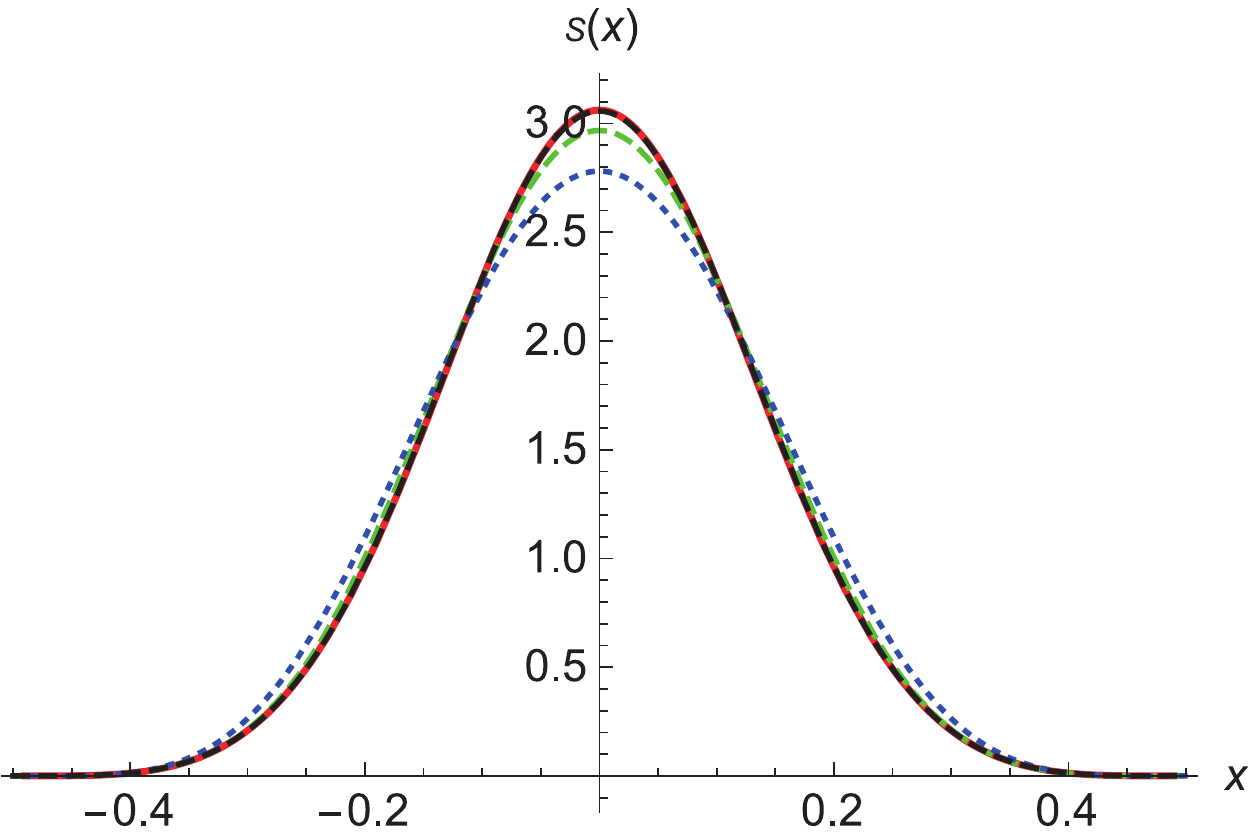}\fig{8.5cm}{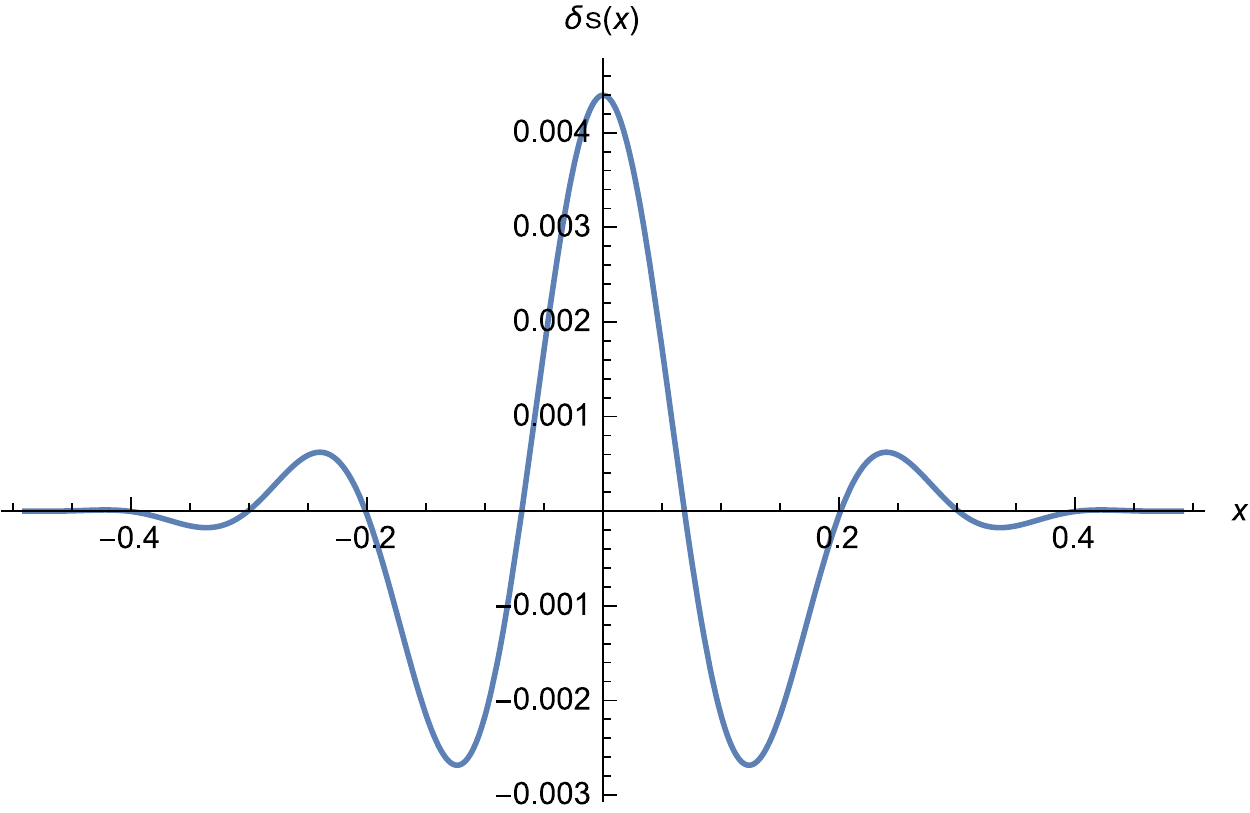}
\caption{Left: The function $s_0(x)=\phi_0^2(x)$, as obtained by solving the differential equation (\ref{SP-eq}) (red solid curve). This is contrasted to the variational ansatz (\ref{var-ansatz}), with one (blue dotted), two (green dashed) and 15  variational parameters (black-dashed, indistinguishable from the solution of the differential equation). 
Right: Difference between the solution of the differential equation, and the best variational solution.}
\label{f:col-shape}
\end{figure}
Alternatively, a variational solution can be used. We make  the ansatz   
\be\label{var-ansatz}
\phi_{\rm var} (x) = {\cal N}_c \left(x^2-{\textstyle \frac14} \right)^2 \left(1+ \sum_{i=1}^{i_{\rm max}} c_i (x^2-{\textstyle \frac14})^i\right) \ , \quad  \mbox{and } s_{\rm var}(x) = \phi_{\rm var}^2( x).
\ee
The behavior at the boundary  $x=\pm \frac12$ is chosen in agreement with the numerical solution of the saddle-point equation. One can also show that this ansatz leads to an  energy which remains finite at the boundary.  
The $\vec c$-dependent normalization ${\cal N}_c$ is chosen s.t.\ $\int_{-1/2}^{1/2} \rmd x\,  \phi_{\rm var} (x) ^2=1$.
For a given vector $\vec c = \{c_1,..., c_{i_{\rm max}} \}$, one then evaluates ${\cal H}[\phi_{\rm var}]$. Using a Monte Carlo algorithm,  the minimum energy is searched  by steepest decent in the space of all $\vec c$ with  given \ $\int_{-1/2}^{1/2}
\rmd x\,  \phi_{\rm var} (x) ^2=1$. In Figure \ref{f:col-shape} we show that for the shape of the avalanche, this procedure rapidly converges against the solution obtained by solving the differential equation (\ref{SP-eq}). 
Our best estimate is for  $i_{\rm max}=15$, where  we find 
\bea\label{c-sol}
\vec c &=& \{ -1.00301, 20.6871, 83.4237, 211.353, -270.898, 179.973, -72.6636, 
16.3962, \nn\\
&&  ~~{-12.2786}, 6.11179, -0.33042, 11.777, 0.750034, -6.77598, 
-4.56253 \}\ .
\eea
This result is compared to the numerical solution of the saddle point on Figure \ref{f:col-shape}.
The energy of this solution gives us, in good agreement with Eq.~(\ref{elastic-energy-col}), the variational bound 
\be
{\cal E }_0\le 2803.96\ .
\ee
In \ref{appshapedisc}  we confront this result to a study of the optimal shape in a discrete setting. There we also show (see also Figure \ref{f:M-spectrum} below) that this saddle-point is stable. Hence, the reduced shape of an avalanche becomes deterministic in the limit of $S/\ell^4 \gg 1$: $s(x) \stackrel{S/\ell^4 \to \infty}{\longrightarrow} s_0(x) = \phi_0^2(x)$ with probability one. Formula (\ref{optimizationreduced}) then shows that ${\cal E }_0$ is measurable in the tail of the distribution of aspect ratios,
\be
{\rm Proba}(S/\ell^4) \stackrel{S/\ell^4 \gg 1 }{\sim} \exp \left( -{\cal E }_0 \frac{S}{\ell^4} \right)
\ee
with possibly some sub-dominant factors,  as e.g.\ a power-law. This is confronted to numerics below.

\subsection{Simulations: Protocol and first results}
\label{s:simul+protocol}

\paragraph{Protocol.} Here we describe the simulation used to numerically study the shape of avalanches. We use a discretization with $N=512$ points of the equation of motion for the velocity in the BFM (\ref{bfmeqn}) using periodic boundary conditions for a system of total size $L=N$. The mass is chosen as $m=10/L$ in order to get a scale-free statistics for a wide range of events. The other parameters are set to unity, $\eta=\sigma=1$. The time is  discretized using a time-step $dt=0.01$ and a discretization scheme identical to \cite{Chate}. Simulations are done via Matlab and results are analyzed using Mathematica. At $t=0$ the system is at rest and we choose to drive it using a kick of size $\delta w=100$ on a single site. This is motivated by the fact that we want to study (single) quasi-static avalanches: the value of $\delta w$ is chosen to be small in adimensioned units $\frac{m^3}{\sigma} \delta w \simeq 7.4.10^{-4}$. Following the discussion of Section \ref{quasi} and \ref{app-levy}, we thus know that an avalanche resulting from our driving protocol can either be a ``small" avalanche $O(\delta w^2)$ or, with a small probability $p_0 = O(\delta w)$ a quasi-static avalanche of total size $S = O(1)$ (we neglect the $O(\delta w^2)$ probability that several quasi-static avalanches have been triggered). Schematically, we write
\be \label{approx-quasi}
P(\vec S ) \simeq (1-p_{0})``\delta"(\vec S) + p_{0} \rho_{i_0} (\vec S)\ ,
\ee
where $i_{0}$ is the driven site. Here  $``\delta"(\vec S)$ is not a true delta distribution since in the BFM the interface always moves, but it rather denotes the PDF of all the small, non quasi-static avalanches, which is expected to depend highly on the driving. This is made more precise below, and in particular we discuss how we identify the quasi-static avalanches and $p_0$ from our data set.

\begin{figure}
\centerline{\fig{8cm}{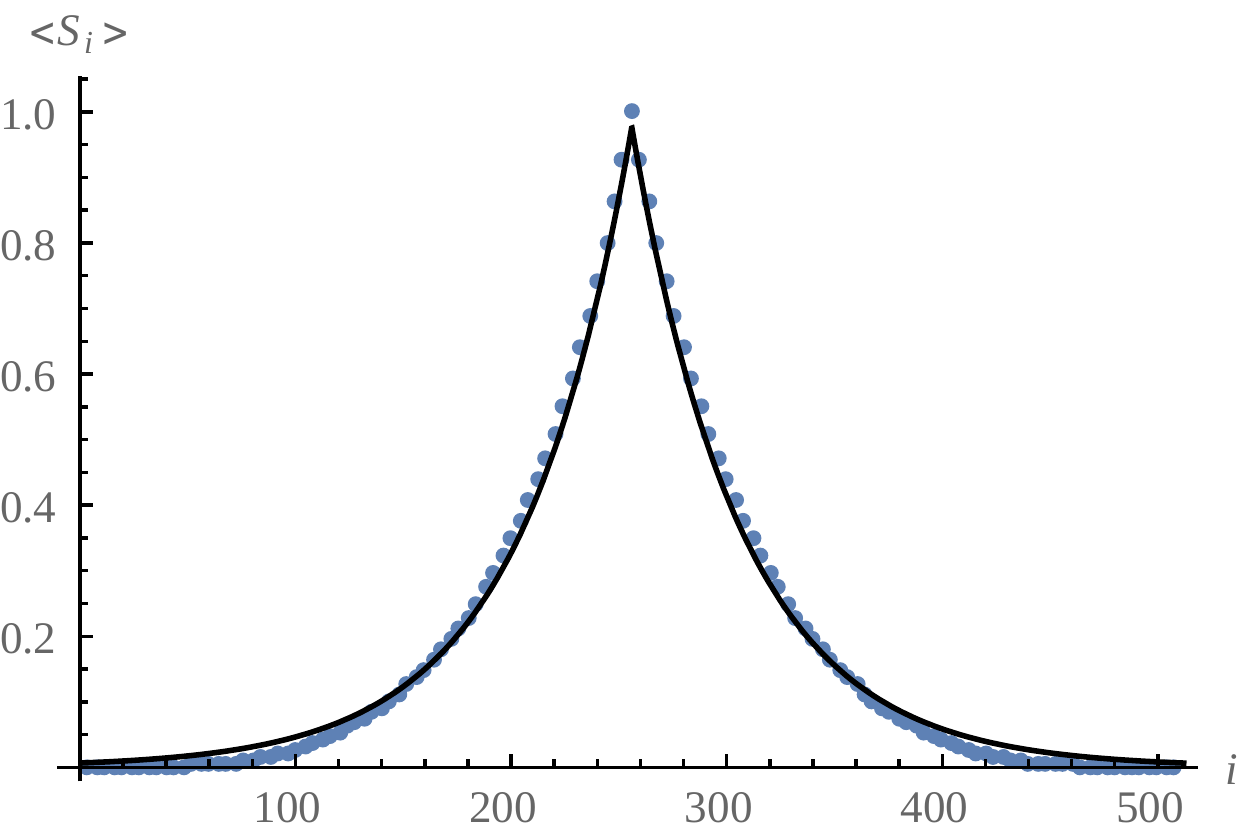}}
\caption{Measurement of $\langle S_i \rangle$ and comparison with the exact result $\langle S_i \rangle = \frac{ m w}{2} e^{ - m|i-i_{0}|}$ with $i_0=256$. The total moment is measured as $\langle S \rangle = 99.461$.}
\label{f:meanS}
\end{figure}

We stop the simulation for the rare events when an avalanche reaches the periodic boundary,  since we are interested in the distribution of shapes on an infinite line. For every generated avalanche, we numerically compute its shape characteristics $S, \ell$ (avalanches are indeed observed as having a finite support) and $s(x)$ (discretized with $\ell$ points). We report results using $n_{\rm it}=2.10^7$ simulations of a kick. As a first verification, we check on Figure \ref{f:meanS} a coarse-grained information on the spatial structure by measuring the mean local avalanche  size. The discrepancy at the boundaries can be attributed to the fact that we stop the simulation when an avalanche reaches the PBCs. This is the only bias expected in our procedure. It is not a problem since for the rest of the article we are interested in observables at large $S/\ell^4$,   automatically excluding the largest $\ell$.

\begin{figure}
\centerline{\fig{6cm}{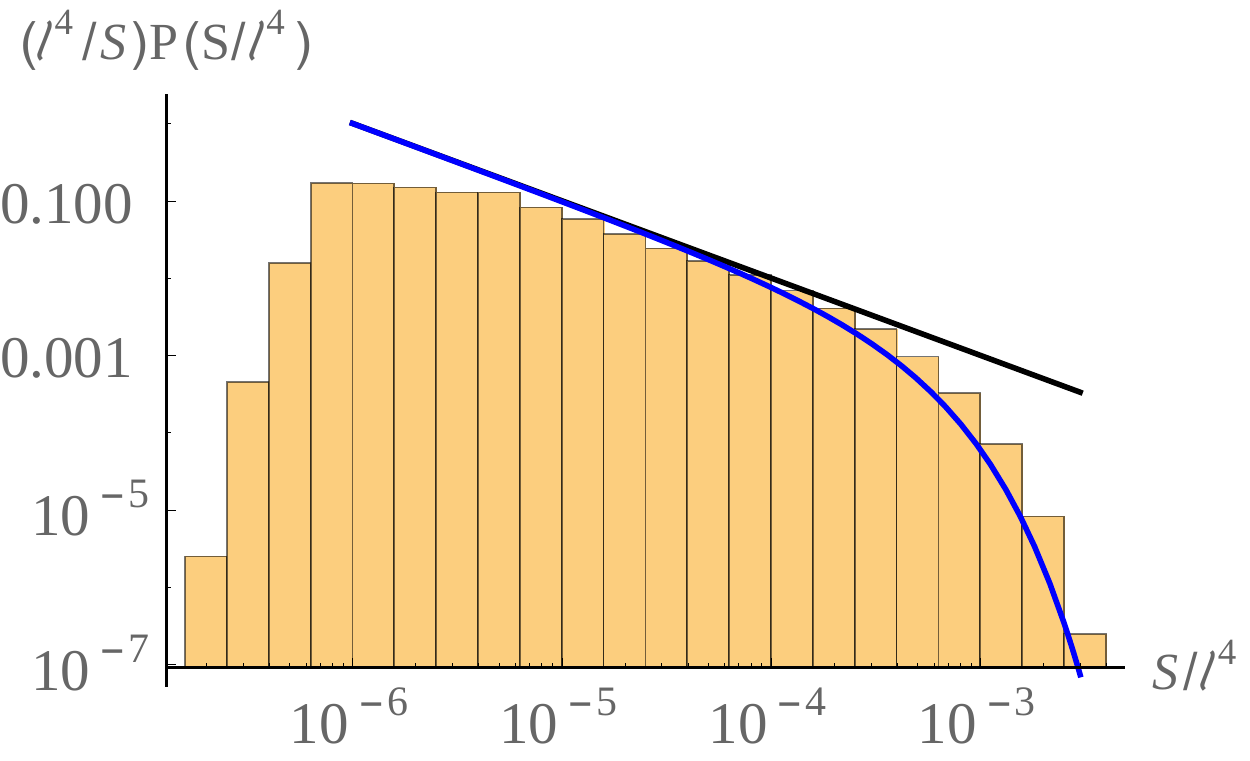} ~~ \fig{6cm}{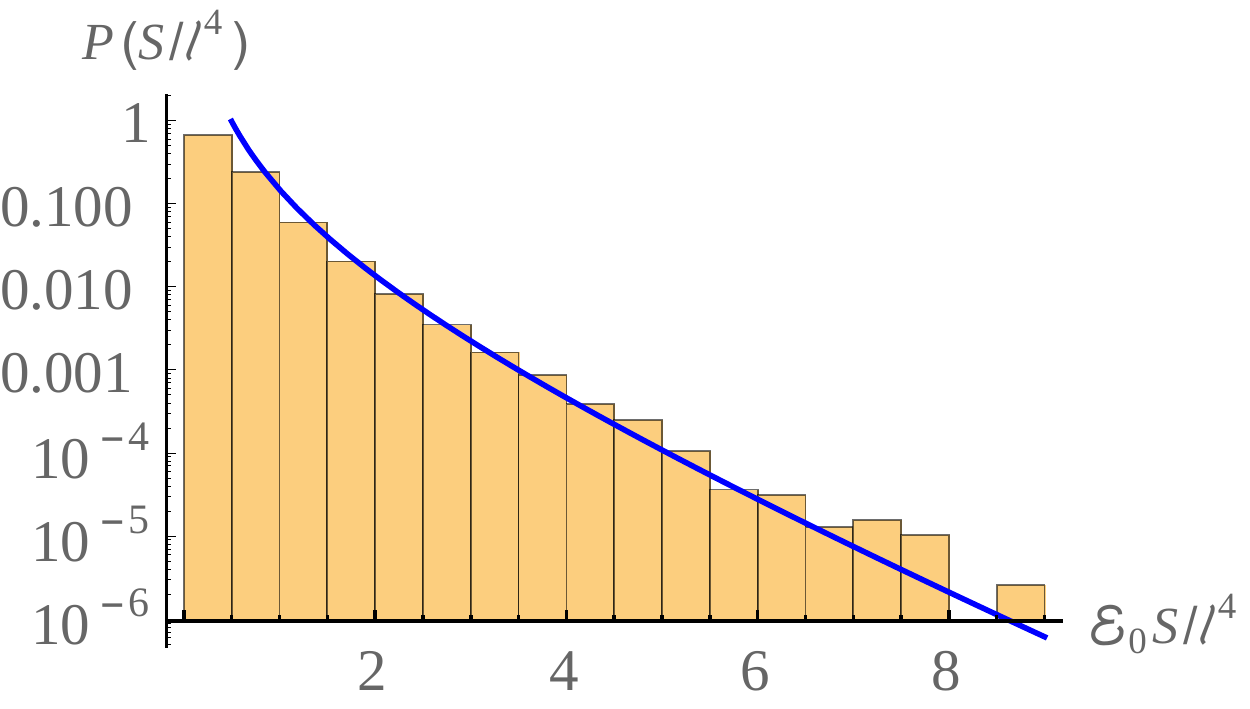} ~~ \fig{6cm}{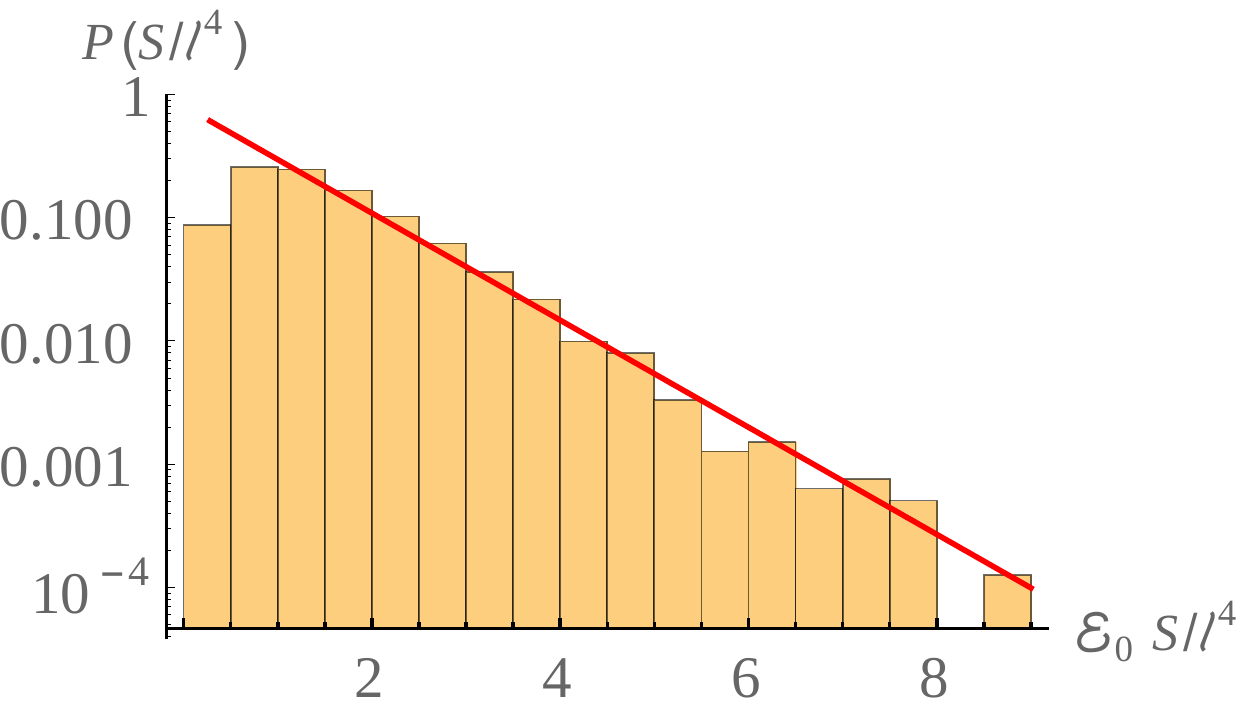}}
\caption{Different histogram of the PDF of $S/\ell^4$ obtained numerically with different binning procedures for the $x$ axis and scale for the $y$ axis. Left: log-log histogram of the full distribution. Center: log histogram of the distribution for aspect ratio $S/\ell^4 \geq 1/{5 \cal E}_0$. Right: log histogram of the distribution for avalanches of extension $\ell \geq 1/m$. The black line on the left emphasizes the observed power-law behavior ${\rm Proba}(S/\ell^4) \sim \ell^8/S^2$. Blue lines are fits using an ansatz of the form ${\rm Proba}(S/\ell^4) \sim \ell^8/S^2 \exp(- {\cal E}_0 S/\ell^4)$. The red line is a fit using only the cutoff function: ${\rm Proba}(S/\ell^4) \sim \exp(- {\cal E}_0 S/\ell^4)$. }
\label{f:PSvsl4}
\end{figure}

\paragraph{Consistency check of ${\cal E }_0=2804$.} We predicted above that ${\cal E }_0$ controls the tail of the  distribution of aspect-ratios.  Numerically, we find that this distribution possesses a power-law part coherent with an exponent of $2$ and an exponential cutoff for large $S/\ell^4$ with a prefactor coherent with ${\cal E }_0=2804$: ${\rm Proba}(S/\ell^4) \simeq \ell^8/S^2 \exp(- {\cal E}_0 S/\ell^4)$  (see left and center of Figure \ref{f:PSvsl4}). We also remark that the exponential cutoff function seems to entirely control the PDF of $S/\ell^4$ for ``massive" avalanches, of extension $\ell\geq 1/m$ (see right of Figure \ref{f:PSvsl4}). Obviously this does not constitute a precise measurement of ${\cal E }_0$, but rather a verification of its non trivial value, which can probably only be understood by studying the complete spatial structure of avalanches as we did.

\begin{figure}
\centerline{\fig{8cm}{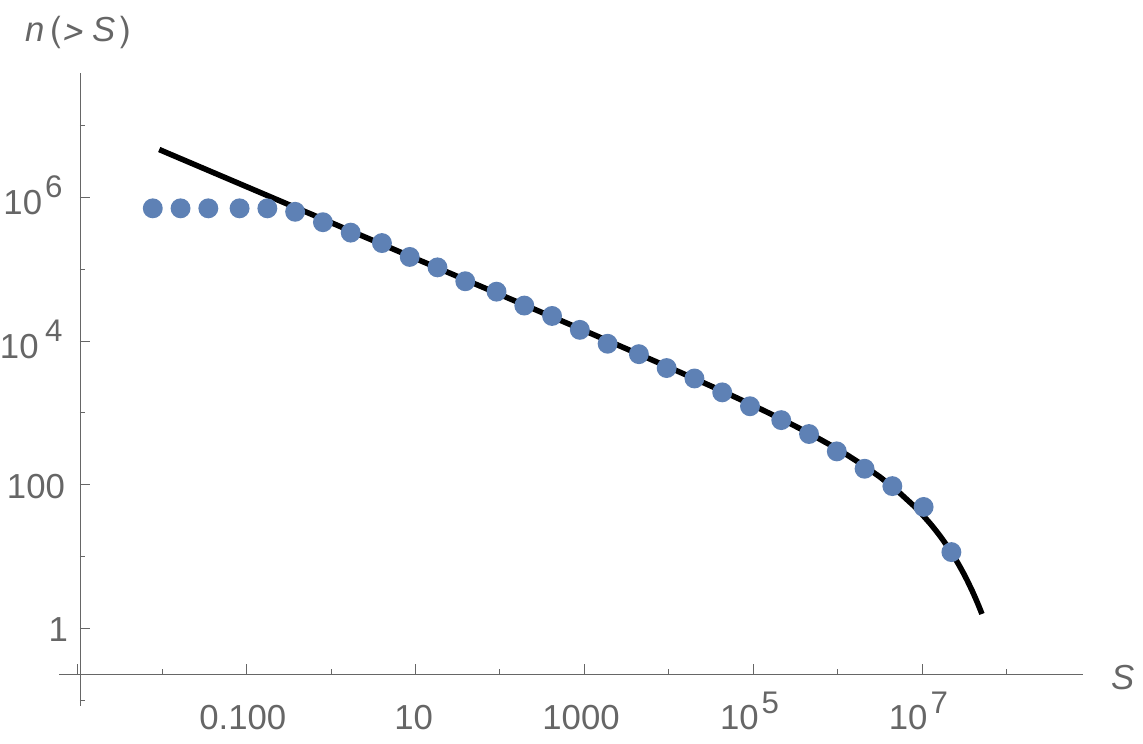}}
\caption{Left: $n_{>S}$ measured from the datas (blue dots) and compared to the quasi-static prediction ((\ref{nofSquasi}), black line) with $S_2 \to S$ ($S_1$ can be chosen anywhere in $[0.5 , 10^5]$ and  $n_{>S_1}$ is measured from the datas).}
\label{f:quasitest}
\end{figure}

\paragraph{Identifying quasi-static avalanches.} From now on we restrict our numerical results to avalanches of extension $\ell \geq 10$ to obtain a decent spatial resolution. This also allows us to isolate quasi-static avalanches. Avalanches with extension larger than $10$ only represents $3.5\%$ of the data. Obviously, this is not a proof that this subset of avalanches  only contains quasi-static avalanches, and one needs to check that it has the statistical properties of a set generated by the quasi-static density. One ``test" is to study the number $n_{>S_1}$ of avalanches of total size $S$ larger than $S_1$, for which the quasi-static hypothesis implies,
\begin{equation}\label{nofSquasi}
n_{>S_2} = n_{>S_1} \frac{\int_{S_2}^{\infty} \rho(S) \rmd S}{\int_{S_1}^{\infty} \rho(S) \rmd S},
\end{equation}
where $\rho$ was defined in (\ref{defrho}). Numerically, we find that this relation holds for all $S_1,S_2$ larger than $S\rm _{min}=0.5$ (see Figure \ref{f:quasitest}). We thus further restrict our set of avalanches to avalanches of total size $S \geq S_{\rm min} $. 
Note that though our reduced set of avalanches now only contains $2.7\%$ of the total number of avalanches, it contributes to $99.44\%$ to the first moment $\langle S \rangle.$ (This gives a precise sense to Eq.~(\ref{approx-quasi}) with $p_0=0.027$). We do not further study the other avalanches here, since their characteristics is highly dependent on the chosen driving.

\begin{figure}
\centerline{\fig{8cm}{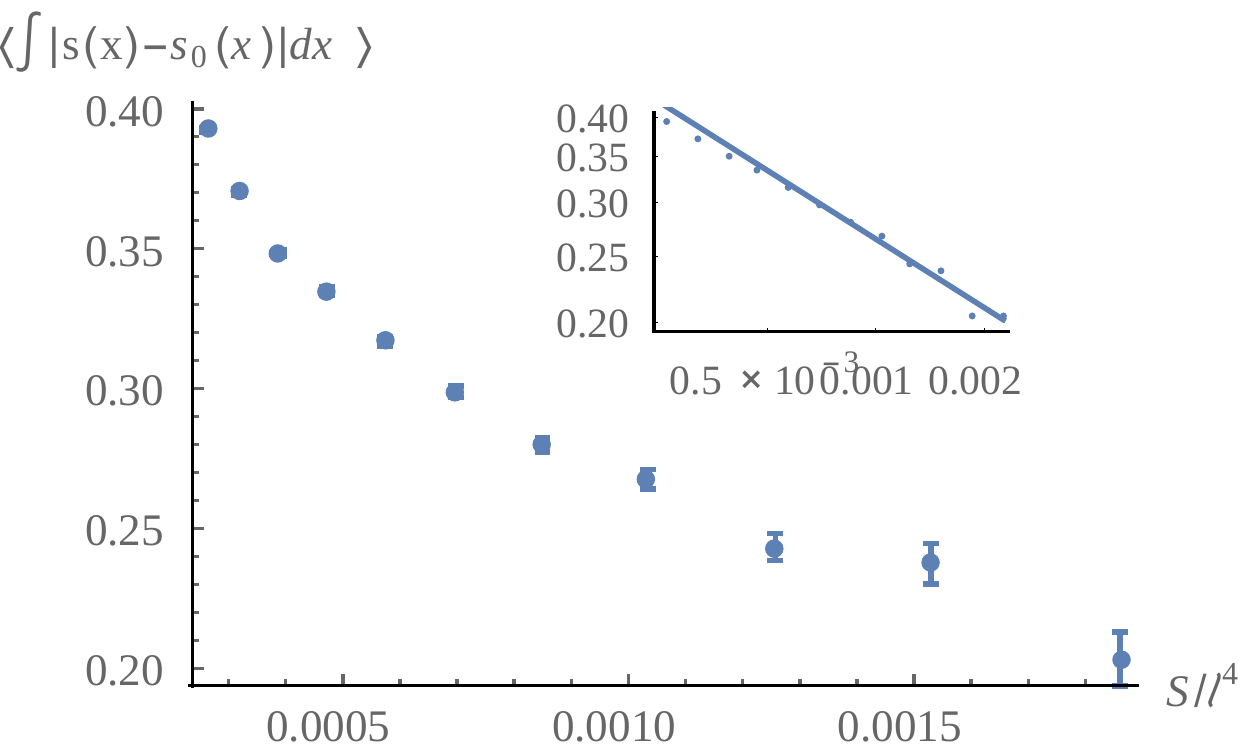} ~~ \fig{8cm}{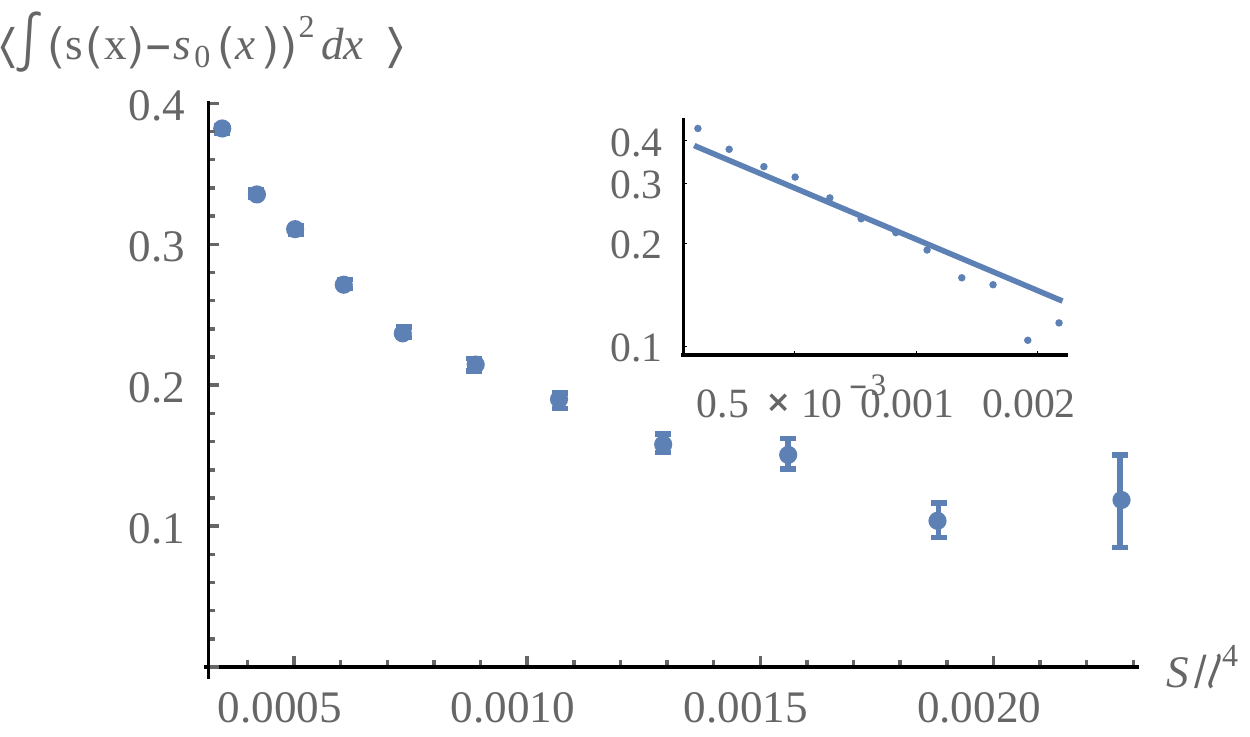} }
\caption{ Left: (resp. Right:) Mean-value at fixed $S/\ell^4$ of the $L^1$ (resp. squared $L^2$) norm between the optimal shape and the simulated shape $\int_{-1/2}^{1/2} \rmd x |s(x)-s_0(x)|$ (resp. $\int_{-1/2}^{1/2} \rmd x (s(x)-s_0(x))^2$). Inset: log-log plot of the same quantity, fitted with a power-law $(\ell^4/S)^{1/3}$ (resp.  $(\ell^4/S)^{1/2}$). Error bars are given using a Gaussian estimate and a numerical measurement of the variance. The fits with power-laws are of low quality, but  sufficient to prove that the convergence is slower than $\ell^4/S$.}
\label{f:cvsdp}
\end{figure}

\paragraph{The convergence to the saddle-point.} We now check the striking prediction that the shape of avalanches becomes {\it deterministic} in the limit of large $S/\ell^4$. To this aim, we measure the distance between the optimal shape $s_0(x)=\phi_0^2(x)$ and the simulated shapes $s(x)$ using either the $L^1$ or the (squared) $L^2$ canonical norms (see Figure \ref{f:cvsdp}). As expected, we find that the mean value of these quantities at fixed $S/\ell^4$ converge to $0$ as $S/\ell^4$ becomes larger. However, we find that the rate of convergence of these quantities is slower than what is expected from perturbation theory (this is developed in the next section), which predicts for both a convergence as $\ell^4/S$. This will be taken into account when comparing the numerical results to the prediction of perturbation theory for the fluctuations around the optimal shape.

\begin{figure}
\centerline{\fig{8cm}{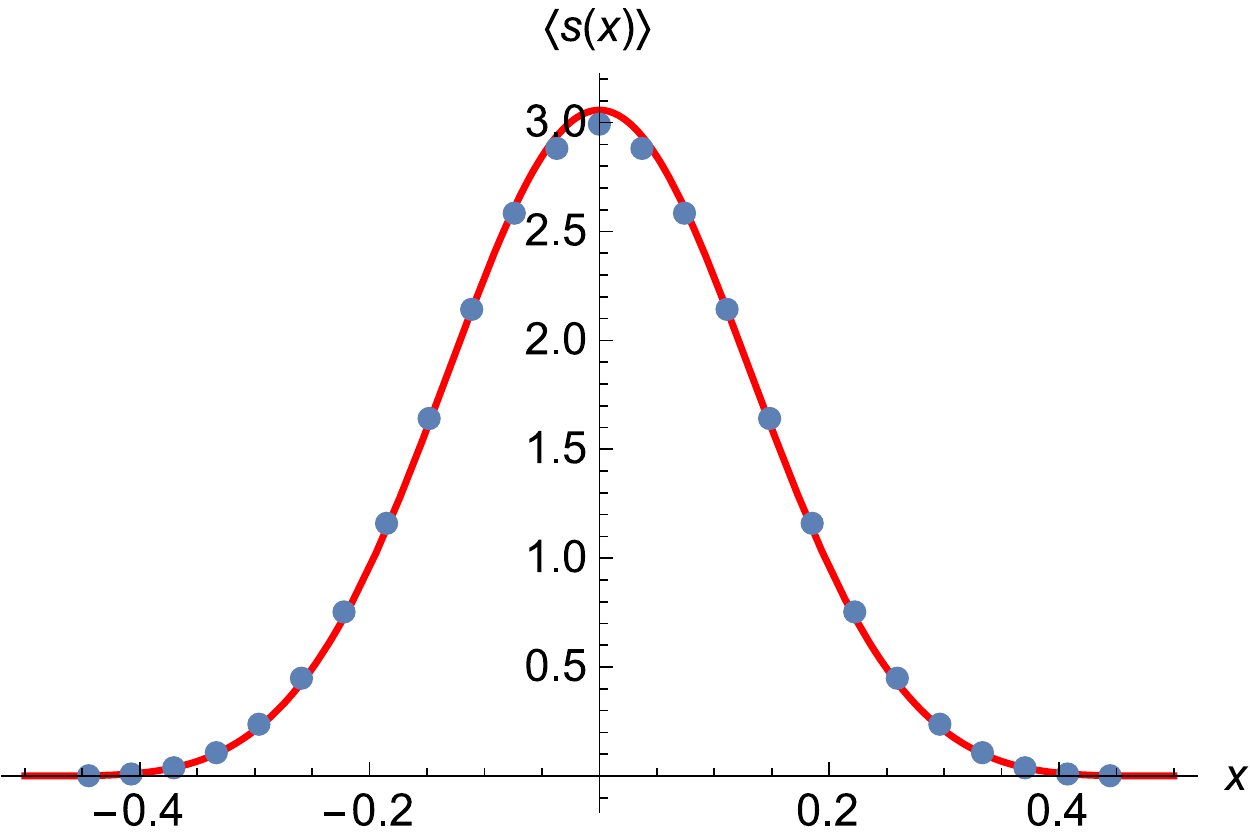} ~~ \fig{8cm}{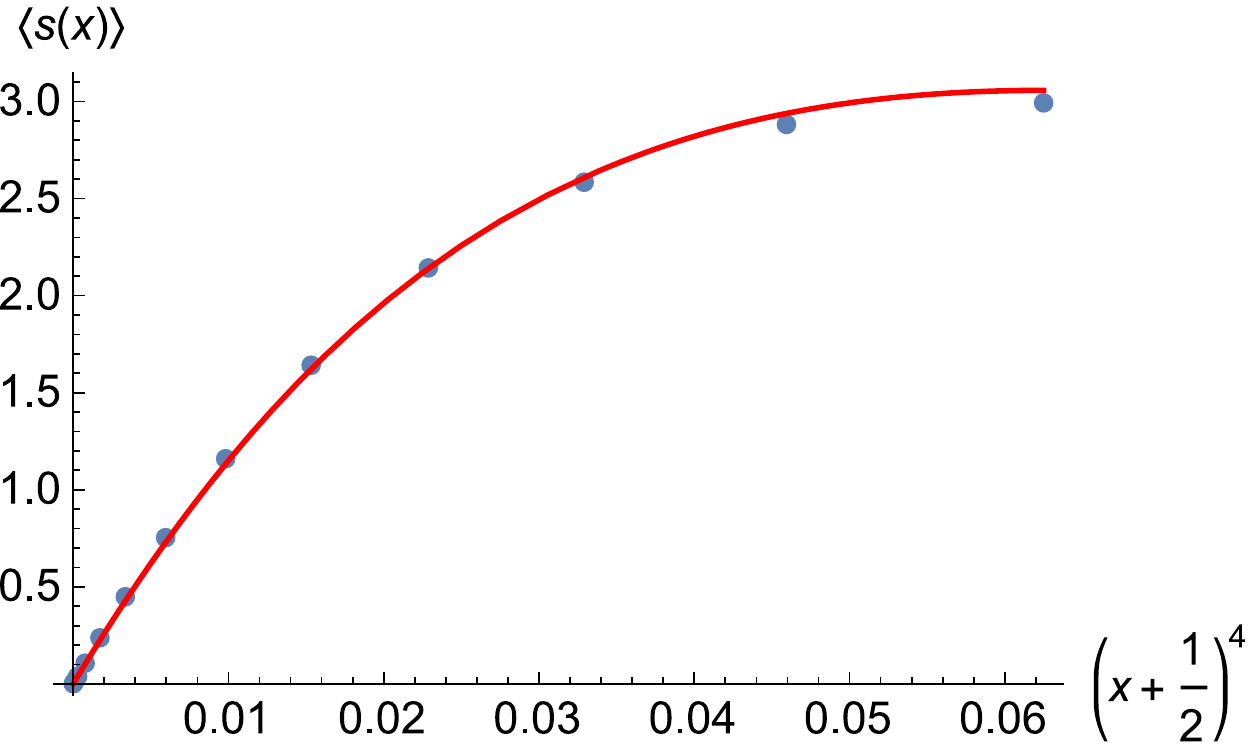} }
\caption{Left: Mean shape obtained by averaging over the $1000$ avalanches with the largest $S/\ell^4$ (blue dots, $0.0011 \leq S/\ell^4\leq 0.0041$), compared to the optimal shape $s_{0}(x)$ (red line). Right: test of the predicted behavior $s(x)\sim (x+1/2)^4$ close to the boundaries.}
\label{f:meanShape}
\end{figure}

\paragraph{The mean shape of avalanches.} Finally, we verify on Figure \ref{f:meanShape} that the mean shape $\langle s(x) \rangle$ is given by the optimal shape $s_0(x)$ for large $S/\ell^4$. We also explicitly check that the mean-shape decays as $(x\pm1/2)^4$ close to the boundaries. The agreement is very good, though one can notice that the numerical mean shape is slightly flatter than expected. This observation motivates a study of the fluctuations of the shape around the optimal shape.

\section{Fluctuations around the saddle point}
\label{s:fluctuations}

\subsection{Field theoretic analysis}
We  now study the fluctuations around the saddle point $\phi_0(x)$. To this aim, we set
\be
\phi(x) = \phi_0(x) + \delta \phi(x)\ . 
\ee
Expanding the action yields
\begin{figure}
\centerline{\fig{8cm}{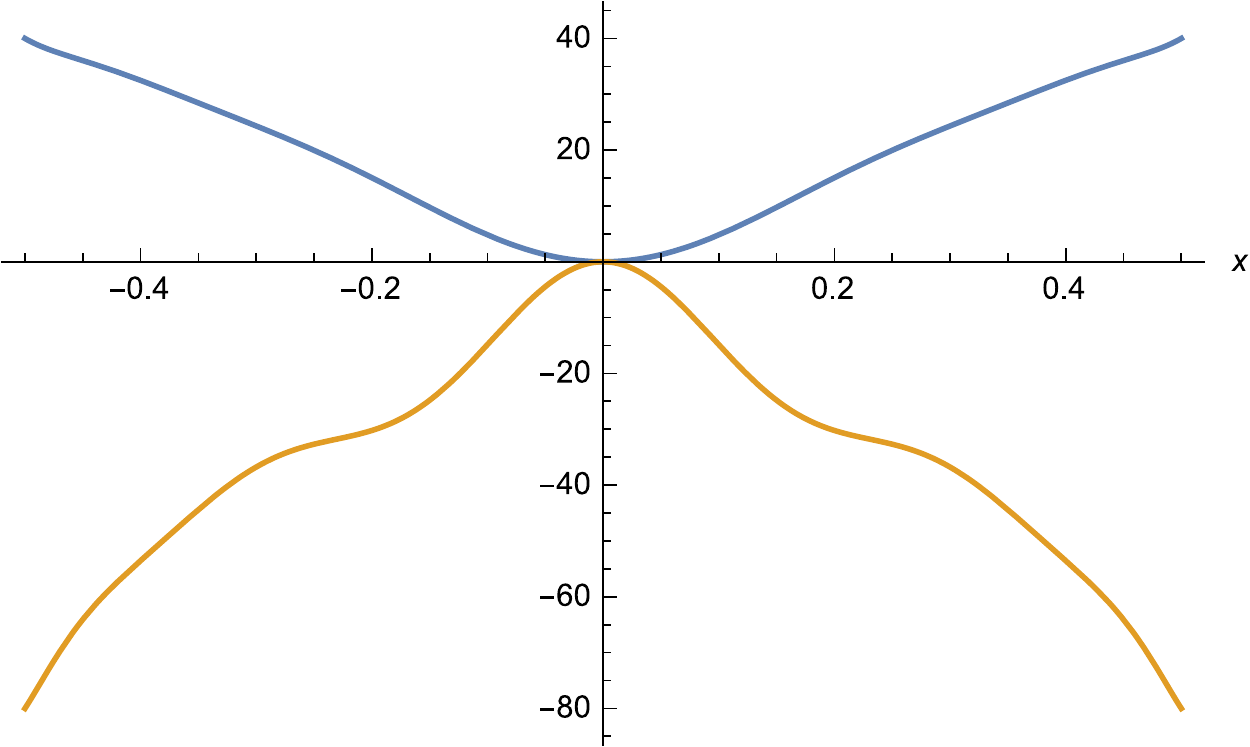}~~~~\fig{8cm}{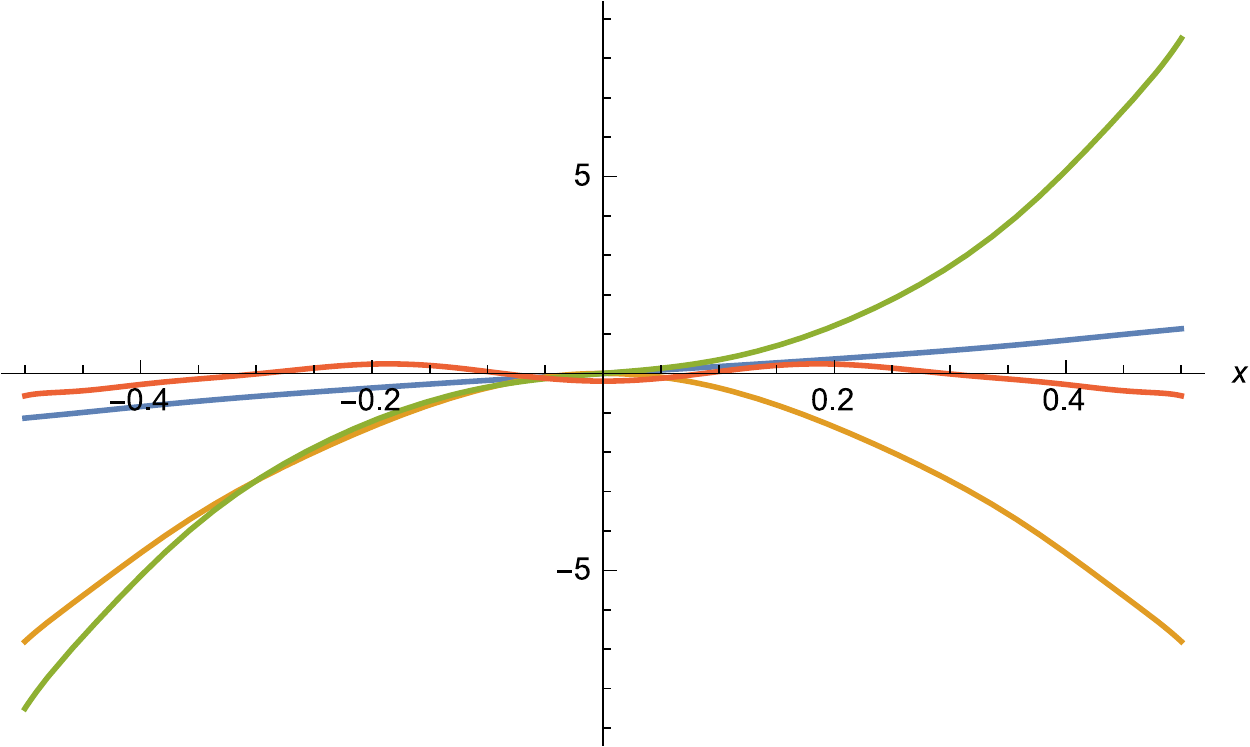}}
\caption{The coefficients multiplying the different terms in ${\cal H}_2[\phi_{0},\delta \phi]$ (left) and ${\cal H}_3[\phi_{0},\delta \phi]$ (right), after replacing $\delta \phi(x)\to (x^{2}-1/4)^{2}$ and $\delta \phi'(x)\to x^{2}-1/4$. This shows that $\delta \phi(x)$ must have the same behavior $ \sim (x^{2}-1/4)^{2}$ as $\phi_{0}(x)$ at the boundary $x=\pm 1/2$.}
\label{f:coeffs}
\end{figure}
\bea \label{67}
\fl {\cal H}_{{\rm el}}[\phi ] = {\cal E}_0  + 2 {\cal E}_0 \int_x \phi_0(x) \delta \phi(x) + {\cal H}_2[\phi_{0},\delta \phi]+ {\cal H}_3[\phi_{0},\delta \phi] + ... \\
\fl{\cal H}_2[\phi_{0},\delta \phi] = \int_{x} \delta \phi(x)^{2}\left[ \frac{20 \phi _0'(x){}^2 \phi _0''(x)}{\phi
   _0(x){}^3}-\frac{15 \phi _0'(x){}^4}{\phi _0(x){}^4}\right] + \delta \phi '(x)^2 \frac{10  \phi _0'(x){}^2}{\phi _0(x){}^2} + \delta \phi ''(x)^2 \\
\fl{\cal H}_3[\phi_{0},\delta \phi] =5 \int_{x} \delta \phi (x)^2 \delta \phi '(x) 
   \frac{3 \phi_0'(x){}^3- \phi _0(x) \phi _0'(x) \phi_0''(x)}{\phi _0(x){}^4}
   -4\delta \phi (x)
   \delta \phi '(x)^2 \frac{  \phi _0'(x){}^2}{\phi
   _0(x){}^3} \nn\\
\fl ~~~~~~~~~~~~~~~~~~~ + \frac43 \delta \phi '(x)^3 \frac{ \phi _0'(x)}{ \phi
   _0(x){}^2}- \frac13 \delta \phi (x)^3 \frac{  \phi
   _0'''(x) \phi _0'(x)+\phi _0''(x)^2 }{ \phi
   _0(x){}^3}
\eea
The first term  in Eq.~(\ref{67}) comes from the saddle-point equation (\ref{SP-eq}) at $\phi= \phi_0$, ${\cal A}_0 \, \phi_0(x)  = \frac12 \frac{\delta {\cal H}_{\rm el}[\phi]}{\delta \phi (x)}|_{\phi(x) = \phi_0(x)}$ together with (\ref{elastic-energy-col}). We have used our freedom to integrate by part to arrive at these expressions: For ${\cal H}_2[\phi_{0},\delta \phi] $ we gave a form in which each term is proportional to the square of a $\delta \phi$-derivative. For  the cubic term, which  is used in perturbation theory our strategy is different: Since derivatives of $\left< \delta \phi(x) \delta\phi(y)\right>_{{\cal H}_2}$ are numerically unstable, we wrote this expression without a second derivative $\delta\phi''(x)$.

\begin{figure}
\centerline{\fig{8.5cm}{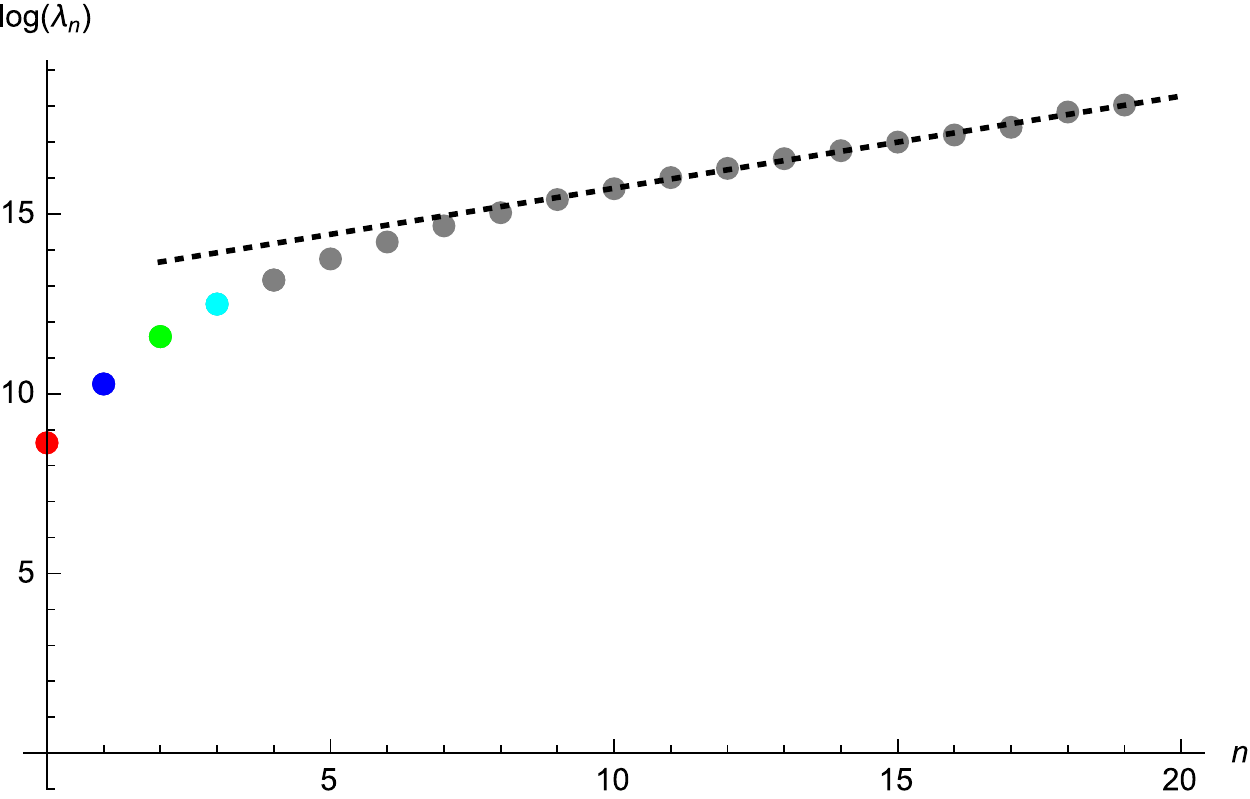} ~~ \fig{8.5cm}{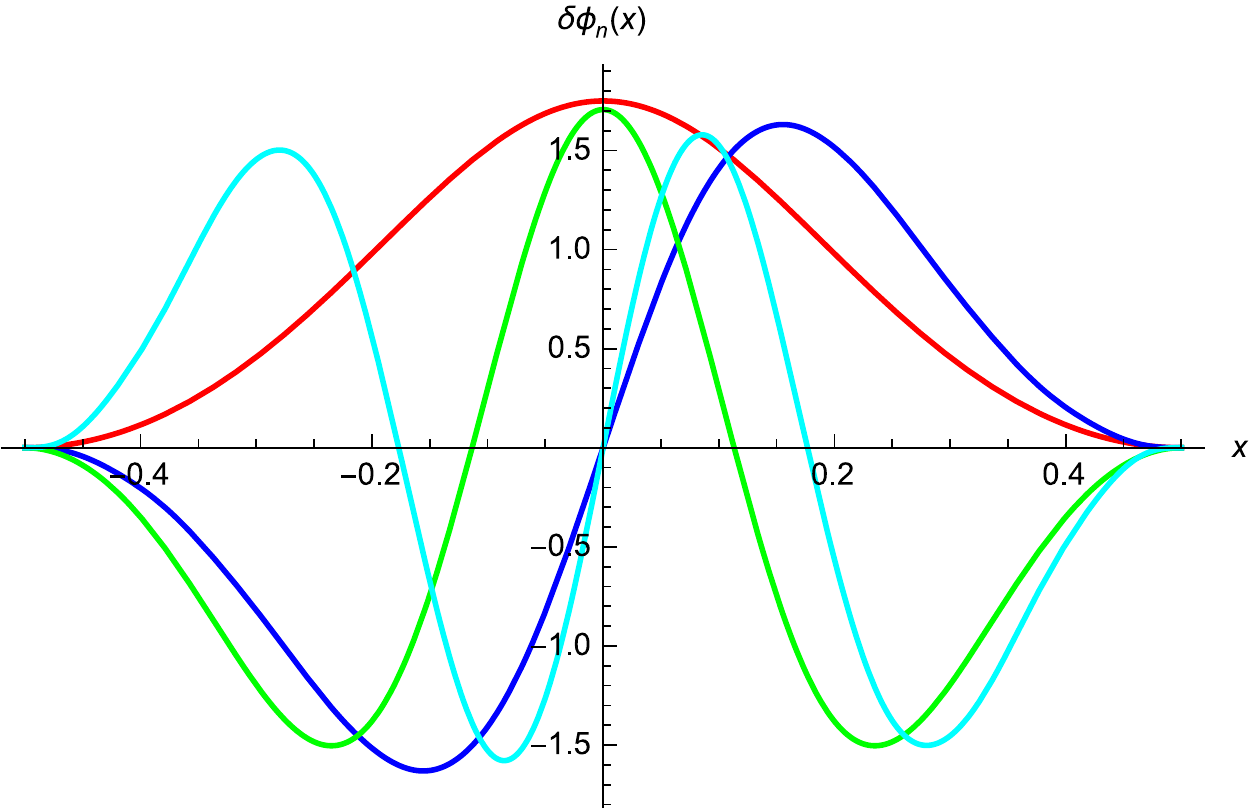}}
\caption{Left: The spectrum of ${\cal M}$. The smallest eigenvalue is $\lambda_0=2 {\cal E}_0$ (given with precision $10^{-4}$ for $n_{\rm max}=10$. The next two eigenvalues are $\lambda_1=5.143 \lambda_0$, and $\lambda_2 = 19.20 \lambda_0$.   Eigenvalues for large
modes grow exponentially with the mode, $\ln \lambda_n \simeq 13.1 + 0.256\, n$ (black dashed line), showing that the spectrum
of fluctuations is massive. The lowest modes are colored in red, blue, orange and cyan. Right: Plot of the first four eigenfunctions in the same colors as the corresponding eigenvalues. $\delta \phi_n(x)$ has $n$ nodes.}
\label{f:M-spectrum}
\end{figure}

To evaluate the coefficients, we use the variational ansatz (\ref{var-ansatz}), with the optimal $\vec c$ of Eq.~(\ref{c-sol}). The plot in Figure \ref{f:coeffs} shows that  $\delta \phi(x)$ should have the same behavior $ \sim (x^{2}-1/4)^{2}$ as $\phi_{0}(x)$ at the boundary $x=\pm 1/2$. We therefore make the  ansatz
\be\label{delta-phi-ansatz}
\delta \phi  (x) = a_0 v_0(x)+ \sum_{n=1}^{n_{\rm max}} \Big[a_{2n-1}v_n(x)   + a_{2n} u_n(x) \Big]  \ .
\ee
The  basis $u_n(x)$, $v_n(x)$ is constructed using Gram-Schmidt out of 
\bea
\bar v_0(x) = \sqrt{\frac{2} 3 } \Big[ 1+\cos( 2\pi x) \Big]\\
\bar v_n(x)= (-1)^{n+1} \cos (2 \pi  (n+1) x)+\cos (2 \pi  x) \, \quad \mbox{for~~}{n\ge 1} \\
\bar u_n(x) = \frac{(n+1) \sin (2 \pi  x)+(-1)^{n+1} \sin (2 \pi (n+1) x)}{\sqrt{\frac{n^2}{2}+n+1}}\ .
\eea
 This basis is  orthonormal.
In this basis, the energy ${\cal H}_2[\phi_{0},\delta \phi]$ can be written as
\be
{\cal H}_2[\phi_{0},\delta \phi] = \frac12 \int_{x,y} \delta\phi(x) {\cal M}(x,y) \delta \phi(y) = \frac12\sum_{i,j} {\cal M}_{ij}a_{i} a_{j}\ .
\ee
This defines ${\cal M}$ which we now diagonalize. Its lowest eigenvalue is $\lambda_0= 2 {\cal E}_0$, with eigenfunction $\delta\phi_0(x)= \phi_0(x)$. 
This can be proven with the help of the saddle-point equation (\ref{SP-eq}). The higher eigenfunctions $\delta \phi_n(x)$ have $n$ knots, see Figure \ref{f:M-spectrum}. Since ${\cal M}$ is symmetric they  form an orthonormal basis. The spectrum is massive (no soft massless modes); we observe that $\ln \lambda_n \simeq 13.1 + 0.256 n$, i.e.\ the eigenvalues grow in geometric progression. This ensures that a truncation at $n_{\rm max}=10$ is sufficient for  practical purposes.

A delicate problem is to obtain results at fixed $\int_x \phi(x)^{2}=1$.
To do so, we write for the  expectation value of an observable ${\cal O}[\phi]$ 
\bea \label{obsO}
\fl
\left< {\cal O}[\phi] \right>  =& \frac1{\left<1\right>} \int {\cal D}[\phi] \, {\cal O}[\phi] \, \delta \Big(\int_{x} \phi^2(x)-1\Big)  \exp\left(-\frac{S}{\ell^{4}} \Big\{ {\cal H}_{\rm el} [\phi_{0},\delta \phi] - {\cal E}_0  \Big\} \right)
\nn\\
\fl\hphantom{\left< {\cal O}[\phi] \right>} =& \frac1{\left<1\right>} \int {\cal D}[\phi] \, {\cal O}[\phi] \, \delta \Big(\int_{x} \phi^2(x)-1\Big) \nn\\
\fl& \times \exp\left(-\frac{S}{\ell^{4}} \Big\{ {\cal H}_{2} [\phi_{0},\delta \phi] -  {\cal E}_0 \int_x  \delta \phi(x)^2  + {\cal H}_{3} [\phi_{0},\delta \phi]+{\cal H}_{4} [\phi_{0},\delta \phi]+...\Big\} \right)
\eea
We subtracted the constant ${\cal E}_{0}$ from the energy in the path integral and used the constraint $\int_x \phi(x)^2 =1$ to rewrite the linear term appearing in (\ref{67}) as a quadratic term: $2 {\cal E}_0 \int_x \phi_0(x) \delta \phi(x) = -{\cal E}_0 \int_x \delta \phi(x)^2$. It ensures that the minimum of the exponential factor  at $\delta \phi(x)=0$ becomes a global saddle point; in addition, the lowest-energy fluctuation $\delta \phi_{0}$ has zero energy.
If we write $\phi(x)$ in the basis of eigenmodes $\delta \phi_{n}(x)$ of ${ {\cal M} }$, i.e.\  
\be\label{73}
\phi(x) = \phi_{0}(x) + \sum_{n=0}^{\infty}  a_{n} \delta \phi_{n}(x) \equiv  (1+a_{0})\phi_{0}(x) + \sum_{n=1}^{\infty}  a_{n} \delta \phi_{n}(x)\ ,
\ee
then
\be
\int_{x} \phi(x)^{2} = \int_{x} \left[ \phi_{0}(x) + \sum_{n=0}^{\infty}  a_{n} \delta \phi_{n}(x)  \right]^{2} = (1+a_{0})^{2}+ \sum_{n=1}^{\infty} a_{n}^{2}\ .
\ee
Solving $\int_{x} \phi(x)^{2} =1$ for $a_{0}$ yields 
\be\label{75}
a_{0}= \sqrt{1 - \sum_{n=1}^{\infty} a_{n}^{2} }-1\qquad \Longrightarrow\qquad a_{0} = - \frac12   \sum_{n=1}^{\infty} a_{n}^{2}   + ...
\ee
With this, the path-integral (\ref{obsO}) can be written using equations~(\ref{73}) and (\ref{75}) as 
\bea
\fl
~~~~~\left< {\cal O}[\phi] \right> =&  \frac1{\left<1\right>} \prod_{n=1}^{\infty} \rmd a_{n} \, {\cal O}[\phi] \, \left( {1 - \sum_{n=1}^{\infty} a_{n}^{2} }\right)^{\!\!-\frac 1 2}\nn\\
\fl &\times \exp\!\left(-\frac{S}{\ell^{4}} \bigg\{ \sum_{n=1}^{\infty} \frac{\lambda_{n} -\lambda_{0}}{2}a_{n}^{2} + {\cal H}_{3} [\phi_{0},\delta \phi]+{\cal H}_{4} [\phi_{0},\delta \phi]+...\bigg\} \right).
\eea
The factor of $\left( {1 - \sum_{n=1}^{\infty} a_{n}^{2} }\right)^{\!\!-\frac 1 2}$ comes from the derivative of the $\delta$-function, which has been used to eliminate the integration over $a_{0}$. Note that the Jacobian of the transformation from $\prod_x \rmd\phi(x)$ to $\prod_{n} \rmd a_{n}$ is 
$
\det \left(\delta \phi_{n}(x) \right)_{x\in [-\frac{1}{2} , \frac{1}{2}], n \in \mathbb{N}} = 1
$,  since the $\delta \phi_{n}(x)$ are orthonormal.

Hence, to leading order in an expansion in $\ell^4/S$, the expectation value of an observable of $\delta \phi(x)$ can be obtained using the decomposition $\delta \phi(x) =  \sum_{i=0}^\infty a_i\, \delta \phi_i(x)$, where $a_0$ is given by (\ref{75}) and the $a_i$ are centered Gaussian variables with correlation matrix ${\cal M}'$ defined for $i,j\geq 1$ by
\be
\left<a_{i}a_{j} \right>_{{\cal M}'} := \frac{\ell^{4}}{S} \frac{\delta_{ij}}{\lambda_{i}-\lambda_{0}} \ .
\ee

One  then uses Wick's theorem for expectation values of  $\delta \phi$. As an example, the 2-point correlation function is
\bea
\left< \delta \phi(y) \delta \phi(z)\right>   _{ {\cal H} }  && = \sum_{i=0}^{\infty} \sum_{ j=0}^{\infty}  \left< a_i a_j \right>_{{\cal M}'} \delta \phi_i (y) \delta \phi_j (z) + O\Big(\frac{\ell^8}{S^2} \Big) \nonumber \\
 && = \frac{\ell^4}{S} \sum_{i=1}^{\infty} \frac{\delta \phi_i ( y) \delta \phi_i (z)}{\lambda_i - \lambda_0}  + O\Big(\frac{\ell^8}{S^2} \Big) \ .
\eea

\begin{figure}
\centerline{\fig{8.5cm}{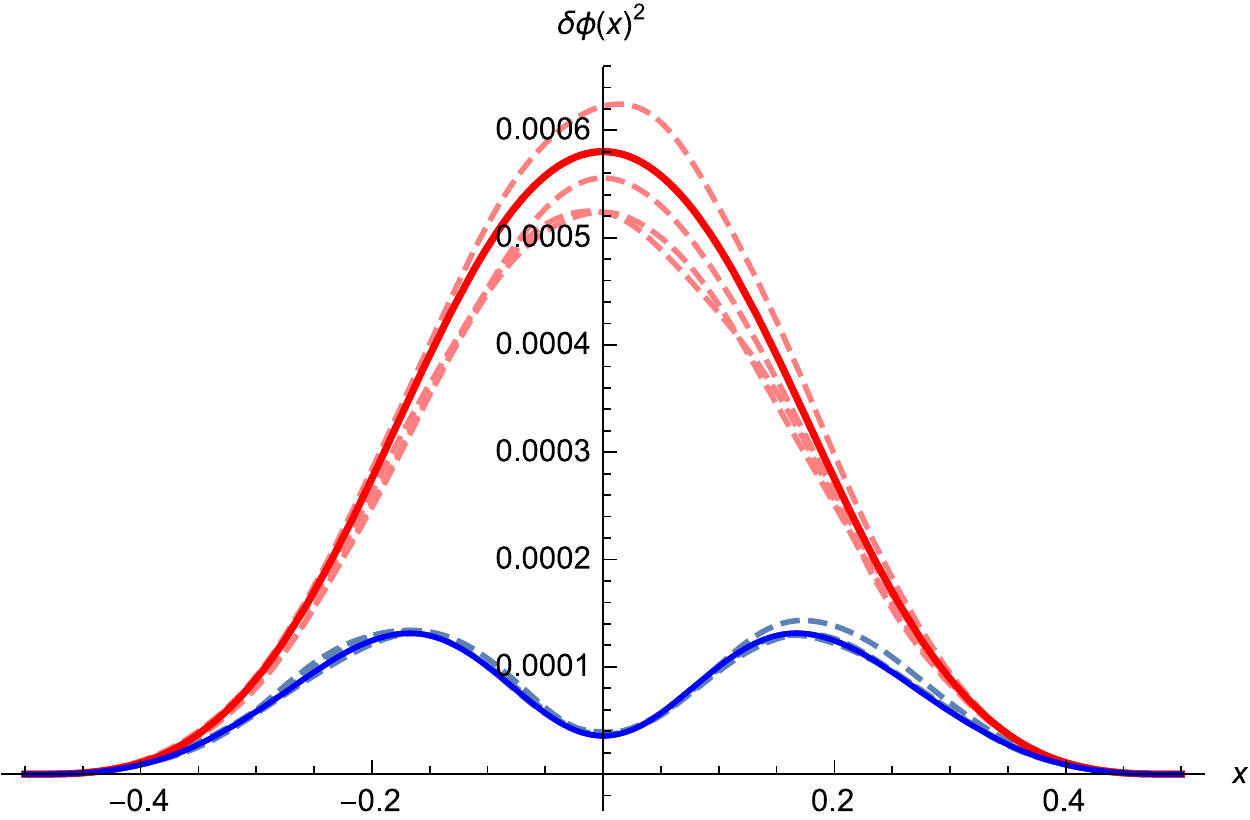}} 
\caption{Left: Plot of the fluctuations $\delta \phi(x)^{2}$ (blue solid line), and including the mode $\delta \phi_0$ (red solid line). The dashed lines are averages over 500 samples using Eq.~(\ref{sample-algo}), including (top pink) or excluding (bottom, blue-gray) this mode.}
\label{f:deltaphi2}
\end{figure}

\subsection{Generating a random configuration, and exact sampling}

Our setting  allows us to generate a random fluctuation with the measure given by the the leading behaviour of ${\cal H}$ for large $S/\ell^4$: Denote by $g_{n}$ a series of uncorrelated Gaussian random numbers with mean zero and variance 1.  Then 
\be\label{sample-algo}
\delta \phi(x)^{\rm rand} = \sum_{n=0}^\infty a_n \delta \phi_n(x)\ , ~~~\mbox{ with }~~~  a_n =\sqrt{ \frac{\ell^{4}}{S}}  \, \frac{g_{n }}{\sqrt{ \lambda_{n}-\lambda_{0} }} \, \mbox {\quad for \quad } n>0\ ,
\ee
and $a_0$ given by Eq.~(\ref{75}). In Figure \ref{f:deltaphi2} (left) we show as an example the expectation of $\delta \phi(x)^{2}$ (solid blue line). This is compared to the average over 500 realizations drawn with the measure  (\ref{sample-algo}), repeated 5 times (the three gray-blue lines, lower set of curves). To illustrate the importance to properly eliminate the mode $\phi_0(x)$, the upper (red) curves are obtained without the constraint on $\int_x \phi^2(x)$, i.e.\ including fluctuations proportional to $\phi_0(x)$ (with amplitude  $\sim1/\sqrt{\lambda_{0}}) $, and not constraining them by Eq.~(\ref{75}).

\begin{figure}
\centerline{\fig{8.5cm}{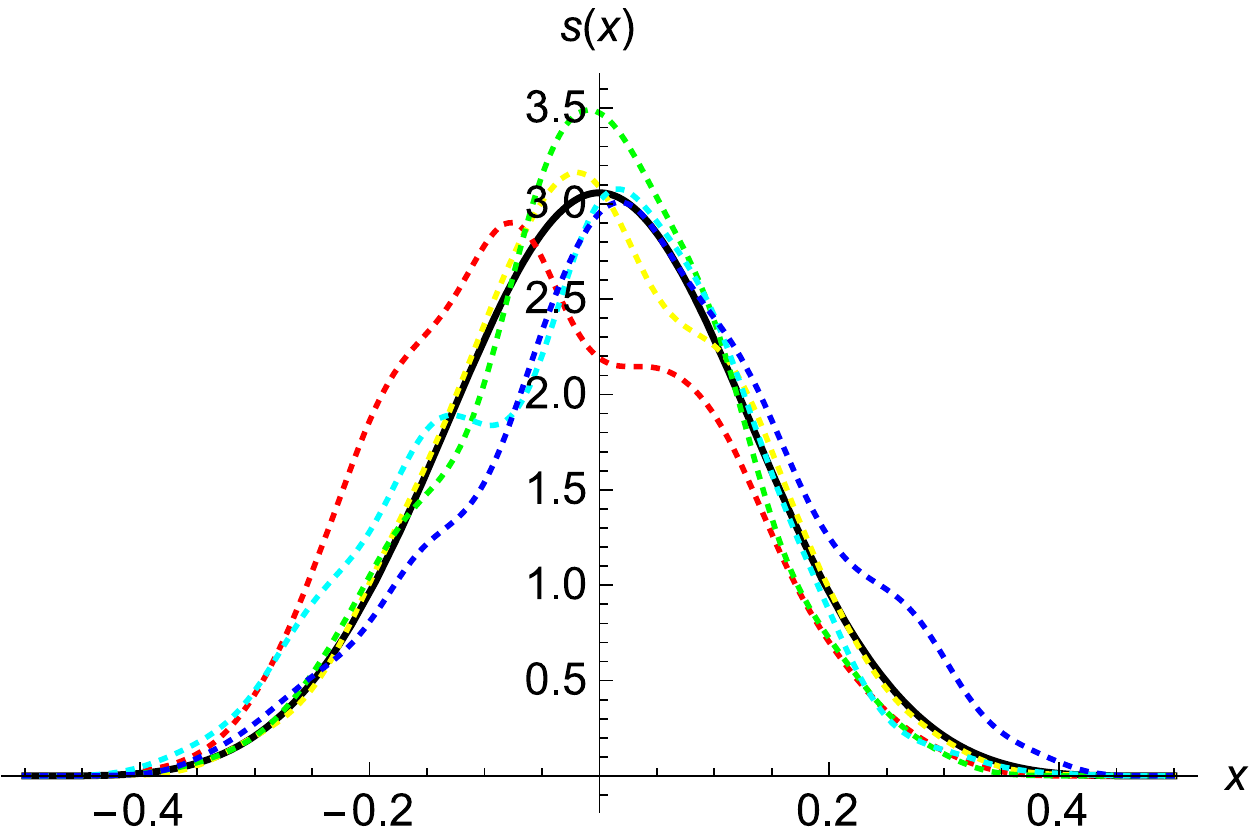}~~\fig{8.5cm}{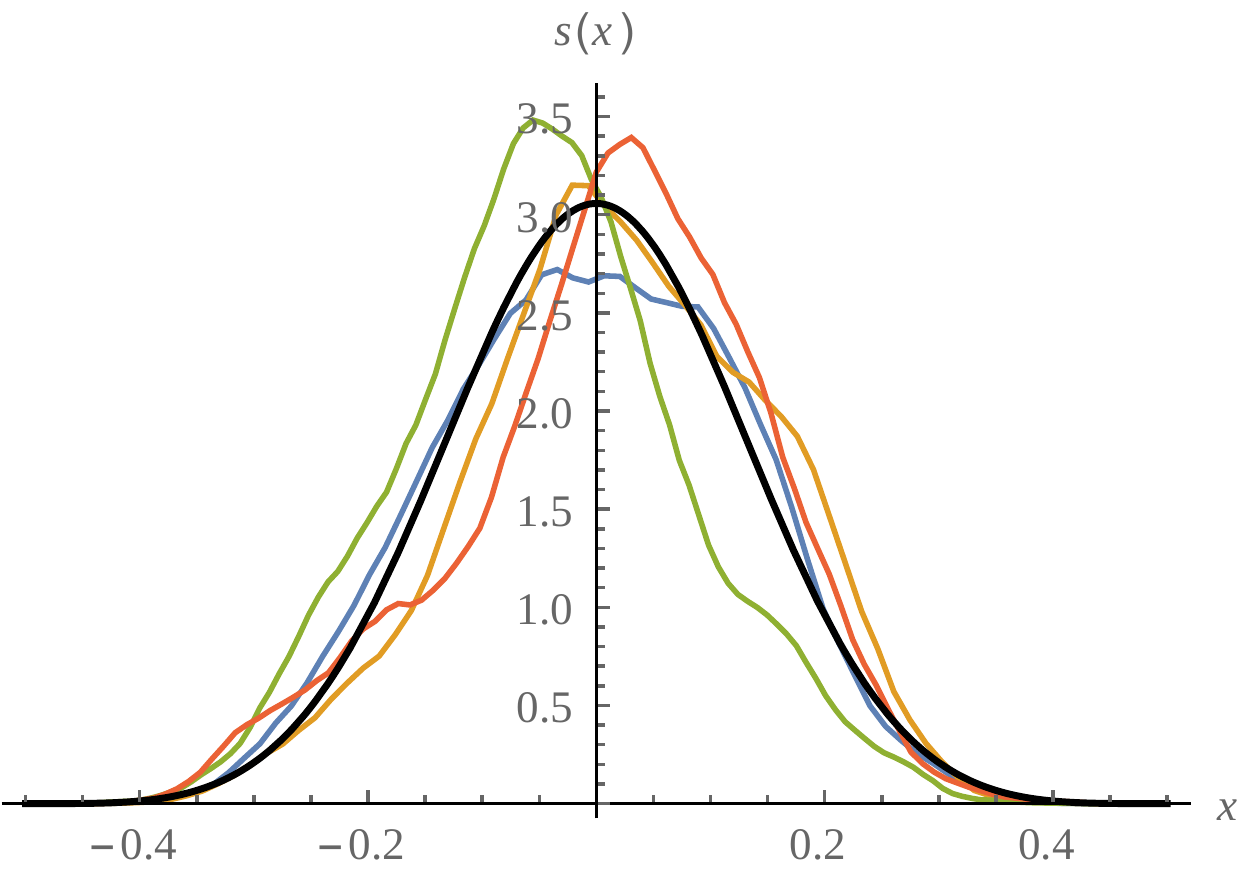}} 
\caption{Left: Plot of the normalized shape $\big[\phi_0(x) +\sqrt{\frac{\ell^4}{{S}}} \delta\phi(x)\big]^2$ for $\frac{\ell^4}{S}=3 \times 10^{-3}$. Right: The same functions  from numerical simulations.}
\label{f:random-shape}
\end{figure}

On Figure \ref{f:random-shape} we show five realizations for the shape drawn from the measure (\ref{sample-algo}), and compare this to numerical simulations at the same ratio $S/\ell^4$. The  agreement is quite good.  
 
We can use this formulation for an efficient algorithm, known in the literature as {\em importance sampling} \cite{Krauth}.  One writes\bea\label{exact-sampling}
\fl
\left< {\cal O}[\phi] \right>  &= \frac1{\left<1\right>}\Bigg< {\cal O}[\phi] \, \left( {1 - \sum_{n=1}^{\infty} a_{n}^{2}
}\right)^{\!\!-\frac 1 2}\exp\!\left(-\frac{S}{\ell^{4}} \Big\{  {\cal H}_{3}[\phi_0,\delta \phi] +{\cal H}_4[\phi_{0},\delta
\phi]+ ...  \Big\} \right) \Bigg>_{\!\!{\cal M}'} \nn\\
\fl &= \frac1{\left<1\right>}\Bigg<
{\cal O}[\phi] \, \left( {1 - \sum_{n=1}^{\infty} a_{n}^{2}
}\right)^{\!\!-\frac 1 2}\exp\!\left(-\frac{S}{\ell^{4}} \Big\{
  {\cal H}_{{\rm el}}[\phi_0 +\delta \phi]-{\cal E}_0- \sum_{n=1}^{\infty}
\frac{\lambda_{n}-\lambda_{0}}{2}a_{n}^{2} \Big\}
 \right)\Bigg>_{\!\!{\cal M}'}\ .
\eea
In the second line we reintroduced the full Hamiltonian ${\cal H}_{{\rm el}}$ using Eq.~(\ref{67}). We will compare to simulations  below. 

\subsection{The leading correction to the shape at large sizes}
For  large $S/\ell^4$, the mean shape is given by  the optimal shape $s_0(x)$. For smaller $S/\ell^4$, this mean shape becomes flatter, an effect which we now investigate using perturbation theory. Consider
\bea \label{delta-s-pert}
\fl \left<  \delta s(x) \right> := \left <s(x) -  \phi_{0}(x)^{2} \right> \nn\\
\fl  \qquad = \left< \Big(  2 \phi_{0}(x) \delta \phi(x)+\delta \phi(x)^{2}  \Big)\left( 1- \frac{S}{\ell^{4}} {\cal H}_{3}[\phi_{0},\delta\phi]+ ...\right) \right>_{\!\!{\cal M}' } \nn\\
\fl  \qquad= \frac{\ell^{4}}S \left[   \Big< \delta \phi(x)^{2}\Big>_{\bar {\cal M}'} -2 \,\phi_{0}(x)
\Big<  \delta \phi(x)  {\cal H}_{3}[\phi_{0},\delta\phi]  \Big>
_{ \!{\bar {\cal M}'} } \right] +O\Big(\frac{\ell^4}{S} \Big)^{\!\!2} \ .\label{74}
\eea
The notation  $\bar {\cal M}'$ indicates  that all expectation values are taken at $S/\ell^4=1$, making the factors of $ \frac{S}{\ell^{4}}$ explicit.

\subsection{Fluctuations of the  shape for large avalanches}
We  now consider the fluctuations of the shape of an avalanche in perturbation theory:
\bea
\fl \left<  \delta s(x)^2 \right>_{ c}  &:= \left<  s(x)^2 \right> - \left<  s(x)\right> ^{2}\nn\\
\fl&=
 \Bigg\{ \left< \left[ { \phi_{0}(x)^2+2
\phi_{0}(x)\delta \phi(x)+\delta \phi(x)^{2}} \right]^{2} 
 \left( 1- \frac{S}{\ell^{4}}
{\cal H}_{3}[\phi_{0},\delta\phi]+ ...\right) \right>_{\!\!{\cal M}' } \nn\\
\fl & ~~~~~~~~~~~~~- \left< \left[ { \phi_{0}(x)^2+2
\phi_{0}(x)\delta \phi(x)+\delta \phi(x)^{2}} \right]
 \left( 1- \frac{S}{\ell^{4}}
{\cal H}_{3}[\phi_{0},\delta\phi]+ ...\right) \right>_{\!\!{\cal M}' } ^{\!\!2} 
\Bigg\} \nn\\
\fl&=4\left( \frac{\ell^{4}}{S}\right) \phi_0(x)^2 \left< \delta \phi (x)^2  \right>_{\!{\bar{\cal M}}'} +O\Big(\frac{\ell^4}{S} \Big)^{\!\!2}  \ .
\label{deltaS2}
\eea
Note that the only term which survives is the contraction between one $\delta \phi (x)$ of each factor $s(x)$.

\subsection{Asymmetry of an avalanche}
Another interesting observable is the asymmetry ${\cal A}$ of an avalanche, defined by 
\be
{\cal A} := 2 \int_x  x  \; \phi^2(x) \ .
\ee
By construction $-1 \le {\cal A}\le 1$. The asymmetry has mean zero $\left< {\cal A}\right>=0$, and variance given in perturbation theory by \be
\left< {\cal A}^2 \right> = 16\left( \frac{\ell^{4}}S\right)
\int_{x,y}xy\phi_0(x)\phi_0(y) \left< \delta \phi (x)\delta \phi(y)  \right>_{\!{ \bar{\cal M}}'} =1.1 \times 10^{-5} { \left( \frac{\ell^{4}}S\right)} \ .
\ee
\begin{figure}
\centerline{\fig{8.5cm}{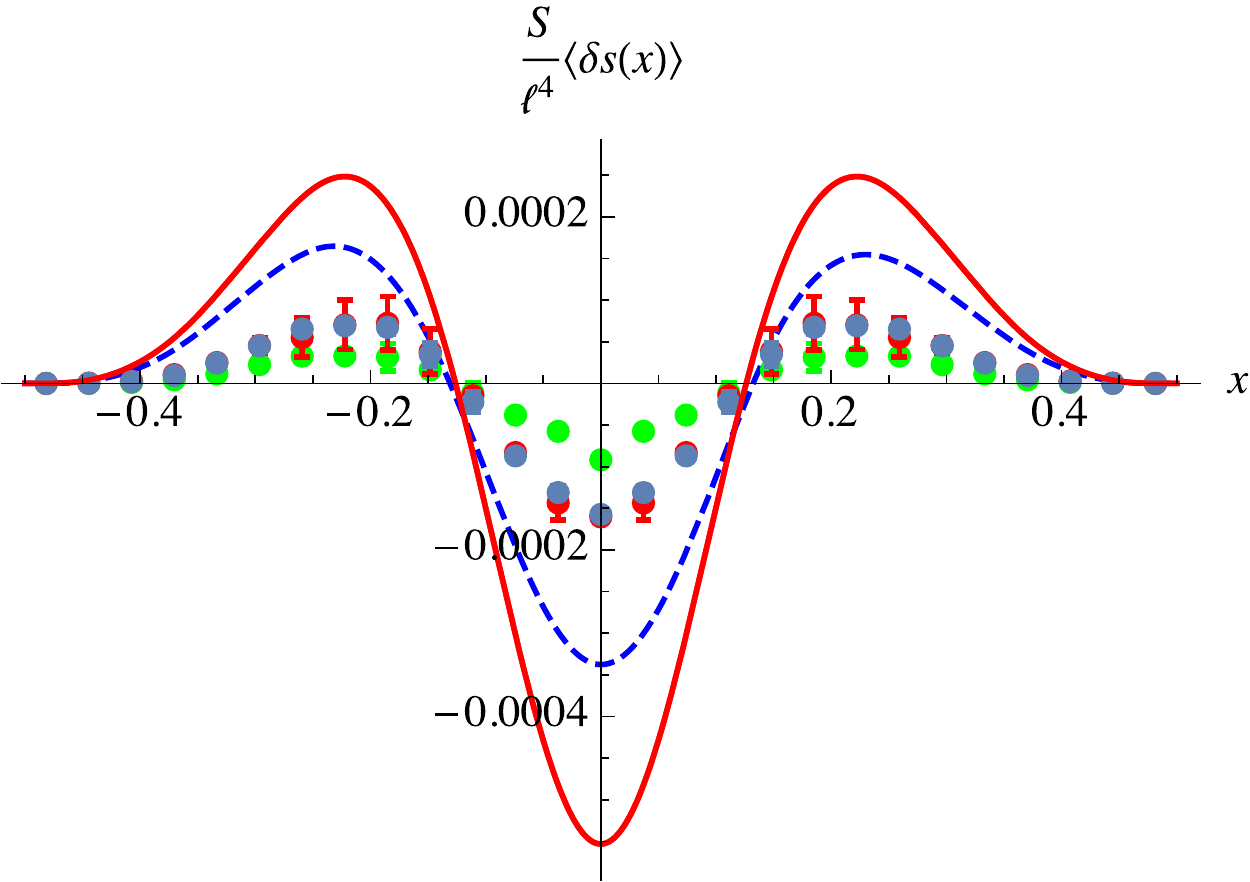} \fig{8.5cm}{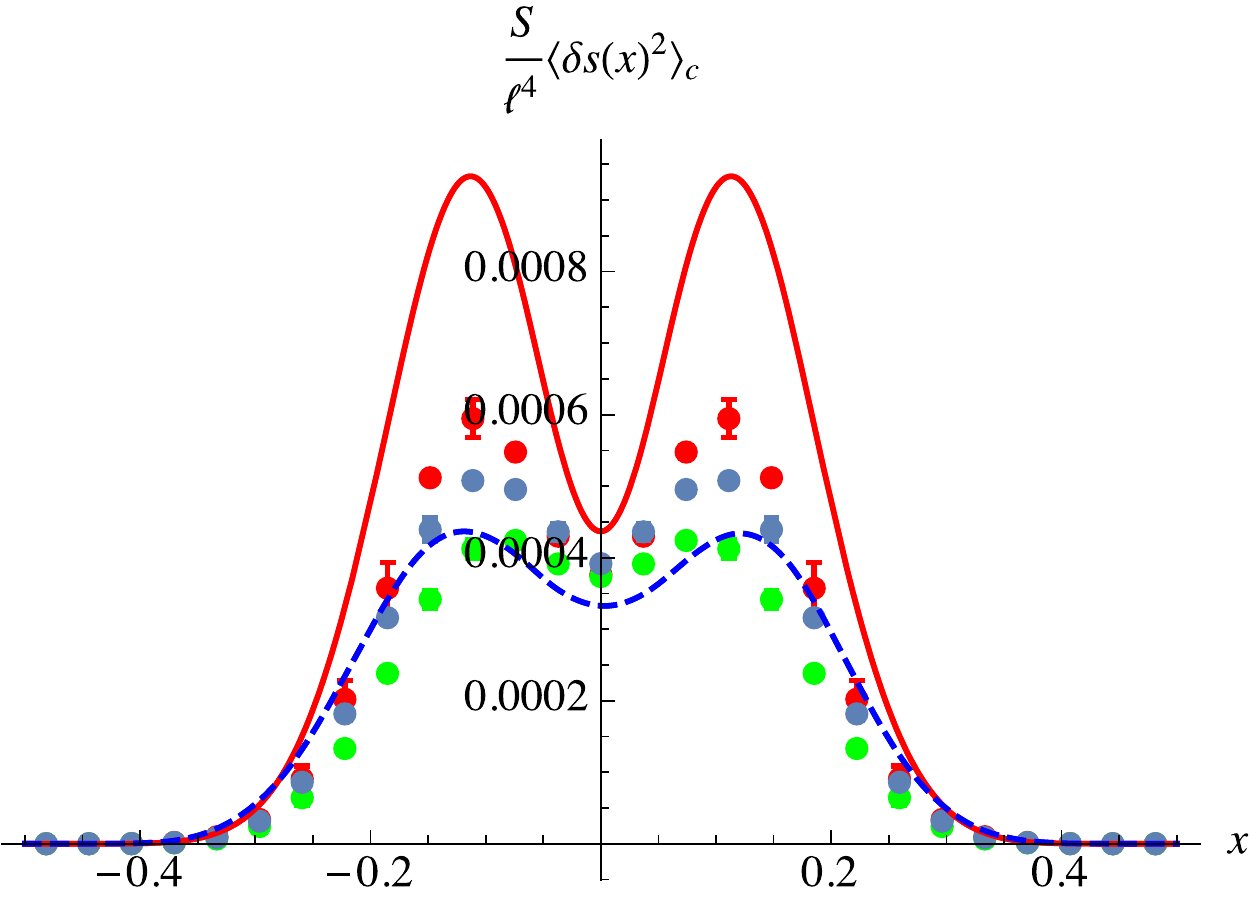}}
\caption{Left: (resp. Right:) normalized mean shape displacement $\langle S/\ell^4 (s(x)-s_{0}(x))\rangle$ (resp. shape fluctuations $\langle S/\ell^4 (s(x)-s_{0}(x))^2\rangle_c$). Red line: result of perturbation theory (\ref{delta-s-pert}) (resp. (\ref{deltaS2})). Dashed-blue line: result from {\em importance sampling} using (\ref{exact-sampling}) for $\ell^4/S = 900$. Dots: results from the simulations for avalanches with aspect-ratio $0.9/1800\leq S/\ell^4 \leq 1.1/1800$ ($7023$ samples, green), $0.9/900\leq S/\ell^4 \leq 1.1/900$ ($946$ samples, blue) and $S/\ell^4 \geq 1.1/1900$ ($734$ samples, red). We take advantage of the symmetry of the observable $\left<s(x)\right> \equiv \left<s(-x)\right>$ to symmetrize the numerical result. We estimate error bars using the difference between the original result and the symmetrized one.}
\label{f:deltaS}
\end{figure}

\subsection{Comparison of the perturbative corrections to the numerics}
We had already shown some results of our numerical simulations above. For large $S/\ell^4$, the perturbation theory developed in the preceding section gives the correction $\left< \delta s(x)\right>$ of the mean shape to the saddle-point solution, as well as the shape fluctuation $\left< \delta s(x)^2\right>_c$ around the saddle-point. However, as already pointed out in section \ref{s:simul+protocol}, the scaling of these  quantities with a factor of $\ell^4/S$ is not seen in the convergence of the numerical simulations to the saddle point, see Figure \ref{f:cvsdp}.
This indicates that, even at $S/\ell^4 \approx 10^{-3}$, the simulations are not yet in the perturbative (first-order) scaling regime. Non-linear corrections are still important, and $\frac {S}{\ell^4} \left< \delta s(x)\right> $ as well as $\frac {S}{\ell^4} \left< \delta s(x)^2\right>_c $ still depend on $\frac {S}{\ell^4} $. 
This is illustrated on Figure \ref{f:deltaS}. 

As can be seen on the left of Figure \ref{f:deltaS} (as well as on the left of Figure \ref{f:meanS}), corrections to the mean shape are very small, of the order of $10^{-4}$,  difficult to measure, and  at the limit of our simulations. The red solid line is the perturbative result (\ref{delta-s-pert}). The points correspond to the same quantity from the numerics with increasing $S/\ell^4$ from green over blue-gray to red (see caption for the precise parameters). The dashed blue line is obtained for $S/\ell^4= 1/900$  via {\em importance sampling}, see equation~(\ref{exact-sampling}) \footnote{For $S/\ell^4 =1/900$, about $44\%$ of the proposed configurations in the importance sampling have a zero-crossing in $s(x)$, and  therefore do not contribute. The measured expectation of the weight is $\left<1\right> = 1.61 \pm 0.012$, showing that averages are not dominated by a few configurations.}. One remarks that the amplitude is lowered as compared to the perturbative result, in  qualitative agreement with the simulations. In view of the difficulty of the numerical simulations, it is very encouraging that at least a qualitative agreement has been obtained, and that {\em importance sampling} explains why the observed corrections are smaller than the perturbative result, in agreement with intuition: the shape has to remain positive. 

The fluctuations around the mean shape, $\frac {S}{\ell^4} \left< \delta s(x)^2\right>_c $, are given on the right of Figure \ref{f:deltaS} with the same color code as previously. One sees that the  numerical results approach the perturbative result for large $S/\ell^4$. In this case,  importance sampling predicts fluctuations slightly smaller than our numerical simulations, which  converge more quickly towards the perturbative result. We remark that numerically the estimation of $\frac {S}{\ell^4} \left< \delta s(x)^2\right>_c $ is less sensitive than the estimation of $\frac {S}{\ell^4} \left< \delta s(x)\right> $. This may be explained by the fact that only the latter quantity involves non-linearities of $\cal H$ at dominant order in $S/\ell^4$.

For the asymmetry we find $\frac {S}{\ell^4} \left< {\cal A}^2 \right>  =  1.1 \times 10^{-5}$ in perturbation theory, and $ 5.97\pm 0.04 \times 10^{-6}$ via exact sampling for $S/\ell^4=1/900$. Numerical simulations give ${\frac {S}{\ell^4} \left< {\cal A}^2 \right> } = (7\pm2)\times 10^{-6}$ for the largest avalanches $S/\ell^4 \geq 0.002$ ($37$ samples), $(5.6\pm0.3)\times 10^{-6}$ for the data with  $1.1/900 \leq S/\ell^4 \leq 0.002$ ($697$ samples), $(4.7\pm0.2)\times 10^{-6}$ for the data with $0.9/900 \leq S/\ell^4 \leq 1.1/900$ ($946$ samples) and $(3.05\pm0.05)\times 10^{-6}$ for the data with $0.9/1800 \leq S/\ell^4 \leq 1.1/1800$ ($7023$ samples). Once again we see that the order of magnitude is correctly predicted (an already non-trivial achievement), and that the numerical results get closer to the perturbative one as $S/\ell^4$  increases.

From a conceptual point of view it is interesting to note that most of the amplitude of the 
``double-peak'' structure observed  on the right of Figure~\ref{f:deltaS}
is   due to the first sub-leading mode $\delta \phi_{1}(x)$ with
one node at $x=0$ (see Fig.~\ref{f:M-spectrum}). The same holds
true for $\left< {\cal A}^2 \right>$.

In conclusion, we have seen that the numerical results agree very well with the theoretical prediction at large $S/\ell^4$, and that the mean shape of avalanches is given by the optimal shape $s_0(x)$ (Figures \ref{f:cvsdp} and \ref{f:meanShape}). The consequence for the tail of the PDF of $S/\ell^4$ was successfully verified (Figure \ref{f:PSvsl4}). For  finite $S/\ell^4$, namely  fluctuations around the optimal shape, we only got a partial, though already satisfying agreement: The discrepancy with the perturbative results  was clearly identified as a consequence of strong non-linearities, even for the largest $S/\ell^4$. This was qualitatively understood by an implementation of {\em importance sampling}, though the remaining discrepancy raises the question of wether our simulations are sufficiently precise to measure  these delicate observables (Figures \ref{f:random-shape} and \ref{f:deltaS}).

\section{Application of our results to realistic interfaces and stationary driving}\label{sec-appli}

Up to now we considered avalanches following a stopped driving (see Section \ref{secBFM}). 
However, as discussed in \cite{LeDoussalWiese2011a,LeDoussalWiese2012a,DobrinevskiLeDoussalWiese2011b} 
this setting also yields the densities for the statistics of quasi-static avalanches in the steady state (Middleton state) for stationary driving in the quasi-static limit ($\dot{w}_{t} =v$ and $v \to 0^+$). These are the avalanche densities defined in Section \ref{quasi}, hence the denomination used in this article.
\smallskip

Furthermore, it was shown in Ref.\ \cite{LeDoussalWiese2012a}, that the BFM is the mean-field theory of an avalanche in the quasi-static limit for an interface in short-ranged disorder with equation of motion
\bea \label{generaloverdamped}
\eta_{0} \partial_t u_{xt} =  \int_y c_{xy}u_{yt} +m^2(w_{xt}-u_{xt}) + F(x , u_{xt}) \ .
\eea
The disorder-force correlator is given by $\overline{F(x , u )F(x' , u' )} = \Delta_{0}(u-u') \delta^d(x-x')$ with $\Delta_0(u)$ a fast decaying function as $|u| \to \infty$ and $c_{xy}$ a convex elastic kernel. The prediction of the functional renormalization group (FRG) for such systems is that, in the quasi-static limit, when $m \to 0$ and for $d = d_{\rm uc}-\epsilon$, $\epsilon\geq 0$ ($d_{\rm uc}=4$ for short-ranged elasticity and more generally $d_{\rm uc}=2 \gamma$ for $g(q) \sim_{q \to \infty} q^{\gamma}$), the physics becomes universal in the small-$m$ limit (e.g.\ independent of microscopic details of the disorder) and entirely controlled by only two relevant couplings, the renormalized friction $\eta_m$ and the renormalized disorder  cumulant $\Delta_m(u)$. The (rescaled and renormalized) second cumulant of the disorder at the fixed point is non-analytic and exhibits a cusp. It is uniformly $O(\epsilon)$, allowing to formulate a controlled perturbative expansion of any observable. For observables associated to a single avalanche, it was shown in \cite{LeDoussalWiese2011a,LeDoussalWiese2012a} that near the upper critical dimension $d_{\rm uc}$ only the behavior of $\Delta_m$ near zero, i.e.\ its cusp, $\Delta_m(u) \simeq_{u \to 0} = -\sigma_m |u|$ plays a role. In this context, the mean-field theory for single-avalanche motion is the BFM studied here, with renormalized parameters $\eta \to \eta_m$ and $\sigma \to \sigma_m$. Hence, the avalanche densities derived in Section \ref{quasi} are exact for interfaces at their upper critical dimension. They also open the way to a perturbative calculation for $d\leq d_{\rm uc}$. Interestingly, some physical systems described by (\ref{generaloverdamped}) are at their upper critical dimension, as  e.g.\ domain walls in certain soft magnets for which $\gamma = 1$ \cite{DurinZapperi2000}. 

\section{Conclusion}
In this article we obtained an exact formula for the joint PDF 
of the local sizes of avalanches in a discrete version of the BFM model. This result is valid
for an arbitrary elasticity matrix and arbitrary monotonous driving. This allowed us to derive the densities describing the quasi-static avalanches in the limit of small driving, and to discuss in depth the physical picture underlying this avalanche process. 
We presented two applications where it was possible to go further in the analytical calculation of
detailed physical properties. For the fully connected model we obtained 
the joint distribution of the local and global jumps. This allowed us to retrieve in a rigorous way the usual
large-$N$ limit, as well as a new regime, and finite-$N$ information.

We then presented another application by analyzing the most probable shape of avalanches of a given size and extension, first for systems 
made of few coupled particles, then in the continuum limit for an elastic line with short-ranged elasticity.
Quantitative results for the optimal shape and the fluctuations around it were obtained and compared to a numerical simulation of the model.

Let us conclude by stressing that, since our formula was obtained in a general setting and contains all the spatial statistics of avalanches, it should be possible to extract from it a variety of new information on their spatial structure
of direct experimental interest. It would also be interesting to compare our results for the shape of avalanches to other models through simulations or experiments, the BFM being the relevant mean-field theory for various more realistic systems.

\bigskip

{\it Acknowledgments:}
We acknowledge support from PSL grant ANR-10-IDEX-0001-02-PSL.
We thank KITP for hospitality and 
support in part by NSF  
Grant No.\ NSF PHY11-25915.

\appendix

\section{Recall of the result for the generating function}\label{appMSR}

For completeness, we recall in this section, the derivation, here in a discrete
setting, of the exact result for the generating function of the BFM (\ref{generatingaval}).
Related derivations 
can be found in \cite{DobrinevskiLeDoussalWiese2011b,LeDoussalWiese2012a}. The original equation of motion, including the quenched noise term $\partial_t F_i(u_{it})$
reads
\begin{equation}
\eta \partial_t \dot{u}_{it} = \sum_{j=1}^N c_{ij} \dot{u}_{jt} - m^2( \dot{u}_{it}-\dot{w}_{it} ) +\partial_t F_i(u_{it}) \ .
\end{equation} 
We use the dynamical field theory formalism \cite{MSR,Jannsen}
which allows to compute the disorder average of any physical observable $O[\dot{u}]$. 
We introduce $N$ response fields $\tilde{u}_{it}$ such that disorder averages can be computed as
\begin{equation}
\overline{O[u]}= \int \mathcal{D}[ \dot{u} , \tilde{u} ] O[u] e^{-S[\dot{u},\tilde{u}]} \ .
\end{equation}
The dynamical action splits into a deterministic, quadratic part and a disorder part: $S[\dot{u},\tilde{u}] = S_0[\dot{u},\tilde{u}]+S_{\rm dis}[\dot{u},\tilde{u}]$, with
\begin{eqnarray}
S_0[\dot{u},\tilde{u}] &&= \sum_{i=1}^N \int_t \tilde{u}_{it} \left( \eta \partial_t \dot{u}_{it}  - \sum_{j=1}^N c_{ij} \dot{u}_{jt} + m^2( \dot{u}_{it}-\dot{w}_{it} ) \right) \nn \\
&&=-\sum_{i=1}^N \int_t m^2  \tilde{u}_{it} \dot{w}_{it}  +\sum_i \int_t \dot{u}_{it} \left( - \eta \partial_t \tilde{u}_{it} - \sum_{j=1}^N  c_{ij}\tilde{u}_{jt}+ m^2 \tilde{u}_{it} \right)
\end{eqnarray}
where in the second line, we made an integration by part assuming $\dot{u}$ vanishes at infinity. The disorder part of the action is
\begin{equation}
S_{\rm dis}[\dot{u},\tilde{u}] = \frac{\sigma}{2} \sum_{i=1}^N \int_{tt'} \tilde{u}_{it} \tilde{u}_{it'} \partial_t \partial_{t'} |u_{it}-u_{it'}| \ ,
\end{equation}
it contains all the correlation of the Gaussian force (\ref{browniancorrel}). As noted in
 \cite{DobrinevskiLeDoussalWiese2011b,LeDoussalWiese2012a}, the action functional can be simplified using the Middleton property recalled in the main text, valid for our setting: $t_2 \geq t_1 \Longleftrightarrow u_{it_2} \geq u_{it_1}$ so that
\begin{equation}
 \partial_t \partial_{t'} |u_{it}-u_{it'}| = \dot{u}_{it} \partial_{t'} sgn(t-t') = - 2 \dot{u}_{it} \delta(t-t') \ .
\end{equation}
This leads to
\begin{equation}
S_{dis}[\dot{u},\tilde{u}] = - \sigma \sum_{i=1}^N \int_{t} \tilde{u}_{it}^2 \dot{u}_{it} \ .
\end{equation}
It is straightforward to check that the replacement $\partial_t F_i(u_{it}) \to \sqrt{2 \sigma \dot{u}_{it}}\xi^i_t$ used in the main text leads to the same action. This shows that both theories are equivalent for this choice of initial conditions. As written, the action is linear in $\dot{u}$: this simplifies the calculation of the generating functional of the velocity field $G[\lambda,w] = \overline{e^{\sum_{i=1}^N \int_t \lambda_{it} \dot{u}_{it}}}$:
\begin{eqnarray}\nn
G[\lambda,w] && =\int \mathcal{D}[ \dot{u} , \tilde{u} ] e^{\sum_{i=1}^N \int_t \lambda_{it} \dot{u}_{it} - S[\dot{u},\tilde{u}]  }  \\
&&= \int \mathcal{D}[\tilde{u} ] e^{m^2 \sum_{i=1}^N \int_t \tilde{u}_{it} \dot{w}_{it} } \prod_{it} \delta\left(  \lambda_{it} + \sigma \tilde{u}_{it}^2 +\eta \partial_t \tilde{u}_{it} +\sum_{j=1}^N  c_{ij}\tilde{u}_{jt}- m^2 \tilde{u}_{it} \right) \nonumber \\
&&= e^{m^2 \sum_{i=1}^N \int_t \tilde{u}_{it}^{\lambda} \dot{w}_i(t)}. \label{generatingresult}
\end{eqnarray}
In the last line, the response field $\tilde{u}_{it}^{\lambda}$ is solution to the ``instanton" equation
\cite{LeDoussalWiese2011a,DobrinevskiLeDoussalWiese2011b,LeDoussalWiese2012a}
\begin{equation}\label{generalinstanton}
\lambda_{it} + \sigma \tilde{u}_{it}^2 +\eta \partial_t \tilde{u}_{it} +\sum_{j=1}^N  c_{ij}\tilde{u}_{jt}- m^2 \tilde{u}_{it}=0 \ .
\end{equation}
It is imposed by the delta functional. Note that this evaluation involves a $w$-independent Jacobian, which equals unity since we have supposed the interface to be at rest and stable for $t \leq 0$, so that if 
$\dot{w}_{it} = 0$ then $\dot{u}_{it} = 0$. 
The above result is thus correctly normalized. Equation (\ref{generalinstanton}) must in general be supplemented by some boundary conditions, depending on the observable (e.g. if $\lambda_{it}=0$ for all $i$ and $t>t_1$, we should also have $\tilde{u}_{it}=0$ for all $i$ and $t>t_1$). Note that a rigorous version (in discrete-time, without path integral) of this result was given in 
\cite{DobrinevskiLeDoussalWiese2011b}.
In the main text we are looking for the statistics of avalanches $S_i$, which is obtained using constant sources $\lambda_{it} = \lambda_i$, and for which one can look for constant solutions $\tilde{u}_{it}= \tilde{u}_i$ of (\ref{generalinstanton}).

\section{Tests of the main formula, computation of moments and numerical checks.} 
\label{appcheckmainformula}

We checked (\ref{mainform}) using two methods: the first one consists in solving exactly the instanton equation for small values of $N$ in an expansion in powers of $c$ for a given elasticity matrix. This gives an approximation of the Laplace transform, which can be inverted to give the joint probability distribution up to a certain order in $c$. This program has been successfully achieved up to $O(c^4)$ for $N=2$,  $O(c^3)$ for $N=3$ and $O(c^2)$ for $N=4$.
The other method consists in numerically computing various moments of the probability distribution, which can then 
be compared to the exact results that use the instanton equation (\ref{instantonaval2}): the cumulants are given by
\begin{eqnarray}
\!\!\!\!\!\!\!\!\!\!\!\!\!\!\!\!\!\! \langle  S_{i_1} \cdots S_{i_n} \rangle^c = \left( \frac{\partial}{ \partial \lambda_{i_1} \cdots  \partial \lambda_{i_n}}  \ln G( \vec \lambda ) \right) _{ \lambda_i = 0} =  \sum_{k=1}^N w_k \left( \frac{\partial v_k}{ \partial \lambda_{i_1} \cdots  \partial \lambda_{i_n}}  \right) _{ v_i = 0} 
\end{eqnarray}
and theses derivatives are numerically computed using  $\frac{ \partial v_i }{ \partial \lambda_j} = J_{ij}^{-1}$ where $J_{ij} = - 2 v_i \delta_{ij} + C_{ij}$, as seen from (\ref{instantonaval2}).

\section{Backward Kolmogorov method for a kick driving}
\label{app-kolmo}

In this section, we provide another verification that (\ref{mainform}) is correct when the system is driven by a kick (i.e.\ $\dot{w}_{it} = w_i \delta(t)$). For simplicity, we directly consider the dimensionless equation of motion
\bea 
\partial_t \dot{u}_{it} && = \sum_{j=1}^N c_{ij} \dot{u}_{jt} -\dot{u}_{it}+\dot{w}_{it}+ \sqrt{2 \dot{u}_{it}}\xi^i_t \nn \\
  && =_{t>0} - \sum_{j=1}^N C_{ij} \dot{u}_j + \sqrt{2 \dot{u}_{it}}\xi^i_t \label{0driving}
\eea
where in the second line we used the definition of $C_{ij}$ (\ref{dimensionlesselast}) and wrote the equation for $t >0$ when $\dot{w}_{it} =0$. For a kick, it is equivalent to consider the equation of motion with $\dot u_{it=0}=0$, 
or to consider the equation without driving for $t>0$ (\ref{0driving}) supplemented with the initial condition $\dot{u}_{i , t=0^+}  = w_i$. The generating function $G$ is still given by $G( \vec \lambda ) = \overline{ e^{ \sum_{i=1}^N \lambda_i \int_{0}^{\infty} \rmd t \dot{u}_{it}}}$. For a kick, we can write it as a conditional expectation value on the process without driving (\ref{0driving}): $G( \vec \lambda ) = \hat{G}( \vec \lambda , \vec w , 0 , \infty)$ where $\hat{G}$ is defined as 
\bea
 \hat{G}(\vec \lambda , \vec w , t_i , t_f) = \mathbb{E} \left(  e^{ \sum_{i=1}^N \lambda_i \int_{t_i}^{t_f} \rmd t \dot{u}_{it}}  \Big| \dot{u}_{it_i} = w_i \right)
\eea
where $\dot{u}_{it}$ evolves according to (\ref{0driving}) for all times and $\mathbb{E}\left(  \dots \Big| \dot{u}_{it_i} = w_i \right)$ denotes the average on the stochastic process without driving (\ref{0driving}) conditioned to the initial condition $\dot{u}_{it_i} = w_i$. We now derive a partial differential equation (PDE) fo $G$, similar to a Backward Kolmogorov equation, using a splitting of $[t_i , t_f]$ into $[t_i, t_i + \delta t] \cup [t_i + \delta t , t_f]$ with $\delta t$ small:
\bea
\hat{G}( \vec \lambda , \vec w , t_i , t_f)&&  = \mathbb{E} \left(  e^{ \sum_{i=1}^N \lambda_i \int_{t_i+\delta t }^{t_f} \rmd t \dot{u}_{it}  + \sum_{i=1}^N \lambda_i \int_{t_i }^{t_i+\delta t} \rmd t \dot{u}_{it} }  \Big| \dot{u}_{it_i} = w_i \right)\nn \\
&& =\mathbb{E} \left(  e^{ \sum_{i=1}^N \lambda_i \int_{t_i+\delta t }^{t_f} \rmd t \dot{u}_{it}  + \sum_{i=1}^N \lambda_i \delta t w_{it} }  \Big| \dot{u}_{it_i} = w_i \right)  + o(\delta t)\label{kolmostep1}
\eea
Where in (\ref{kolmostep1}) we used that $\dot{u}_{it}$ is continuous. The expectation value in (\ref{kolmostep1}) can now be split in two parts. We can first average over the noise for $t \in [t_i , t_i +\delta t]$, with $\delta t$ small, or equivalently on 
the velocity variation $\delta w_i := \dot{u}_{i,t_i+\delta t} - \dot{u}_{i,t_i}= \dot{u}_{i,t_i+\delta t} - w_i$, as obtained from the equation
of motion (\ref{0driving}). Secondly, we average over the noise in $[t_i + \delta t , t_f]$ (these are independent) knowing that the velocity at $t_i + \delta t $ is $\dot{u}_{i,t_i+\delta t} = w_i + \delta w_i$, i.e.
\bea
\!\!\!\!\!\!\!\!\!\!\!\!\!\!\!\!\!\!\!\!\!\!\!\!\!\!\!\!\!\! \hat{G}( \vec \lambda , \vec w , t_i , t_f)&& =\mathbb{E}_{\{\delta w_i\}} \left( \mathbb{E} \left(  e^{ \sum_{i=1}^N \lambda_i \int_{t_i+\delta t }^{t_f} \rmd t \dot{u}_{it}   }\Big| \dot{u}_{i,t_i + \delta t_i} = w_i +\delta w_i \right)  \Big| \dot{u}_{i t_i} =w_i  \right)e^{\sum_{i=1}^N \lambda_i \delta t w_{it}}+ o(\delta t) \nonumber \\
&& = \mathbb{E}_{\{\delta w_i\}} \left(  \hat{G}(\vec \lambda , \vec w + \delta \vec w, t_i + \delta t_i , t_f )  \Big| \dot{u}_{i t_i} =w_i \right) e^{\sum_{i=1}^N \lambda_i \delta t w_{i}} + o(\delta t)
\eea
The average over $\{\delta w_i\}$ can be computed at first order in $\delta t$ using Ito's lemma (we use $\overline{ \delta w_{\alpha} } =- \delta t\sum_{j=1}^N C_{\alpha j} w_j $ and $\overline{ \delta w_{\alpha}^2}=  2 w_{\alpha} \delta t + O(\delta t^2)$). This leads to
\bea \label{kolmostep2}
\!\!\!\!\!\!\!\!\!\!\!\!\!\!\!\!\!\!\!\!\!\!\!\!\!\!\!\!\!\! \hat{G}(\vec \lambda , \vec w , t_i , t_f)&& =\left( \hat{G} + \sum_{\alpha =1}^N  \delta t \left(  \frac{ \partial \hat G}{\partial w_{\alpha} } ( - \sum_{j=1}^N C_{\alpha j} w_j ) + \frac{1}{2}\frac{ \partial^2 \hat G}{\partial w_{\alpha}^2 } (2 w_{\alpha} ) \right)   + \delta t \frac{ \partial \hat G}{\partial t_i }   \right) \\
&& \hspace{2 cm} \times \left( 1 + \sum_{i=1}^N \lambda_i \delta t w_{t} \right) + o(\delta t). \nonumber
\eea
We also expanded the last term at first order in $\delta t$. In the r.h.s. of (\ref{kolmostep2}), all generating functions are taken at the same position $\hat{G}( \vec \lambda , \vec w , t_i , t_f)$. Now the l.h.s. is of order $O(\delta t^0)$ and in the l.h.s., we exactly computed the $O(\delta t)$ term. This shows that the generating function $\hat{G}$ solves the following PDE:
\bea \label{kolmostep3}
-\frac{ \partial \hat G}{\partial t_i }  = \sum_{\alpha =1}^N \left( -   \frac{ \partial \hat G}{\partial w_{\alpha} } \sum_{j=1}^N C_{\alpha j} w_j + \frac{ \partial^2 \hat G}{\partial w_{\alpha}^2 } w_{\alpha} + \lambda_{\alpha} w_{\alpha} \right)
\eea
which is also equal to $\frac{ \partial \hat G}{\partial t_f } $ as a consequence of the time translation invariance of the Brownian motion. The initial condition is $\hat{G}( \vec \lambda , \vec w , t_i , t_i)=1$. 

To study avalanche sizes, we consider the long-time behavior of $\hat{G}$ to obtain $G = \hat{G}( \vec \lambda , \vec w , t_i , \infty)$. In this case we can assume that $\hat{G}$ reached the stationary state, i.e.\ 
\bea
\sum_{\alpha =1}^N \left( -   \frac{ \partial G}{\partial w_{\alpha} } \sum_{j=1}^N C_{\alpha j} w_j + \frac{ \partial^2 G}{\partial w_{\alpha}^2 } w_{\alpha} + \lambda_{\alpha} w_{\alpha} \right) = 0 \ .
\eea
This is automatically satisfied if $G$ is given by (\ref{generatingaval}) and if
the $\tilde u_i$ satisfy the instanton equation (\ref{instantonaval}). This provides 
a connection between the two methods.

An interesting feature of this method is that one can now write a PDE directly for the probability distribution $P(\vec w, \vec S)$ of avalanche sizes in the BFM model following arbitrary (positive) kicks $\dot{w}_{it}=w_i \delta(t)$. This equation reads: 
\bea\label{kolmo}
\sum_{\alpha =1}^N  \left( - \frac{ \partial P}{\partial w_{\alpha} } \sum_{j=1}^N C_{\alpha j} w_j + \frac{ \partial^2 P}{\partial w_{\alpha}^2 } w_{\alpha} - w_{\alpha} \frac{ \partial P}{\partial S_{\alpha} } \right) =0 \ .
\eea
We need to find a solution which satisfies the following boundary condition:
\bea \label{bcP} 
P(\vec w = \vec 0 , \vec S ) = \prod_{i=1}^N \delta(S_i) \ .
\eea
Let us now discuss its solution. Inspired by our result (\ref{mainform}), we make the change of variable $P (\vec w , \vec S) = F( \vec x, \vec S)$ with $\vec x = \vec w -C \cdot \vec S$. The equation for $F$ then takes a very simple form: 
\bea \label{kolmoF}
\sum_{\alpha=1}^N  w_{\alpha}  \left( \frac{ \partial^2 F}{\partial x_{\alpha}^2 } - \frac{ \partial F }{\partial S_{\alpha} } \right) =0
\eea
where $w_{\alpha} = x_{\alpha} +\sum_{j=1}^N  C_{\alpha j}  S_j  $ and we used that $C$ is a 
symmetric matrix.
In this new variables, we write our main result (\ref{mainform}) using the following decomposition:
\bea \label{fdef}
&& F( \vec x, \vec S ) = \det\left(  M_{ij}  \right)_{N \times N} \tilde F(\vec x, \vec S) \quad , \quad  M_{ij} = C_{ij} + \delta_{ij} \frac{x_i}{ S_i}   \\
&& \tilde F( \vec x, \vec S ) =(\frac{1}{2\sqrt{\pi}})^N (\prod_{i=1}^N  S_i ) ^{-\frac{1}{2}} \exp \left(-\frac{1}{4} \sum_{i=1}^N \frac{x_i^2}{ S_i} \right) \ . \label{ftilde}
\eea
This decomposition sheds some light on the structure of (\ref{mainform}), here rewritten as $F$ in (\ref{fdef}): it is simple to see that $\tilde{F}$ defined in (\ref{ftilde}) already solves (\ref{kolmoF}), $\tilde{F}$ can indeed be interpreted as the PDF of the position $x_i$ at "time" $S_i$ of $N$ independent particles diffusing from the origin at time $S_i=0$. However
the result $F= \tilde F$ would not satisfy the boundary conditions (\ref{bcP}). 
We now check that the extra factor $\det (M)$ provides the proper solution.
In order for (\ref{fdef}) to also solve (\ref{kolmoF}), the determinant must verify
\bea
\sum_{\alpha=1}^N w_{\alpha} \left(  \frac{ \partial^2 \det(M)}{\partial x_{\alpha}^2 }  \tilde F + 2 \frac{ \partial \det(M)}{\partial x_{\alpha}}  \frac{ \partial \tilde F}{\partial x_{\alpha}} - \frac{ \partial \det(M)}{\partial S_{\alpha}} \tilde F \right) = 0.
\eea
Using $ \frac{ \partial \tilde F}{\partial x_{\alpha}}  = -\frac{x_{\alpha}}{2 S_{\alpha}} \tilde F $, this implies an equation for $\det(M)$
\bea
\sum_{\alpha=1}^N w_{\alpha} \left(  \frac{ \partial^2 \det(M)}{\partial x_{\alpha}^2 }  -\frac{x_{\alpha}}{ S_{\alpha}}  \frac{ \partial \det(M)}{\partial x_{\alpha}} - \frac{ \partial \det(M)}{\partial S_{\alpha}} \right) = 0.
\eea
The first term $ \frac{ \partial^2 \det(M)}{\partial x_{\alpha}^2 }$ is equal to $0$, since $x_{\alpha}$ only appears in the $\alpha$-th column of $M$. The remaining terms vanishes since $M$ depends on $x_\alpha$ and $S_\alpha$ only through the combination $\frac{x_\alpha}{S_{\alpha}}$. This completes the proof that our result (\ref{mainform}) indeed solves the PDE (\ref{kolmo}). The boundary condition is now satisfied since $P_{\vec w}(\vec S)$ is a continuous PDF on positive variables and we know (see Section \ref{secMain} and \ref{appcheckmainformula}) that $\langle S_i \rangle = \sum_{j=1}^N C^{-1}_{ij} w_j$ vanishes when $w_i \to 0$.

\section{Poisson-Levy process for normalizable jump densities}\label{app-levy}

\paragraph{Center of mass} We already discussed in the main text the infinite divisibility property (\ref{infinitedivisibilityABBM}) of $P_{\sf w}(S)$. Given this property, one would like to interpret an avalanche $S$ as the sum of $n$ iid elementary avalanches $s_i$ with $n$ drawn from a Poisson distribution and $s_i$ drawn from a given distribution (this defines a Poisson-Levy jump process, see e.g.\cite{duchon}). This interpretation is valid at the level of the moments of $P_{\sf w}(S)$  (see (\ref{decomp1})) but we now show that it does not extend to the probability itself. Let us first assume that the jump density $\rho$ appearing in (\ref{decomp1}) 
is normalizable (see also the discussion in \cite{LeDoussalWiese2011b}, Appendix J). Then one can write $\rho(s) = \rho_0 p(s)$ with $p$ a regular function normalized to unity $\int \rmd s p(s) =1$ and $\rho_0$ the density of avalanches; i.e.\ the mean number of quasi-static avalanches occuring in response to the total driving ${\sf w}$ is $\rho_0 {\sf w}$.  
Using the following identity:
\bea \nn
\fl && \int \rmd s_1 \cdots \rmd s_n ( e^{\lambda s_1} -1) \cdots ( e^{\lambda s_N} -1)\rho(s_1) \cdots \rho(s_n) \\
&& =  \sum_{m=0}^n \frac{(\rho_0 \sf{w})^m}{m!}  \frac{(-\rho_0 \sf{w})^{n-m}}{(n-m)!}\int \rmd s_1 \cdots \rmd s_{m} e^{\lambda (s_1+ \cdots + s_{m})} p(s_1) \cdots p(s_{m}) 
\eea
 (\ref{decomp1}) can be rewritten as (performing the sum over $n>m$):
\bea
\int \rmd S e^{\lambda S} P_{\sf w}(S) = \sum_{m=0} ^{\infty}  \frac{(\rho_0 \sf{w})^m}{m!} e^{-\rho_0 \sf{w}} \int \rmd s_1 \cdots \rmd s_m e^{\lambda (s_1 +  \cdots + s_m)} p(s_1) \cdots p(s_m) \ .
\eea
This leads to a formula for the
probability, $P_{\sf w}(S) =  \sum_{m =0} ^{\infty}  \frac{(\rho_0 \sf{w}) )^m}{m !} e^{-\rho_0 \sf{w} } (p*)^m(S)$. 
Here $ (p*)^m$ denotes $m$ convolutions of $p$ with itself, making the interpretation in terms of a Poisson jump process transparent. 
One can define the ``complete" avalanche-size density as
\bea
\tilde \rho (S ) =\frac{d P_{\sf w}(S)}{d \sf w} |_{\sf w=0} = - \rho_{0} \delta(S) + \rho(S) \ .
\eea
Where here the first equality holds in the sense of distributions. This total density appears as the sum of the regular density $\rho(S)$ (defined in the main text) and of a delta singularity that accounts for the finite probability that the interface does not jump. As a consequence, $\frac{d G_{\sf w}(\lambda)}{d \sf w} |_{\sf w=0} =Z(\lambda)=\int \rmd S \tilde e^{\lambda S} \tilde \rho(S) = \int \rmd S \left( e^{\lambda S} -1 \right)\rho(S)$. For the ABBM model, the scale invariance of the Brownian motion leads to an accumulation of small avalanches of arbitrary small sizes, leading to $\rho_0 =\infty$ (in particular for any $\sf{w}>0$, $P_{\sf w}(S=0) = e^{- \rho_0 {\sf w}} \to 0$) and one can not define $\tilde \rho$. The formula $\frac{d G_{\sf w}(\lambda)}{d \sf w} |_{\sf w=0} = \int \rmd S \left( e^{\lambda S} -1 \right)\rho(S)$ is however still valid and allowed us to prove (\ref{decomp1}).

\paragraph{Levy Process for the interface}
The generalization to the interface is immediate: in this case, the LT of  $P_{\vec w}(\vec S)$ reads
\bea
\int \rmd ^N \vec S  e^{\vec \lambda \cdot \vec S } P_{\vec w}(\vec S)  =  e^{\vec w \cdot \vec v}  = \sum_{n=0}^{\infty} \sum_{(i_1 ,\dots , i_n )} \frac{w_{i_1} \dots  w_{i_n}}{n!} v_{i_1}, \cdots v_{i_n}
\eea
where the second sum is for all $(i_1 ,\dots , i_n ) \in \{1 , \dots , N \}^n$ and the $v_i$ variables are functions of $\vec \lambda$ solutions of (\ref{instantonaval2}). Using our conjecture (\ref{vjconj}), we obtain
\bea
\!\!\!\!\!\!\!\!\!\!\!\!\!\!\!\!\!\!\!\!\!\!\!\! \int \rmd^N \vec S  e^{\vec \lambda \cdot \vec S } P_{\vec w}(\vec S) = \sum_{n=0}^{\infty} \sum_{(i_1 ,\dots , i_n )}\frac{w_{i_1} \dots  w_{i_n}}{n!}  \prod_{l=1}^n  \rmd^N \vec s_{i_{l}}( e^{ \lambda \vec s_{i_l} } -1)  \rho_{i_1} (\vec s_{i_1}) \dots \rho_{i_n} (\vec s_{i_n}) 
\eea
which is the multidimensional generalization of (\ref{decomp1}) and shows that the densities $\rho_{j}(\vec S)$ entirely control the moments of $P_{\vec w}(\vec S)$. It is also in agreement with the interpretation of an avalanche $\vec S$ as a superposition of independent avalanches, as already discussed in the main text.

\section{Details on the fully connected model}\label{app-fully}

Here we detail the calculations leading to the results of Section \ref{secfullyc}, and give some results for the fully-connected model driven by a single site.
\paragraph{Marginals distributions for uniform driving}
For uniform driving, the matrix $C$ and $M$ entering in (\ref{mainform}) admit the following simple expressions, allowing us to evaluate $\det M$ in a concise way:
\bea
&& C_{ij} = (1+c) \delta_{ij} - \frac{c}{N} \quad  , \quad  M_{ij} = \delta_{ij} \frac{1}{S_i} (w + c S/N) - \frac{c}{N} \nonumber \\
&& \det M =  w (w+ c S/N)^{N-1}  \prod_{i=1}^N \frac{1}{S_i} \quad , \quad  S = \sum_{k=1}^N S_k
\eea 
This leads to (\ref{pfully}). Various marginals of this PDF can be computed by noting that the Laplace transform of $p_{w , S/N}(s)$ entering into (\ref{pfully}) reads
\bea
\int_0^\infty \rmd s p_{w , S/N}(s) e^{-p s} = e^{ \frac{1}{2} (1+c) (w + c S/N)(1 - \sqrt{1 + \frac{4 p}{(1+c)^2})}}.
\eea
We write the joint PDF of local and total size as
\bea
P(\vec S, S) = \delta \left( S - \sum_{i=1}^N S_i \right) P( \vec S).
\eea
For any $1\leq m \leq N-1$, the marginal $P(\{S_1, \dots ,S_m\} , S)$ can be computed as
\bea \label{fullycmarg}
\!\!\!\!\!\!\!\!\!\!\!\!\!\!\!\!\!\!\!\!\!\!\!\!\!\!\!\!\!\! P( \{S_1 , \dots , S_m \} , S ) && = \frac{w}{ w + c S/N}\prod_{i=1}^m p_{w , S/N} (S_i) \int_{\sum_{i=m+1}^N S_i = S- \sum_{i=1}^m S_i} \prod_{i=m+1}^N p_{w , S/N}(S_i) \nonumber \\
&& =  \frac{w}{ w + c S/N} \prod_{i=1}^m p_{w , S/N} (S_i)  p_{(N-m)w , (N-m)S/N} (S- \sum_{i=1}^m S_i).
\eea
Where the multiple convolution of $p_{w , S/N}(s)$ has been easily calculated as a consequence of the simple structure of it's Laplace transform. In particular, this leads to the formula (\ref{fullyjoint1}) of the main text. 

\paragraph{Single-site driving} Taking $w_i$ to be non-uniform breaks the permutation invariance $i \leftrightarrow j$ of the problem, making the computation more complicated than for the uniform case. Another solvable case is $w_i =0$ for $i \neq 1$, for which the PDF (\ref{mainform}) takes the form
\bea
P(\vec S) =  \frac{S_1 w_1}{S(w_1 + c S/N)} p_{w_1, S/N} (S_1) \prod_{j=2}^N p_{0, S/N} (S_j).
\eea
The computation of marginals involving an integration over some $S_j$ for $j>1$ is identical to the uniform driving case and leads, for $1 \leq m \leq N-1$, to
\bea
\fl P( \{S_1 , \dots , S_m \} , S ) =  \frac{S_1 w_1}{S(w_1 + c S/N)} p_{w_1, S/N} (S_1) \prod_{j=2}^m p_{0, S/N} (S_j) p_{0 , (N-m)S/N} (S- \sum_{i=1}^m S_i)
\eea
In particular, we obtain 
\bea \nn \label{fullyjoint2}
\! \! \! \! \! \!  \! \! \! \! \! \!  \! \! \! \! \! \!  P(S_1 , S) =&& \frac{w_1}{2 \sqrt{\pi} S_1^\frac{3}{2}}(N-1) \frac{c S_1/N}{2 \sqrt{\pi} (S -S_1)^{3/2}} \exp\left(-  \frac{ \left(w_1 + c S/N - (1+c) S_1  \right)^2}{4 S_1} \right)   \\ 
&& \times \exp\left(- \frac{ \left((N-1)\left(c S/N \right) - (1+c) (S -S_1)  \right)^2}{4 (S-S_1)} \right)  \theta(S-S_1)  .\nonumber \\
\eea
In this case $S = \sum_{i=1}^N S_i$ is typically of order 1 and is distributed according to
\bea
P(S)=\frac{ w_1 }{2 \sqrt{\pi} S^{\frac{3}{2}}} \exp\left( -\frac{ (S-  w_1)^2}{4 S } \right) .
\eea
The large-$N$ limit now exhibits a single non-trivial regime, with $w_1=O(N^0)$, and for which (\ref{fullyjoint2}) admits the limit
\bea
\! \! \! \! \! \! \! \! \! \! \! \!  \! \! \! \! \! \!  P(S_1 , S) = &&\frac{w_1}{2 \sqrt{\pi} S_1^\frac{3}{2}} \frac{c S_1}{2 \sqrt{\pi} (S -S_1)^{3/2}} 
\exp\left(-  \frac{ \left(w_1 - (1+c) S_1  \right)^2}{4 S_1} \right)   \nn\\ 
&& \times \exp\left(- \frac{ \left( c S - (1+c) (S -S_1)  \right)^2}{4 (S-S_1)} \right)  \theta(S-S_1).  
\eea 
Remarkably, in this case one can even integrate over the total size to find the marginal PDF $P(S_1)$ in the large-$N$ limit,
\bea
\! \! \! \! \! \!  && \! \! \! \! \! \!  \! \! \! \! \! \!  P(S_1) = \frac{w_1}{2 \sqrt{\pi} S_1^\frac{3}{2}} 
\exp\left(-  \frac{ \left(w_1 - (1+c) S_1  \right)^2}{4 S_1} \right) .
\eea 
In agreement with the physical intuition, this is the ABBM result for a particle with driving $m^2(w_1 - u)$ 
and $c(\bar S-u)$, as discussed above, and $\bar S=0$, since the center of mass has not
moved appreciably.

\section{Shape for small $N$ at finite driving} \label{app-smallN}
Here we briefly discuss what becomes of the shape transition observed in the quasi-static PDF of avalanche shape at fixed total size $S$ of the linear chain with PBCs  (see Section \ref{discrete}) when one is interested in the full PDF for finite $w_i =w$ as given in (\ref{probashape}). For $N=2$ and $w <  \frac{3}{16 c}$, there is now an additional regime with two transitions instead of one:
\begin{itemize}
\item{$S < \frac{-8 c w-\sqrt{3} \sqrt{3-16 c w}+3}{8 c^2}$ : the distribution of $s$ is peaked around $\frac{1}{2}$.}
\item{$ \frac{-8 c w-\sqrt{3} \sqrt{3-16 c w}+3}{8 c^2} < S <\frac{-8 c w+\sqrt{3} \sqrt{3-16 c w}+3}{8 c^2}$: the distribution possesses two symmetric maxima around $s = \frac{1}{2}$.}
\item{$S > \frac{-8 c w+\sqrt{3} \sqrt{3-16 c w}+3}{8 c^2}$, one retrieves a single maximum at $s=\frac{1}{2}$.}
\end{itemize}
The first regime is new, and was not captured by the study of $\rho$. For small $w \to 0$ it corresponds to avalanches smaller than the lower-scale cutoff $S < \frac{4}{3} w^2$, which are not described by $\rho$ as we know from Section \ref{quasi}. In this regime, the fact that the saddle-point again corresponds to uniform avalanches with $s=1/2$ is not a consequence of elasticity (as noted in Section \ref{quasi}, local avalanche sizes are even independent in this limit), but is related to the fast decay of $p_0(s)$ at its lower cutoff (see Section \ref{quasi}).
For larger $w>\frac{3}{16 c}$, the intermediate regime disappears, and the most probable avalanches are homogeneously distributed. Indeed, as $w$ increases, the motion of the interface becomes mostly deterministic and the remaining fluctuations become negligible.

The case $N=3$ is identical. For $w<\frac{1}{4 c}$ the finite $w$ probability distribution exhibits the same three different regimes with boundaries $0, \frac{1 - 2 c w - \sqrt{1 - 4 c w}}{2c^2}$ and $ \frac{1 - 2 c w + \sqrt{1 - 4 c w}}{2c^2}$. The interpretation is identical to $N=2$.

\section{Stability of infinite, uniform avalanches.} \label{appuniform}
In this appendix, we compute the value $S_c(N)$ such that avalanches uniformly distributed over all the system, and of total size $S>S_c(N)$ are stable. We do this for the fully-connected model and for the linear chain with PBC s, for which uniform avalanches uniformly distributed are always an extremum of the quasi-static density $\rho$ (for uniform driving $f_i = 1$). As such, $S_c(N)$ is the value of $S$ above which all the eigenvalues of the hessian of the quasi-static distribution at this uniform saddle-point are negative. Since this saddle-point and the elasticity matrix are translationally invariant, the Hessian of the logarithm of the probability at the saddle point is a circular matrix given by
\bea
H_{\alpha \beta} = \frac{ \partial^2 \log \rho(\vec s| S)}{ \partial s_{\alpha} \partial s_{\beta} }|_{ s_i = s } = - \frac{S}{2 s}(c^2)_{\alpha \beta} +\frac{1}{2 s^2} \delta_{\alpha \beta} +  h_{\alpha \beta}.
\eea
$c$ is the elasticity matrix of the model (here $m^2=1$), $s=1/N$ is the uniform local avalanche size at the saddle-point and $h_{\alpha \beta}$ depends on the chosen model as $h_{\alpha  \beta} = - \frac{4}{N^2 s^2 } + \frac{1}{ Ns^2} ( 4 \delta_{\alpha \beta} + \delta_{\alpha, \beta-1}+\delta_{\alpha, \beta+1})$ for the linear chain with periodic boundary conditions, and $h_{ \alpha  \beta} = - \frac{(N-2)}{ (N s)^2} + \frac{1}{s^2} \delta_{\alpha \beta}$ for the fully connected one. The eigenvalues of these matrices can be computed using a discrete Fourier transformation, showing that they are indexed by a wave-vector $q = \frac{2 \pi k}{N} $ with $k = 1, ... , N-1$. The $q=k=0$ mode does not intervene since it corresponds to a uniform displacement of the interface, which is forbidden by the fact that we work at fixed $S$: $\sum_{i} d s_i =0$. The eigenvalues of the Hessian are all identical for the fully-connected model: $\lambda_{f.c.} = -\frac{S}{2 s } c^2 + \frac{1}{2 s^2} + \frac{1}{s^2}$. For the linear model they are given by   $\lambda_q = - \frac{2 S}{s} [1 -\cos(q) ]^2 + \frac{1}{2 s^2} + \frac{4}{N s^2} [ 4 + 2 \cos(q)]$. In the latter case, the most unstable mode is $q = \frac{2 \pi }{N}$, leading to the following critical values
\bea
S_c^{\rm fc} (N ) && = \frac{3 N}{c^2}, \\
S_c^{\rm PBC }(N)  && = \frac{N}{2  c^2 (1 - \cos( \frac{2 \pi}{N} ))^2} ( \frac{1}{2} + \frac{1}{N} (4 +2 \cos(\frac{2 \pi}{N}) ) ) \nn \\
 && \simeq_{N \to \infty} \frac{1}{16 c^2 \pi^4} (N^5 + 12 N^4  + O(N^3) ) \ . 
\eea

\section{Continuum limit}\label{appcontinuum1}
Here we detail the scaling that allows to find the probability distribution of the dimensionless continuum avalanches $P[S_x]$ knowing the probability distribution of the discrete case $P(\vec S)$. We denote for clarity the continuum field as $u_t(x)$, $x\in [0,L]$, and its $N$-point discretization as $u_{it} = u_t( i \frac{L}{N})$. We will add indices $c$ and $d$ to distinguish between physical quantities of the continuum and discrete models. An easy way to ensure that the statistic of the discrete case corresponds to the statistic of the continuum one is to compare the different terms in the dynamical action (see \ref{appMSR}) :
\begin{itemize}
\item{The disorder term:} $\sum_{i=1}^N \int_t \sigma_d \tilde{u}_{it}^2 \dot{u}_{it}   \equiv \int_0^L \rmd x \int_t \sigma_c \tilde{u}_{t}(x)^2 \dot{u}_{t}(x) \simeq \sum_{i=1}^N \frac{L}{N} \sigma_c \int_t \tilde{u}_{t}(i \frac{L}{N})^2 \dot{u}_{t}(i \frac{L}{N}) $
\item{The elastic term:} $\sum_{i=1}^N \int_t \tilde{u}_{it} c_d( \dot{u}_{i+1t}-2\dot{u}_{it}+\dot{u}_{i-1t} )  \equiv \int_0^L \rmd x \int_t  \tilde{u}_{t}(x) c_c \Delta{u}_{t}(x) \simeq \sum_{i=1}^N \frac{L}{N} \int_t \tilde{u}_{t}(i \frac{L}{N}) 
c_c \frac{\dot{u}_{t}((i+1) \frac{L}{N}) -2 \dot{u}_{t}(i \frac{L}{N}) + \dot{u}_{t}((i-1) \frac{L}{N})}{\frac{L^2}{N^2}}$
\item{The driving term:} $\sum_{i=1}^N \int_t m_d^2 \tilde{u}_{it} \dot{w}_{it}   \equiv \int_0^L \rmd x \int_t m_c^2 \tilde{u}_{t}(x) \dot{w}_{t}(x) \simeq \sum_{i=1}^N \frac{L}{N} m_c^2 \int_t \tilde{u}_{t}(i \frac{L}{N}) \tilde{w}_{t}(i \frac{L}{N}) $
\end{itemize}

This indicates that the quantity of the discrete model should be $m_d^2= \frac{L}{N} m_c^2$, $c_d = 
\frac{N}{L} c_c$ and $\sigma_d = \frac{L}{N} \sigma_c$. In particular, 
the rescaled quantities which appear in the text, in the formula for the 
dimensionless discrete distributions are $\frac{c_d}{m_d^2} = \frac{N^2}{L^2} \frac{c_c}{m_c^2}$  
and $S_m^d = \frac{N}{L} S_m^c$. Note that we will choose everywhere in the main text $c_c=1$. 
This implies that the probability distribution of the dimensionless rescaled continuum 
avalanches denoted by $P_c$ is given in terms of its discrete analog $P \equiv P_d$ 
given in (\ref{mainform}) as (introducing the explicit dependence in the driving):
\begin{equation}
P_c[S(x),w(x)] = \lim_{N \to \infty} \left( \frac{L}{N} \right) ^N P_d \left( \frac{L}{N} \vec S  , \frac{L}{N} \vec w \right)
\end{equation}
where here $\vec S = (S(Li/N))_{i=1,\cdots,N}$ and $\vec w = (w(Li/N))_{i=1,\cdots,N}$. 
This leads to the formula of the main text. 
Note also that for $\eta$-dependent observables, one should choose $\eta_d = \frac{L}{N} \eta_c$.

\section{Optimal shape in the discrete model}\label{appshapedisc}

Here we compare the results on the continuum optimal shape with the discrete case. This is not only a consistency check, but also allows us to compare the results of the optimization when we include boundary conditions, and to investigate the stability of the shape. 
We choose to work on the discrete model with an elastic coefficient set to unity, which corresponds to a $N$-point approximation of the continuum model with a line of length $L=N$, i.e.\ the index $i$ of the discrete model is the coordinate of the continuum line (see \ref{appcontinuum1}).  In the continuum, the optimal reduced shape $s_0$ is obtained for total size $S$ and extension $\ell$ fixed, and contains all the probability when $S/\ell^4 \gg1$. To compare this result with the discrete model we used two different optimization procedures on the discrete probability. We always impose the total size $S$ and optimize on the shape variables $s_i = S_i/S$ with
\begin{enumerate}
\item{ either the two central points tuned to coincide with the optimal continuum result: we note $n_{mid}$ the integer part of $N/2$ and impose $s_{n_{mid}} = s_{n_{mid}+1} = \frac{1}{\ell} s_0(0.5 / \ell)$.}
\item{ either $N -l$ successive shape variables fixed to be small (below we use $s_i = 10^{-5}$ )}
\end{enumerate}
Procedure (i) is an indirect way to impose the extension by imposing that the avalanche shape is peaked around some region, whereas procedure (ii) is closer to the continuum setting where we directly imposed the finite extension. In both cases we impose $S \gg \ell^4 $ to obtain a true maximum. The optimal shape is always found to be symmetric, which allows us to impose this condition to study reasonably large $N$. The result of the optimization is then compared with the prediction from the continuum theory: $s_i = \frac{S(x=i)}{S} =_{S\gg\ell^4} \frac{1}{\ell} s_0( i /\ell)$. One can then
\begin{itemize}
\item Verify that the optimization on $\rho$ (including boundary conditions) or ${\cal H}$ alone (defined in the continuum in (\ref{defH})) give the same results. It is already obvious for $\ell \ll N$ and Figure \ref{rhovsh} explicitly shows that it is always true for $S \gg \ell^4$, even if $\ell \simeq N$. This validate the hypothesis made in the continuum that boundary conditions do not play a role for large $S/\ell^4$.
\begin{figure}
\centerline{
  \fig{.33\textwidth}{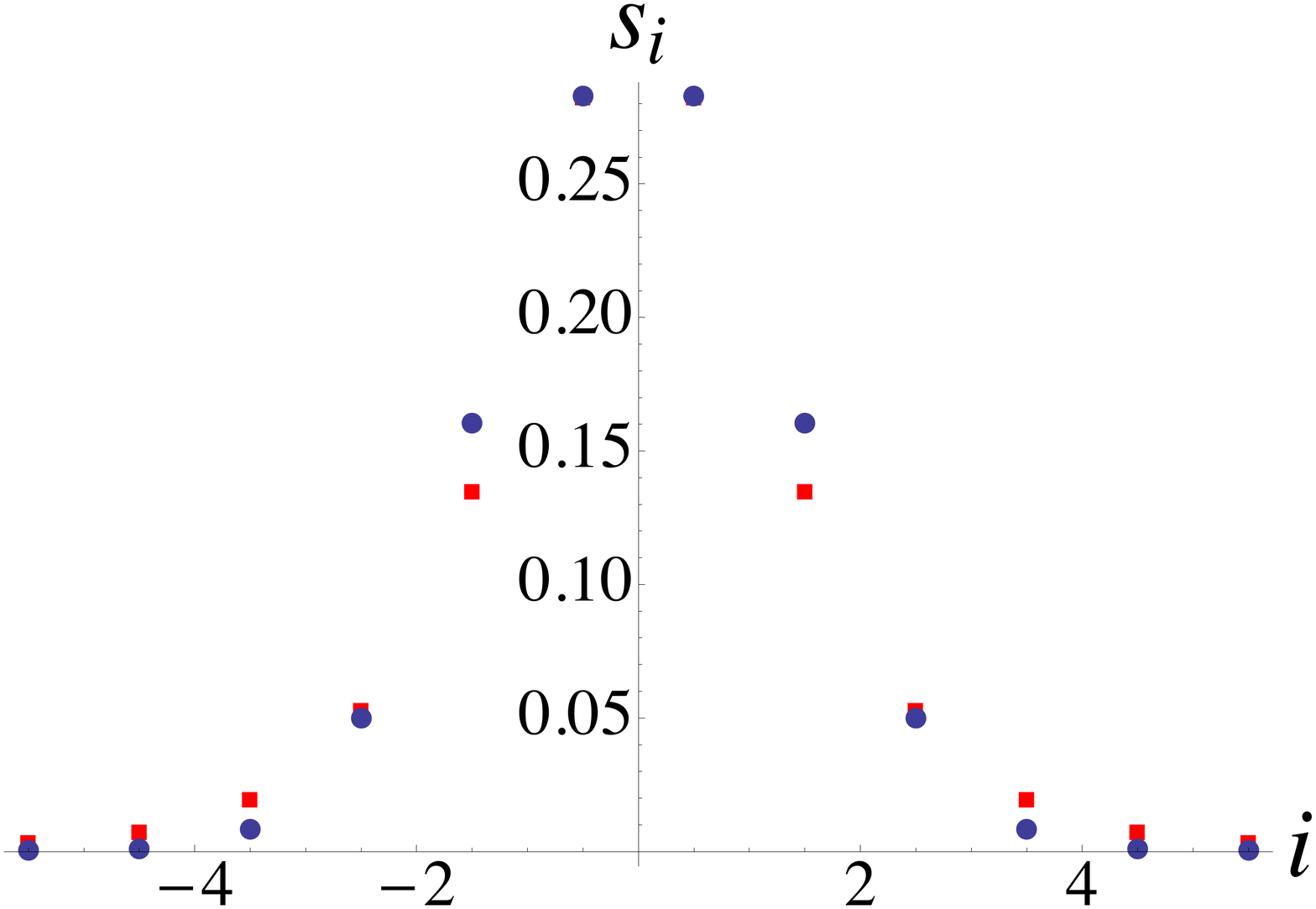}
   \fig{.33\textwidth}{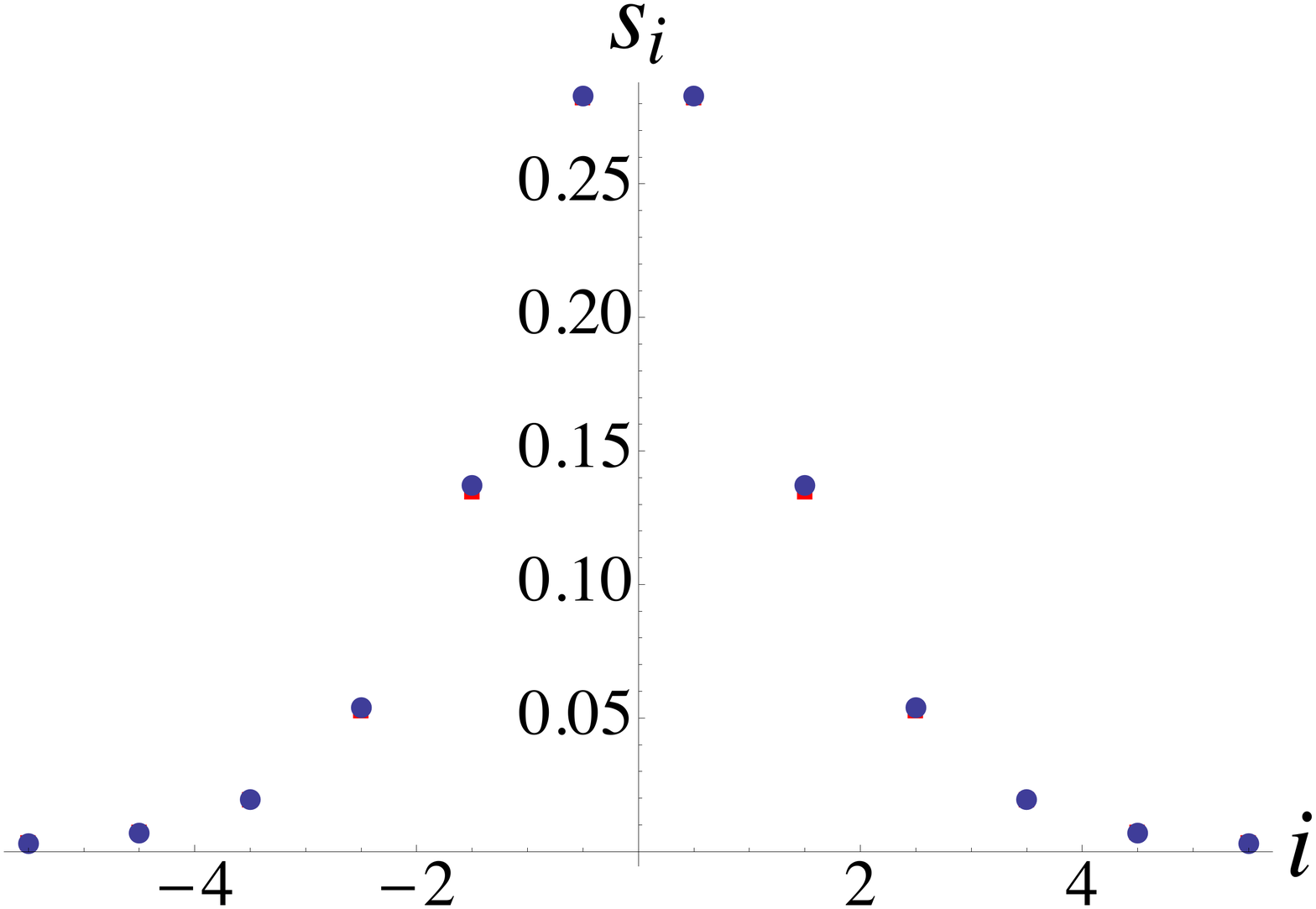}
    \fig{.33\textwidth}{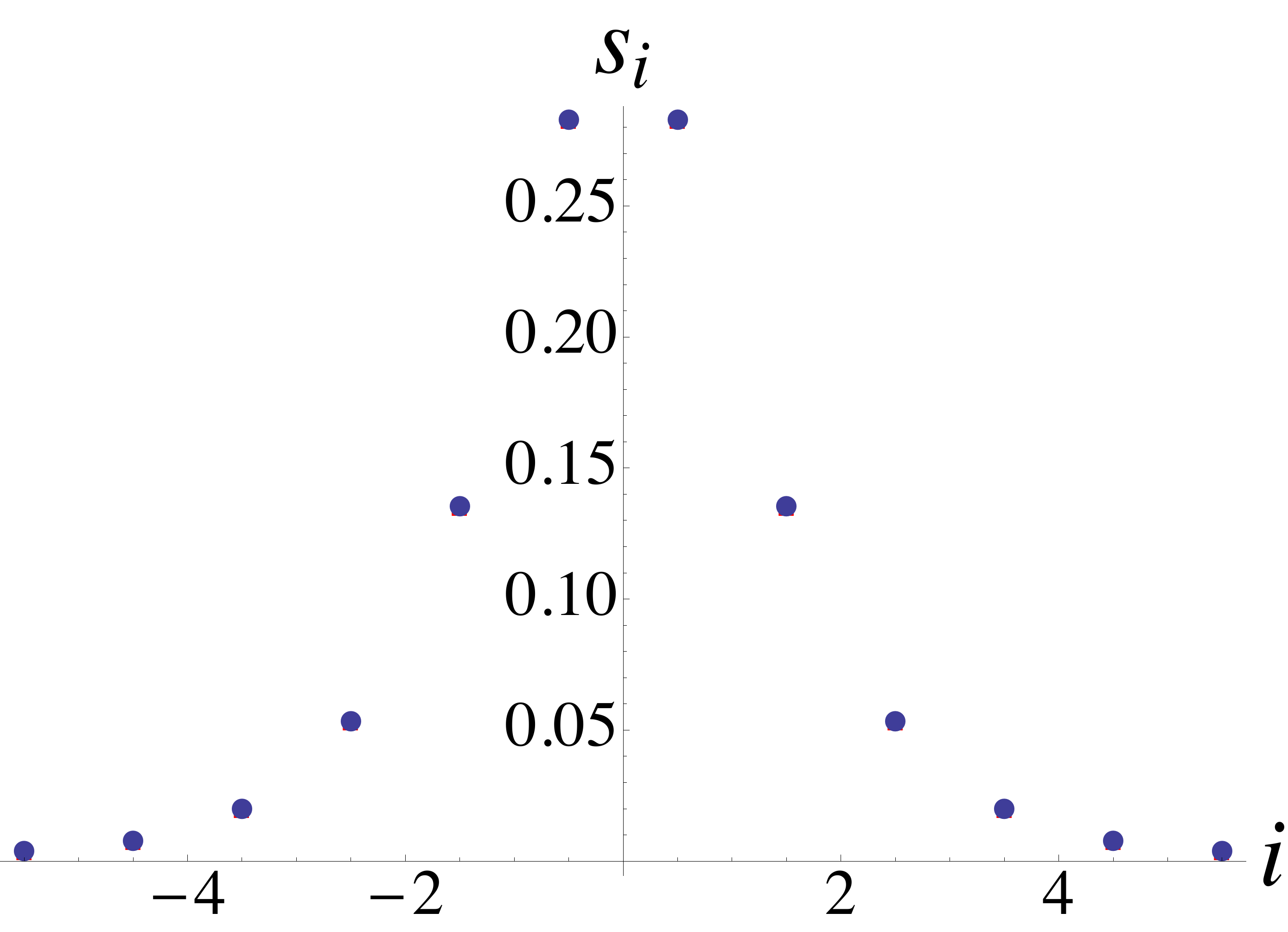}}
 \caption{Comparison between the most probable shape of length $\ell =10$ with $N=12$ computed using optimization on ${\cal H}$ (blue dots) or $\rho$ (red dots), using procedure (i),  and for different total sizes $S$ from left to right: $\frac{S}{\ell^4} = 10^{-2} , 10^{-1} , 1$. The influence of boundary conditions quickly decreases as $S/\ell^4$ is increased.}
\label{rhovsh}
\end{figure}
\item Using an optimization on ${\cal H}$, we can verify that the discrete optimal shape coincides with the continuum one. The results are shown in Figure \ref{discretevscontinuum}. One can see that, apart from some discretization artefacts, procedure (ii) give results in agreement with the continuum result. On the other hand, procedure (i) leads to a shape with an effectively larger extension. This is in agreement with the idea that the property that avalanches have a strictly finite extension is only a feature of the continuum limit, as explained in Section \ref{sec-shape}, and is coherent with the idea that procedure (i) only imposes a ``characteristic" extension in the discrete setting.
\begin{figure}
\centerline{
  \fig{.4\textwidth}{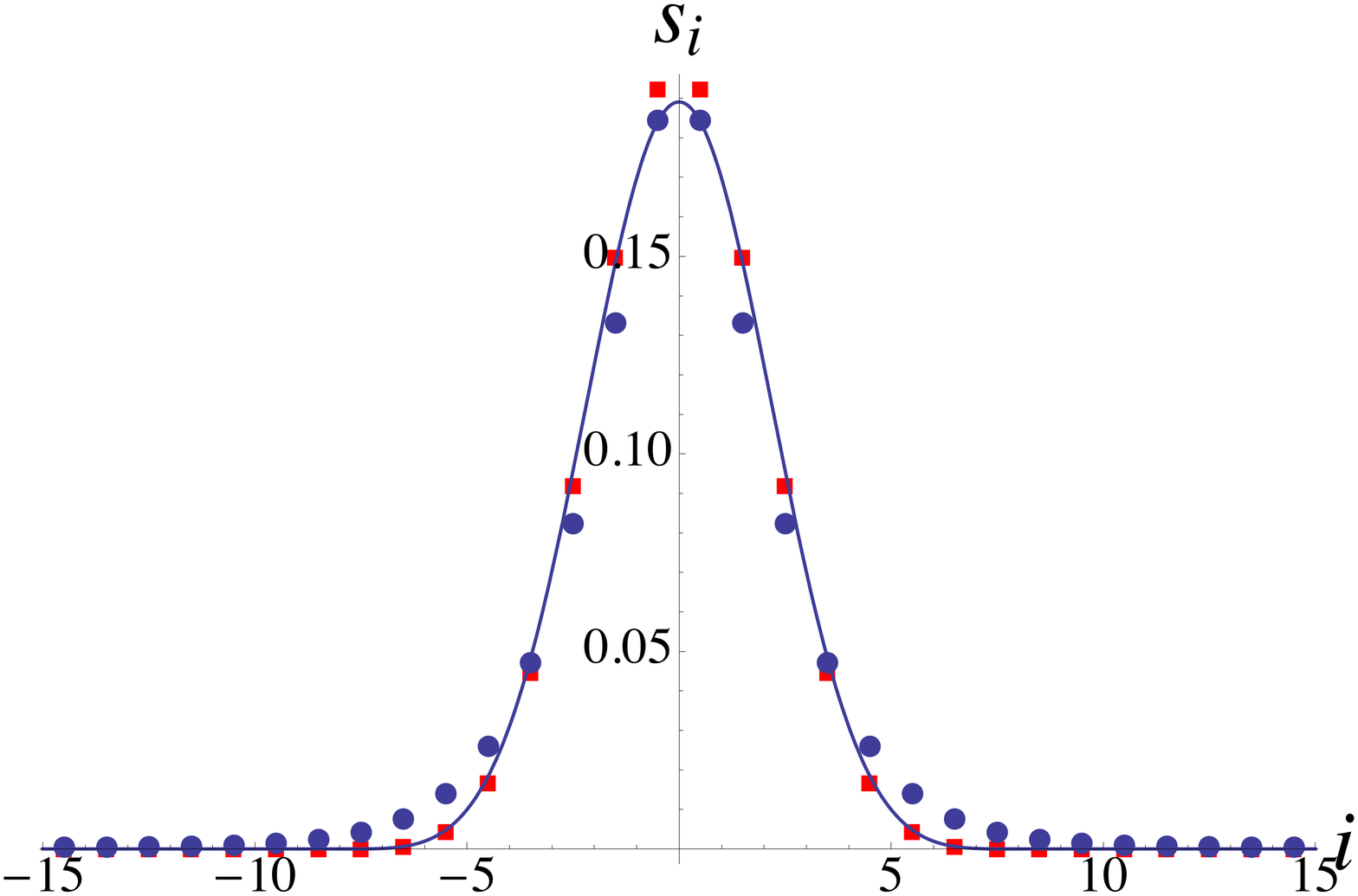}
   \fig{.4\textwidth}{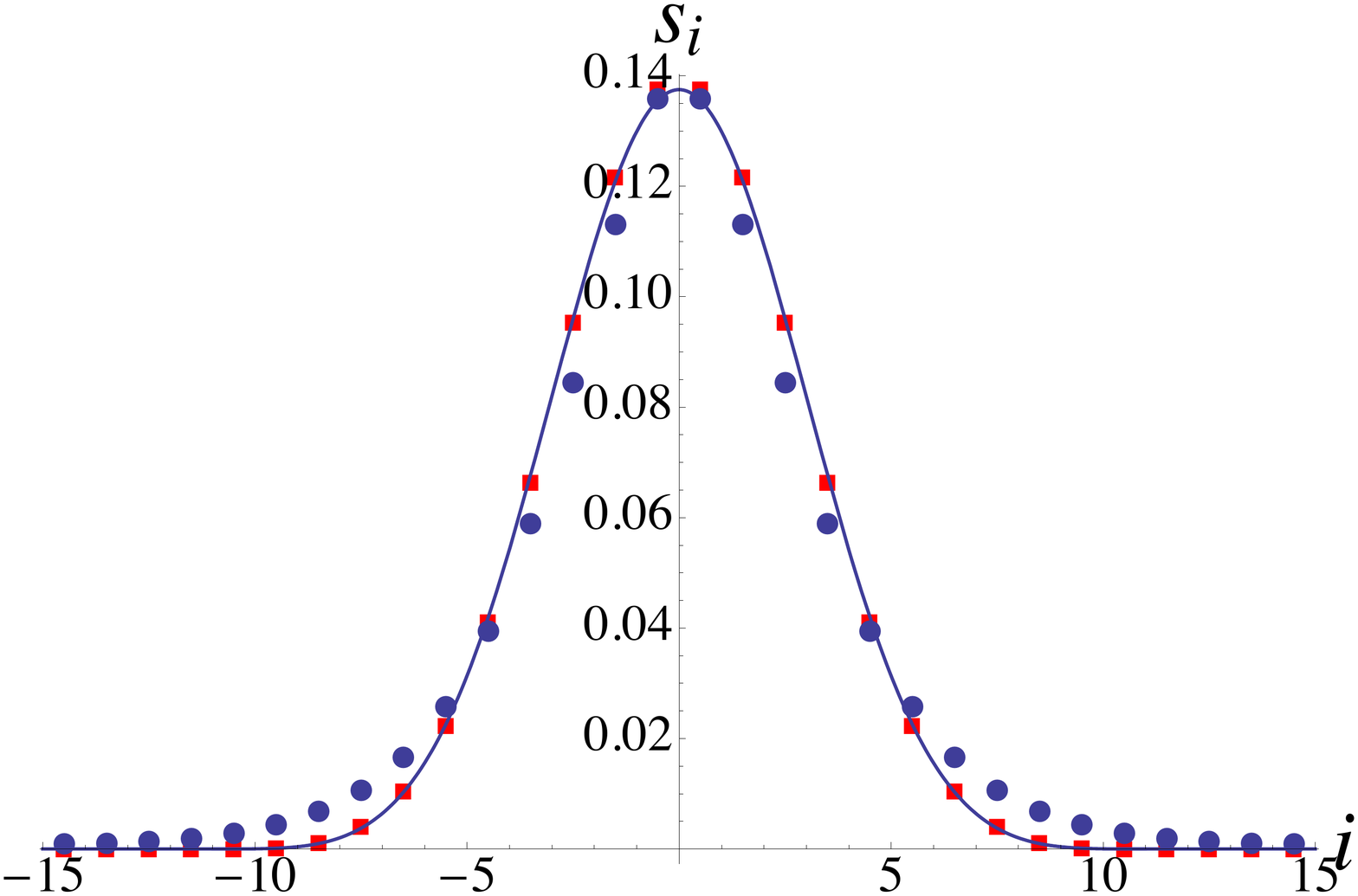}}
 \caption{Most probable shape in the discrete model obtained using numerical optimization on ${\cal H}$ with procedure (i) (blue dots) or procedure (ii) (red square) with $N=30$ and $\ell=16$ (left) or $\ell=22$ (right), compared to the continuum saddle-point prediction $s_0(x/\ell)/\ell$ (straight line).}
\label{discretevscontinuum}
\end{figure}
\item Finally, we can study the behavior of the maximum eigenvalue $\lambda_{max}$ of the Hessian of the discrete Hamiltonian ${\cal H}$ at the most probable shape (since the eigenvalues are negative it is the maximum one that is the closest to $0$ and that controls the stability of the saddle-point) using procedure (i). The behavior of the eigenvalues of the Hessian with $S$ is trivial: since $S$ can be factorized in front of the Hamiltonian, they are proportional to $S$. However, in the discrete case, there is no way to see the scaling $\frac{1}{\ell^4}$ emerge from the Hamiltonian. Still, we clearly numerically find (see Figure \ref{lambdamax}) that $\lambda_{max}$ scales with $1/\ell^4$ for $\ell \to 0$. This thus provides an alternative verification that the saddle-point is  stable, and that it's stability is controlled by $S/\ell^4\gg1$.
\begin{figure}
\centerline{
  \fig{8.5cm}{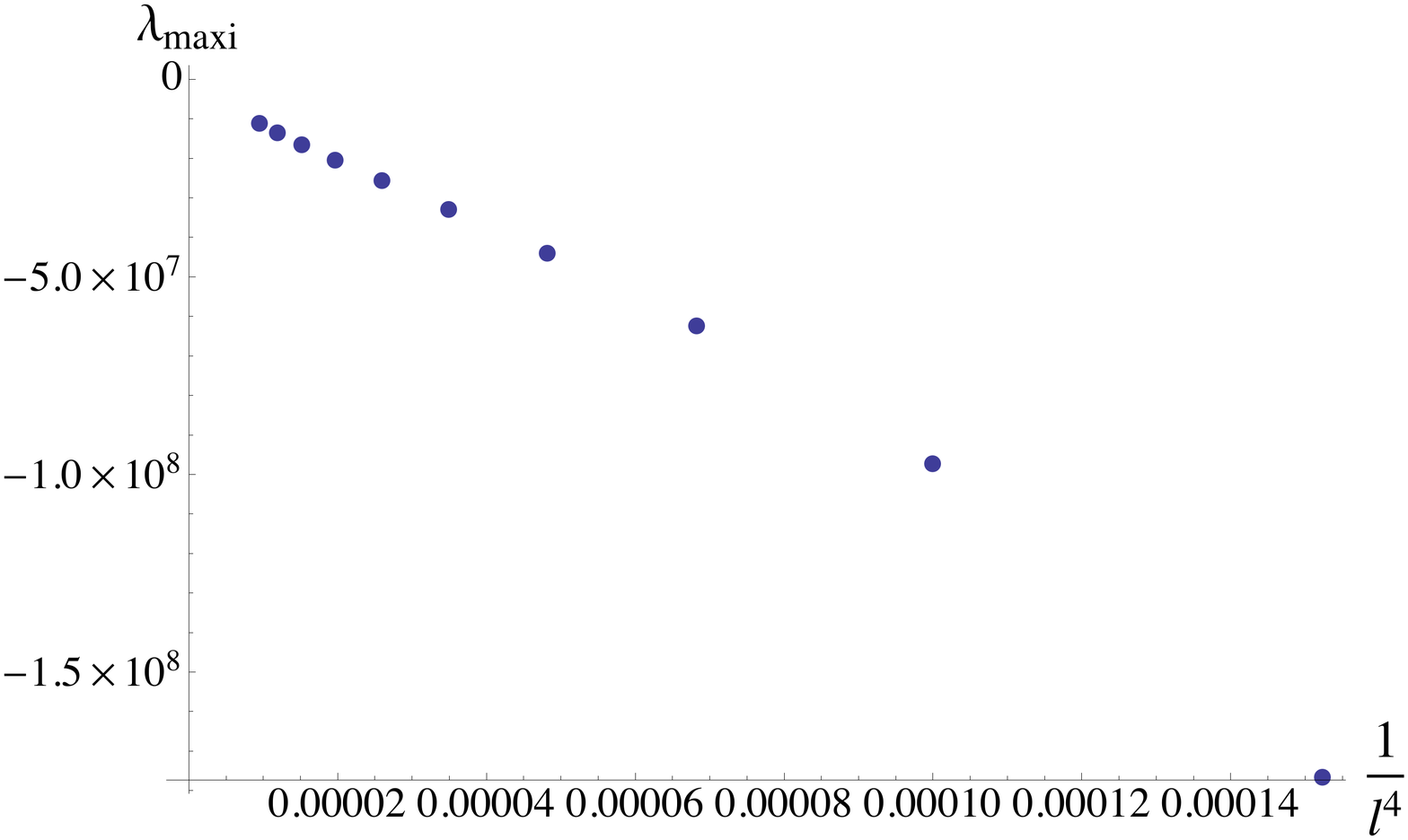}}
 \caption{Maximum  eigenvalue of the hessian of the hamiltonian at the numerical optimum as a function of $\frac{1}{\ell^4}$ for large, fixed $S$ with procedure (i). 
  }
\label{lambdamax}
\end{figure}

\end{itemize}

\end{document}